%% file: 2f.tex
\documentclass[12pt,epsfig,twoside,a4paper]{article}
\usepackage[dvips]{graphicx}

\usepackage{amssymb,rotating}
\newcommand{\lblcaption}[2]{\caption{#2\label{#1}}}

\newcommand{\BFIG}     {\begin{figure}\centering}
\newcommand{\EFIG}     {\end{figure}}

\newcommand{\mr}[1] {\mbox {${\mathrm {#1}}$}}

\newcommand{\ee}       {${\mathrm e}^+{\mathrm e}^-$}
\newcommand{\mumu}     {${\mathrm {\mu^+\mu^-}}$}
\newcommand{\tautau}   {${\mathrm {\tau^+\tau^-}}$}

\newcommand{\qq}       {${\mathrm q} \bar{\mathrm q}$}
\newcommand{\cc}       {${\mathrm c} \bar{\mathrm c}$}
\newcommand{\bb}       {$\mathrm{b}\bar\mathrm{b}$}

\newcommand{\pp}       {${\mathrm p} \bar{\mathrm p}$}
\newcommand{\sps}      {$\sqrt{s^\prime/s}$}

\newcommand{\Pb}[1]  {\mbox{$ {{\cal  {P}}_{#1}}$}}

\newcommand{\Qfb}           {\mbox{$\langle Q_{\mathrm {FB}}\rangle$}}

\newcommand{\lam}  {\mbox{$\Lambda$}}
\newcommand{\eps}  {\mbox{$\epsilon$}}

\newcommand{\half}   {\mbox{$\scriptscriptstyle\frac{1}{2}$}}
\newcommand{\SL}[1]  {$\mathrm{S_{#1}(L})$}
\newcommand{\SR}[1]  {$\mathrm{S_{#1}(R)}$}
\newcommand{\SBR}[1] {$\mathrm{\tilde{S}_{#1}(R)}$}
\newcommand{\SBL}[1] {$\mathrm{\tilde{S}_{#1}(L)}$}
\newcommand{\VR}[1]  {$\mathrm{V_{#1}(R)}$}
\newcommand{\VL}[1]  {$\mathrm{V_{#1}(L)}$}
\newcommand{\VBL}[1] {$\mathrm{\tilde{V}_{#1}(L)}$}
\newcommand{\VBR}[1] {$\mathrm{\tilde{V}_{#1}(R)}$}

\newcommand{\plmo}  {\mbox{$\rm \pm \ $}}

\begin{document}
\begin{titlepage}
\large{EUROPEAN ORGANISATION FOR NUCLEAR RESEARCH (CERN)}

\vspace*{1.5cm}

\begin{flushright}
{\small CERN-EP/2006-xxx} \\
{\small 26 September 2006} \\
%{\small DRAFT 4.14} \\
\end{flushright}

\vspace*{3.cm}
\begin{center}
 \Large{Fermion pair production in \mr{e^+e^-} collisions at
189-209 GeV and constraints on physics beyond the Standard
Model}
\end{center}

\vspace*{1.cm} \begin{center}
 \large{The ALEPH Collaboration$^*)$}
 \end{center}
\vspace*{1.5cm}
\begin{abstract}
Cross sections, angular distributions and forward-backward
asymmetries are presented, of two-fermion events produced
in \ee collisions at centre-of-mass energies from 189 to 209~GeV at LEP,
measured with the ALEPH detector. Results
for \ee, \mumu, \tautau, \qq, \bb\ and \cc\ production are in
agreement with the Standard Model predictions. Constraints are set
on scenarios of new physics such as four-fermion contact
interactions, leptoquarks, Z$^\prime$ bosons, TeV-scale quantum gravity
and R-parity violating squarks and sneutrinos.
\end{abstract}

\vspace*{2.cm}
\centerline{\it \small {To be published in The European Physical Journal C}}
\vspace*{1.cm}
\centerline{\small {Dedicated to the memory of John Strong who died on 
July 31, 2006}}
\noindent
--------------------------------------------\hfil\break
{\small {$^*)$ See next pages for the list of authors}}

\end{titlepage}
\newpage

\input{authb.tex}

\newpage
\section{Introduction}

  In the years 1995-2000, the LEP collider
  delivered \ee collisions at centre-of-mass
 energies from 130 to 209~GeV.
  Measurements of the {\mr {e^+e^- \rightarrow f\bar f}} process
 with the ALEPH detector up to $\sqrt{s}$\,=\,183~GeV
 have been published in~Ref.~\cite{aleph-183}. The
 results obtained at seven additional energy values
 are presented in this paper with analyses largely unchanged with respect
to Ref.~\cite{aleph-183}.
  The seven centre-of-mass energies are listed in Table~\ref{tab:lum},
  together with the corresponding luminosities. In the
 year 2000 the luminosity was delivered in a
 range of energies.
 The 2000
 data are divided into two energy bins, from 202.5~GeV to 205.5~GeV
 and above.

This paper is organized as follows. Section~\ref{sec:detector} gives
a brief description of the ALEPH detector,  Section~\ref{sec:MC}
presents the event generators
used for the simulation of the signal and backgrounds, and
  Section~\ref{sec:signal} recalls some useful definitions.
   Measurements of hadronic, leptonic and heavy-flavour final states
  are discussed in Sections~\ref{sec:hadronic},~\ref{sec:leptonic}
  and~\ref{sec:hf},
   respectively. The results are used to set constraints
  on new physics in Section~\ref{sec:newphys}.

\section{The ALEPH detector\label{sec:detector}}
 The ALEPH detector and
performance are described in Refs.~\cite{aleph-det,aleph-perf},
 and only a short summary is given here.

Charged particles are detected in the central part, comprising a
precision silicon vertex detector, a cylindrical drift chamber and
a large time projection chamber, embedded in a 1.5\,T axial
magnetic field. The momentum $p$ of charged particles is measured
with a resolution of ${\sigma(p)/p\ =\ 6 \times 10^{-4}p_T \oplus
0.005}$ (where $p_T$ is the momentum component perpendicular to
the beam axis in GeV/$c$).  The three-dimensional impact parameter
is measured with a resolution of ${(34 + 70/p ) \times (1 + 1.6
\cos^4\theta)\mu{\rm m}}$ (where $p$ is measured in GeV/$c$ and
$\theta$ is the polar angle with the beam axis). In addition, the
time projection chamber provides up to 344 measurements of the
specific energy loss by ionisation $dE/dx$.

In the following, only charged particle
tracks reconstructed with at least four hits in the time
projection chamber, originating from within a cylinder of length
20~cm and radius 2~cm coaxial with the beam and centred at the
nominal collision point, and with a polar angle fulfilling ${|\cos
\theta|<0.95}$ are considered.

Electrons and photons are identified by the characteristic
longitudinal and transverse developments of the associated showers
in the electromagnetic calorimeter (ECAL), a 22~radiation length thick
sandwich of lead planes and proportional wires chambers with fine
read-out segmentation. The relative energy resolution achieved is
$0.18/\sqrt{E (\rm {GeV})}$ for isolated electrons and photons.

Muons are identified by their characteristic penetration pattern
in the hadron calorimeter (HCAL), a 1.5~m thick iron yoke interleaved with 23
layers of streamer tubes, together with two surrounding
double-layers of muon chambers. In association with the
electromagnetic calorimeter, the hadron calorimeter also provides
a measurement of the hadronic energy with a relative resolution of
$0.85/\sqrt{E ({\rm GeV})}$.

The total visible energy, and therefore the event missing energy, is
measured with an energy-flow reconstruction algorithm~\cite{aleph-perf} which combines all the
above measurements, supplemented by the energy detected down to
34~mrad from the beam axis by two additional electromagnetic
calorimeters, used for the luminosity determination~\cite{aleph-lcal,aleph-sical}. 
The relative
resolution on the total visible energy is ${0.60/\sqrt{E ({\rm GeV})}}$
for high-multiplicity final states. This algorithm also provides
a list of reconstructed energy-flow objects,
 classified as charged particles, photons and neutral hadrons.

The luminosity is determined with small-angle Bhabha events, detected with the
lead-wire luminosity calorimeter (LCAL), using the method described in 
Ref.~\cite{aleph-lcal}. The Bhabha cross section in the LCAL acceptance varies
from $4.3\,{\rm nb}$ at $189\,{\rm GeV}$ to $3.6\,{\rm nb}$ at 
$207\,{\rm GeV}$. The uncertainty on the measurement is smaller than $0.5\%$.

\section{Event simulation and Standard Model predictions\label{sec:MC}}

 Samples of simulated events are produced as follows. The
generator {\tt BHWIDE} version~1.01~\cite{MC-bhwide} is used for the
electron pair channel, and {\tt KK}  version~4.14~\cite{MC-kk2f}
for di-quark, di-tau and di-muon events. Interference between 
 initial-state (ISR) and final-state (FSR)
radiation is included in {\tt KK} generator for the leptonic
channels, whereas for the \qq\  channel the FSR
 is introduced by {\tt PYTHIA} in the parton shower and
therefore interferences with ISR are not
included. {\tt PYTHIA} version~6.1~\cite{MC-pythia} is used
for  ZZ and \mr{Ze^+e^-} production.
 Two-photon interactions (\mr{e^+e^- \rightarrow
e^+e^-X}) are generated with {\tt PHOT02}~\cite{MC-phot02} and
{\tt HERWIG}~\cite{MC-herwig}. 
Finally, backgrounds from W-pair
production are simulated with the {\tt KORALW} 
generator version~1.51~\cite{MC-koralw}.
Single-W processes are simulated with
{\tt EXCALIBUR}~\cite{MC-excalibur}.
Hadronic final states were generated with hadronisation and fragmentation parameters described in Ref.~\cite{QCDmega}.
 
Standard Model (SM) predictions
 in the electron-pair
channel are obtained from {\tt BHWIDE}. The
analytic program {\tt ZFITTER}~\cite{zfitter} is used in all other
cases, with the steering flags and main input parameters listed in
the Appendix.

\section{Cross section definition\label{sec:signal}}

Cross section results are provided for inclusive and exclusive
processes. The inclusive processes include events with hard ISR,
which correspond to about 85\% of the selected events,
while the exclusive processes exclude these final states.

The inclusive processes correspond to a cut  \sps \mr{>} 0.1,
 where $\sqrt{s}$ is the centre-of-mass energy and
$\sqrt{s^\prime}$ is defined as the  invariant mass of the outgoing
 lepton pair for leptonic final states, and  as
the mass of the $s$ channel propagator for hadronic final states.
 Differently from Ref.~\cite{aleph-183}, exclusive processes are defined 
by a cut  \sps\mr{>} 0.85 .

 When selecting events in the analysis, the measured variable $\sqrt{s^\prime_m}$,
  which provides a good approximation to $\sqrt{s^\prime}$ when
only one photon is emitted, is used to isolate exclusive
processes:
$${ {s^\prime_m}\ =\ \frac{\sin\theta_1 +
\sin\theta_2 - |\sin(\theta_1 + \theta_2)|}{\sin\theta_1 +
\sin\theta_2 + |\sin(\theta_1 + \theta_2)|}\ \times s}.$$ Here
\mr{\theta_{1,2}} are the angles of the outgoing fermions
measured with respect to the direction of the incoming electron
beam or with respect to the direction of the most energetic photon seen in the
apparatus and consistent with ISR~\cite{aleph-183}.
In order
to reduce the uncertainties
related to interferences between  ISR and FSR,
the exclusive cross sections and asymmetries are not
extrapolated to the full acceptance. They are evaluated over the reduced
angular range corresponding to $|\cos \theta| < 0.95$, where $\theta$
is the polar angle of the outgoing fermion for the hadronic cross section measurements.
For the leptonic cross section 
and the forward-backward asymmetry measurements $|\cos \theta| < 0.95$ cut applies to both
outgoing fermion and anti-fermion polar angles.

%\section{Hadronic Final States. \label{sec:hadronic}}
% SECTION  HADRON
\input{hadron.tex}

% SECTION LEPTON

\input{dileptons_cor.tex}
\section{Heavy-flavour production\label{sec:hf}}

 Measurements with heavy-flavour final states are described in this section. 
The ratios of the \bb\ and \cc\ cross sections
to the hadronic cross section, indicated as $R_{\rm b}$
and $R_{\rm c}$ respectively, are discussed in Sections~\ref{sec:Rb}
and~\ref{sec:Rc}. The charge forward-backward asymmetry is
 measured on a b-enriched ($A_{\rm FB}^{\rm b}$) and a b-depleted
(\Qfb ) event sample, as presented in Section~\ref{sec:afbb}.
%The latter yields information on the average properties of non-b events
%and can constrain the relative fraction of up-type to down-type
%quarks.

Results are given for the signal definition as
in Ref.~\cite{aleph-183}, with $\sqrt{s^\prime/s}\,>\,0.9$ and an
angular range restricted to \mr{|\cos \theta |<\ 0.95}.
An additional acceptance
cut requiring that both jets have
${\mathrm |\cos \theta _{\rm jet}|<0.9}$\
is applied to ensure that they are contained in the
vertex detector acceptance. Table~\ref{tab:had-events} gives the
number of selected hadronic events at each centre-of-mass energy.
The resulting  efficiency is typically 78\%, with a dependence on
the quark flavour of less than \mr{\pm 1\%}. The background from
\qq\ events produced outside the acceptance, but reconstructed
inside, is of the order of 2.6\% and varies within 0.5\% depending
on the quark flavour. This variation is 
 taken as systematic uncertainty on the
contribution of the radiative background.
 The total uncertainty of the hadronic selection efficiency in the
 considered angular range
 is of the order of 1\%.

 The long lifetime and large decay multiplicity of b hadrons allow
the separation of \bb\ final states from other quarks. The same
tagging variables, complemented by  additional variables,
can be used to separate \cc\ final states from light quarks.
 These selections have a moderate dependence on b-quark and c-quark
 physics modeling
uncertainties~\cite{pdg,lep-hf,aleph-cmod},
 listed in Tables~\ref{tab:bcmod1} and~\ref{tab:bcmod2}.

\subsection{Measurement of $R_{\rm b}$\label{sec:Rb}}

Events containing b hadrons are tagged using the procedure
developed by ALEPH at LEP1~\cite{aleph-rbz1}. For each charged
track, a probability (\Pb{T}) that it originates at the primary vertex is
evaluated using the measured impact parameter significance. This
is defined as the signed distance of closest approach of the track to the
interaction point divided by the uncertainty on that distance. By
taking all tracks or by grouping them according to which
hemisphere  or which jet they populate, the probability that the
event (\Pb{E}), the hemisphere (\Pb{H}) or the jet (\Pb{J}) contains
only light-quark jets can be determined. A low value of the
probability indicates the presence of long-lived states, which
arise dominantly from b-quark production. The b tagging therefore
corresponds to a cut on the appropriate probability.

\begin{figure}[htbp]
\begin{center}
\includegraphics[width=15cm]{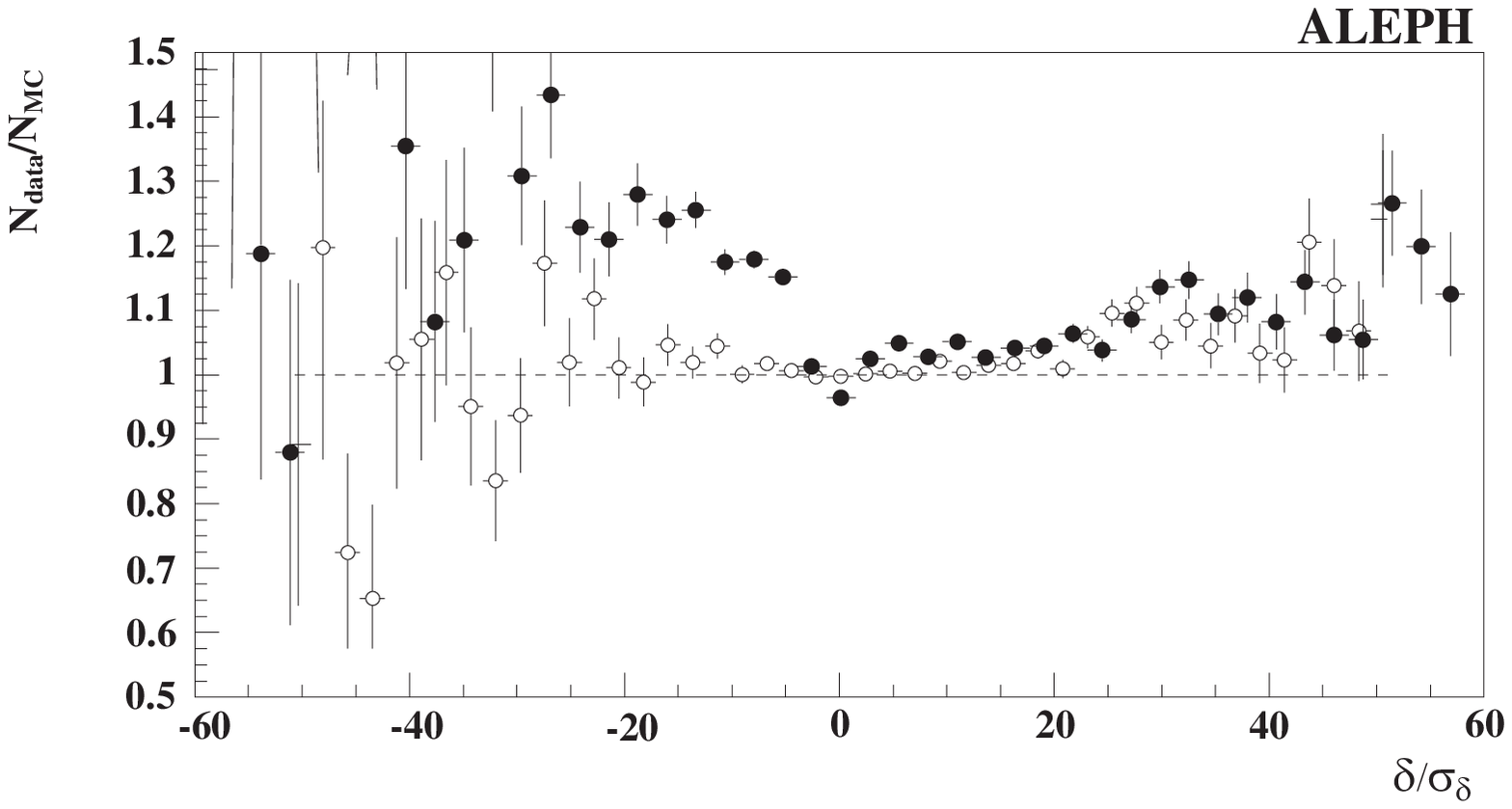}
\caption{\label{fig:ip-smear}{Ratio between Z-peak data and Monte Carlo simulation,
 as a function of the impact parameter significance, before (dots)
 and after (open symbols) smearing the Monte Carlo resolution.}}
\end{center}
\end{figure}

In order to reproduce the detector resolution  in the
simulation, the procedure to smear the impact parameter
significance used for the LEP1
analyses~\cite{aleph-rbz2} is employed. This is based on the \mr{\sim \
3~pb^{-1}} calibration data taken at the Z peak each year,
 in order to optimize the smearing parameters for that year's data
 (Fig.~\ref{fig:ip-smear}).

The crucial factor in the determination of $R_{\rm b}$ is
the b-tag efficiency. The highly accurate measurements of $R_{\rm b}$
at LEP1 were made possible  by the use of a double-tag method,
 relying on the fact that b-quarks are produced in pairs which
populate opposite hemispheres~\cite{aleph-rbz2}. The use of single-
and double-hemisphere tags enables the efficiency, as well as the rate
of \mr{b\bar b} production, to be determined from the
data, leaving only the level of background to be obtained from the
simulation. Furthermore, 
 uncertainties arising from the background knowledge can be
minimized with hard cuts.

%
% from Patrick
%
Unlike at LEP1, the double-tag method is not practical at LEP2 because 
of the much smaller statistics. For this reason, previous ALEPH 
measurements of $R_{\rm b}$ at 130-183 GeV~\cite{aleph-183} were made with a single overall 
event tag. The efficiency was then determined either directly from 
the simulation, or by correcting the simulated efficiency by the ratio 
of the $R_{\rm b}$ value measured with each year Z peak data to the world average. 
Neither method was satisfactory as they both require an extrapolation 
(either from the basic simulation or from the Z to LEP2 energies), 
with mostly unknown related systematic uncertainties.

The full LEP2 data sample, however, has become sufficiently large for 
an average $R_{\rm b}$ value to be measured with the double-tag procedure, with 
reduced and controlled systematic uncertainty. An overall efficiency 
correction can therefore be obtained by taking the ratio of the average 
values of $R_{\rm b}$ over all centre-of-mass energies, measured with the 
double- and single-tag methods respectively, so that

$$ R^{k} = R^{k}_{s} \frac{\overline{R_{d}}}{\overline{R_{s}}} $$

\noindent where $R^{k}$ is the final value of $R_{\rm b}$ at
    energy $k$,
    $R^{k}_{s}$ is the value of $R_{\rm b}$ determined
    by the single-tag procedure at energy $k$, and
    ${\overline{R_{d}}}$ and ${\overline{R_{s}}}$ are the values
    of $R_{\rm b}$,
    averaged over all energy points,
    as measured with the double- or  single-tag
     method respectively.
 The above correction, which amounts to about 1.05, assumes that the ratio 
between the double- and single-tag efficiencies
 is energy independent,
which is true as long as the cuts are not changed on an
energy-by-energy basis. It does not require  the b-tag cut to
be the same for both methods. 
The optimal selection cut for both the event and hemisphere tags is taken to be the point
where the total fractional error on $R_{\rm b}$ is minimized. The b-tag cut corresponds to
a b-selection efficiency of 49\% (69\%) and to a purity of 80\% (72\%) for the event
(hemisphere) tag. The correlation
between the single- and double-tag methods is estimated to be 0.95
from the simulation.

 The final statistical uncertainty  is dominated by the statistical
precision on ${\overline{R_d}}$. To
determine  the systematic uncertainty,
it is assumed that both the uncertainty for each method and
the correlation between them are energy independent. It can then
be shown that the relative systematic uncertainty at each
energy is given to a good approximation by the relative
systematic uncertainty on the average double-tag determination.
%, i.e
%
%$$ \mathrm{\frac{\sigma_{f}^{syst,k}}{R^{k}_{f}} = \frac{\sigma^{syst}}{\overline{R_{d}}}}  $$
%
 The systematic uncertainties for the double-tag method are
calculated over the full data set, and the contributions
are given in Table~\ref{tab:rb-syst}. The dominant sources
come from the b-tagging parameters (used to define the track selection and the
jet reconstruction) and from the smearing procedure~\cite{richard-phd}.
 In addition, by comparing the average
efficiency obtained with the double-tag method between data and
simulation, the uncertainty on the uds
and c backgrounds is found to be smaller than 11\%.

The measured average value of $R_{\rm b}$ is

$$ R_{\rm b} = 0.151 \pm 0.012 ({\rm stat}) \pm  0.007 ({\rm syst})
\ \ \ \ (\langle\sqrt{s}\rangle = 198\,{\rm GeV}) $$

\begin{figure}[htbp]
\begin{center}
\includegraphics[width=15cm]{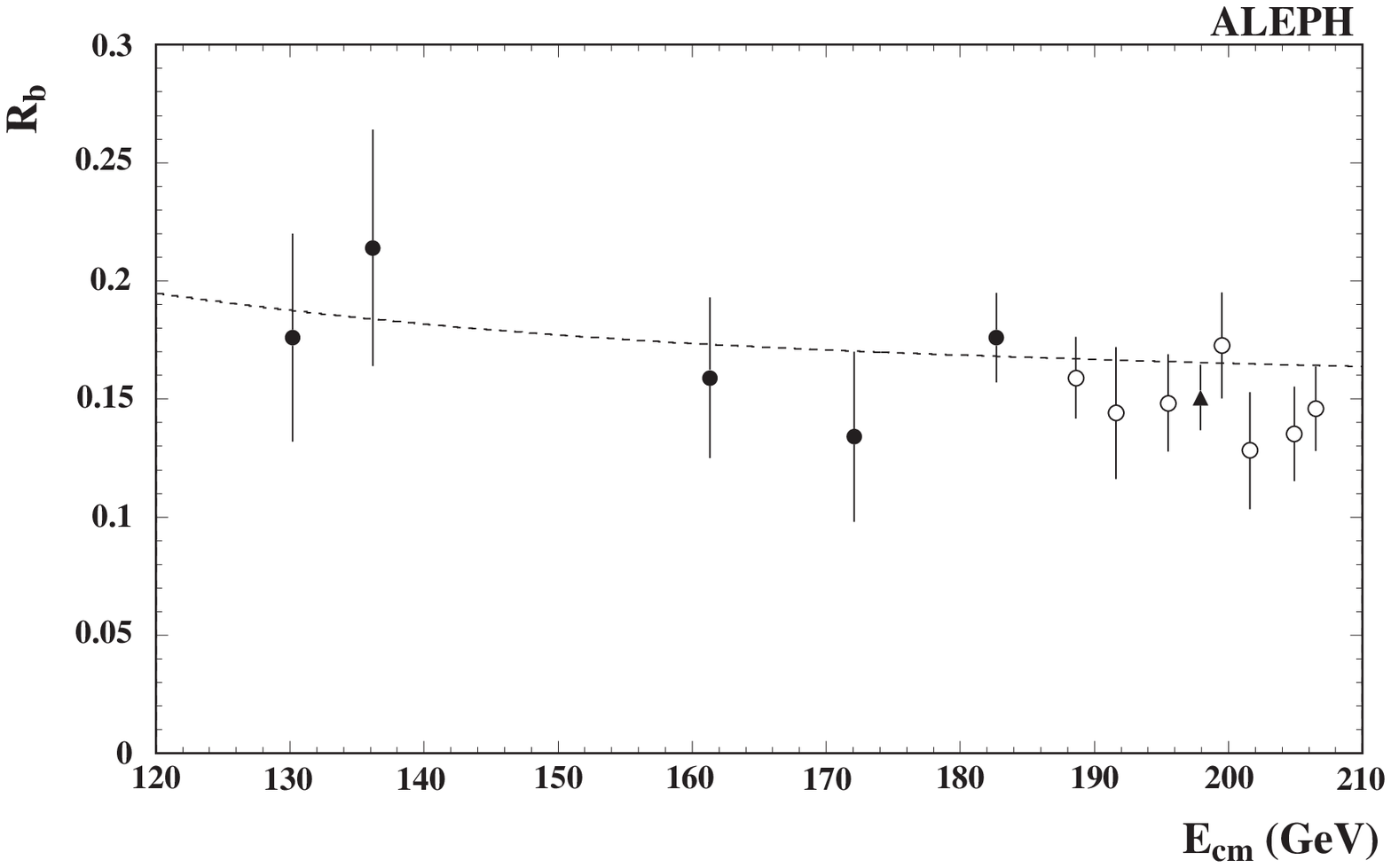}
\caption{\label{fig:rb-sqrt}{Measured values of $R_{\rm b}$  (open symbols), as
 a function of centre-of-mass energy, together with the SM prediction (dashed curve).
 The dots show previous results~\cite{{aleph-183}} and 
 the triangle the average value obtained over
 the 189-209~GeV energy range.}}
\end{center}
\end{figure}

The individual values determined at each energy point
are presented in Table~\ref{tab:rb-final} and in Fig.~\ref{fig:rb-sqrt}.

\subsection{Measurement of $R_{\rm c}$\label{sec:Rc} }

The ratio of the \cc\ cross section to
the hadronic cross section, $R_{\rm c}$, is measured from
the hadronic sample pre-selected
as described above.

 In a first step, the background from \bb\ events is reduced to 4\%
of the hadronic sample with a cut on \Pb{E} ($\log{\cal P}_E > -2$),
which retains 86\% of
the \cc\ events and close to 100\%\ of the light-quark events.
The efficiency of this cut is controlled on a sample of WW
events, and the resulting systematic uncertainty is about 1\%. %  (1.5\% at 189 GeV) .

 The final selection of  \cc\ events uses a Neural Network (NN)
 algorithm trained to
separate jets originating from c quarks from jets originating from light quarks.
 The nine input variables, exploiting the lifetime of  D  mesons,
 their masses and their decays into leptons or kaons, are:
\begin{itemize}
\item{\Pb{J}, as defined in Section~\ref{sec:Rb}.}
\item{The probability that tracks having a high rapidity with
 respect to the jet axis originate from the primary vertex.}
\item{The decay length significance of a reconstructed secondary vertex.~\cite{aleph-bsosc}}
\item{The $p_T$, with respect to the jet axis,
of the last track used to build a
2~GeV/$c^2$ mass system, tracks being ordered by increasing \Pb{T}.}
\item{The sum of the rapidities, with respect to the jet axis, of all energy-flow
 objects within 40 degrees of this axis.}
\item{The total energy of the four most energetic energy-flow 
objects in the jet.}
\item{The missing energy per jet defined as the difference between the beam 
energy and the reconstructed jet energy.}
\item{The largest rapidity of lepton candidates with respect to
 the jet axis.}
\item{The largest momentum of kaon candidates. 
Here a charged particle track is identified as a kaon candidate 
if its ionization energy loss ($dE/dx$) is compatible with that 
expected from a kaon within three standard deviations, 
and more compatible with that expected from a kaon than with that expected from a pion.
}
\end{itemize}

\begin{figure}[htbp]
\begin{center}
\includegraphics[width=14cm]{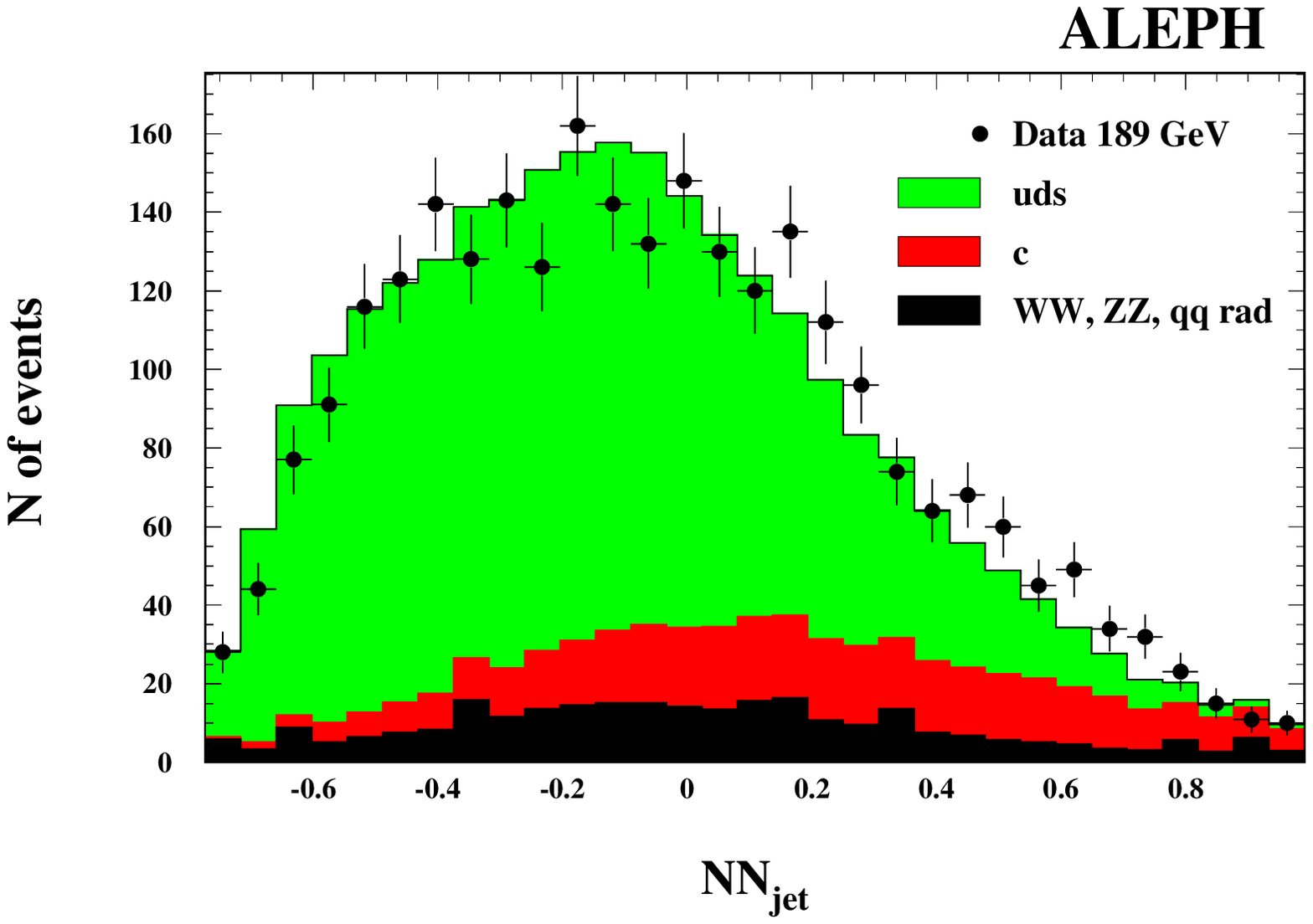}
\includegraphics[width=14cm]{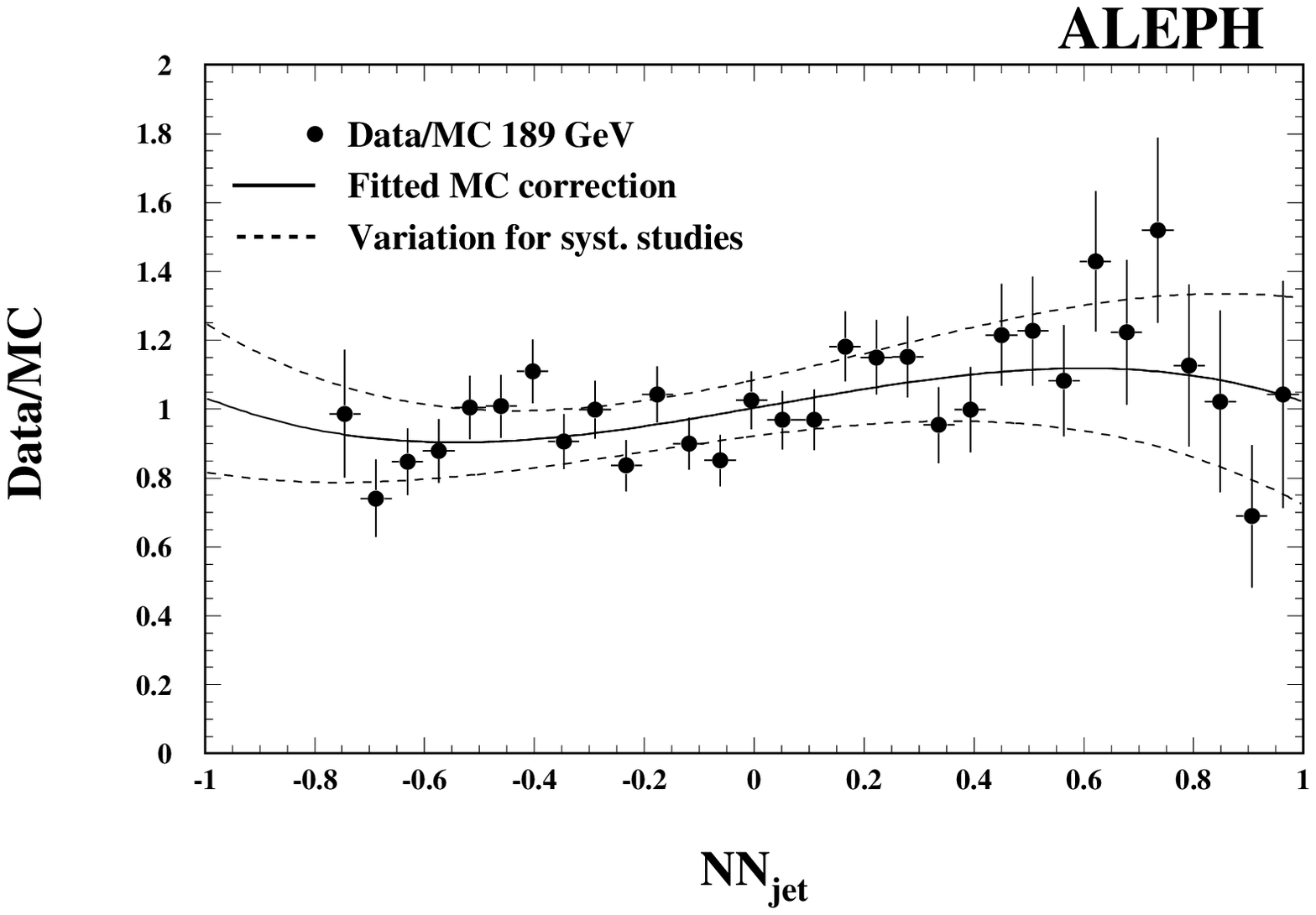}
\end{center}
\caption{\label{fig:rc-corr}{Top: output of the c-tag Neural Network
for a uds-enriched sample  from the data (dots)  and the simulation (histograms).
 The Monte Carlo distribution is normalized to the data.
 Bottom: ratio between data and Monte Carlo as a function
of the Neural Network output for a uds-enriched sample. The third degree polynomial fit
to this distribution (full curve with the
error band) is used to correct the
simulation (see text).}}
\end{figure}

 The distribution of the NN output for
 light-quark jets  in the simulation is corrected
with the data by comparing enriched samples of light-quark jets
selected with a cut applied to the opposite hemisphere. 
The correction is applied energy by energy. The statistical
error on this correction is taken as systematic uncertainty; an
additional uncertainty originates from the residual \cc\
background in the selected sample. An example of the distributions
used to derive the correction and the correction itself are
shown in Fig.~\ref{fig:rc-corr}. A {\bb}-enriched sample is used
to control the fraction of \bb\ background in the final
event sample to 5\%. Other sources of systematic
uncertainties come from the limited statistics of the
Monte Carlo samples, the knowledge of the luminosity,
detector effects (smearing
and momentum corrections), the hadronic pre-selection, and
the modeling of c-quark physics.
 They are listed in Table~\ref{tab:rc-syst}.

\begin{figure}[htbp]
\begin{center}
\includegraphics[width=15cm]{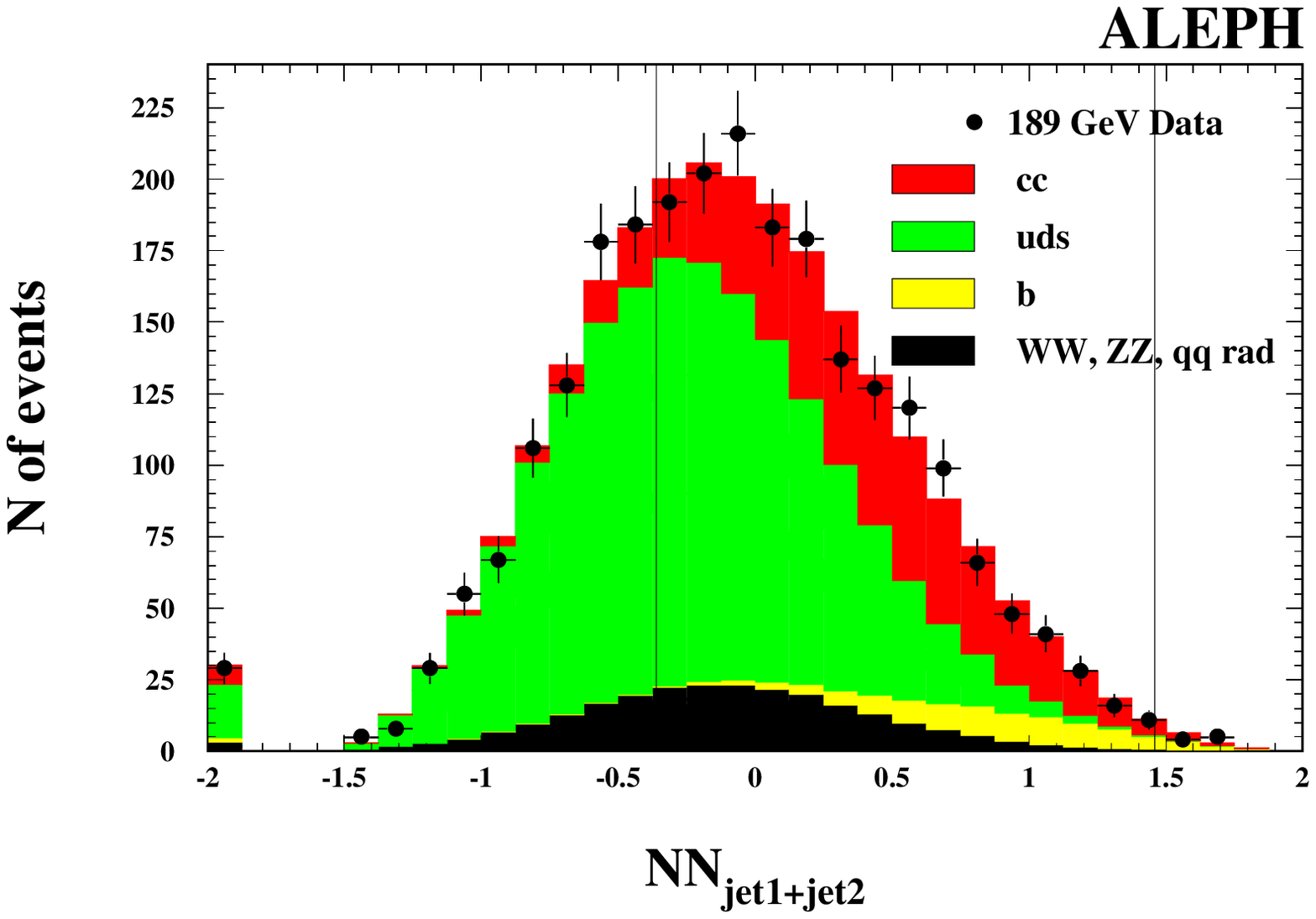}
\end{center}
\caption{\label{fig:rc-dist-189} {Distribution of the sum of the c-tag
Neural Network outputs for the two jets in the event for the data (dots) and
the simulation (histograms).
 The region between the two vertical lines is selected for the $R_{\rm c}$ measurement. The
bin at NN output$=-2$ contains events for which some of the NN input variables are not 
available.}}
\end{figure}

 The distribution of the sum of the NN outputs for both
jets in the event is shown in Fig.~\ref{fig:rc-dist-189}.
 At each energy point, the NN cuts (indicated
 in Fig.~\ref{fig:rc-dist-189} for the $\sqrt{s}$=~189~GeV case)
 are chosen so as to minimize the total uncertainty. The upper cut
suppresses about 5\% of the remaining b background with a signal loss of
less than 1\%. The typical efficiency is 75\% with a signal-to-background ratio of
50\%. The resulting $R_{\rm c}$  measurements are listed in
Table~\ref{tab:rc-meas}.

\subsection{Measurements of $A_{\rm FB}^{\rm b}$ \textbf{and \Qfb\label{sec:afbb}}}

The $A_{\rm FB}^{\rm b}$ and \Qfb\ measurements are both extracted from hadronic
events pre-selected as described above.
 A b-enriched
sample and a b-depleted sample are obtained using appropriate cuts on
\Pb{E} ($\log{\cal P}_E < -4.3$ and $\log{\cal P}_E > -2$, respectively). 
The cuts are set with the aid of the simulation, and
correspond to a b content of the order of 90\% and 4\% for the
two samples, respectively. The selection efficiencies vary only slightly with
the centre-of-mass energy.

The jet charge ${Q_{\rm jet}}$ of each jet is defined as

$$Q_{\rm jet} = \frac{\sum_i q_i\cdot p^{\kappa}_{\|,i}}{\sum_i p^{\kappa}_{\|,i}}$$

\noindent where the sums extend over the reconstructed
charged tracks in the jet and $q_i$ and $p_{\|,i}$ are the track charge
and track momentum parallel to the jet axis, respectively. The
parameter \mr{\kappa} is optimized with simulated events so as to
maximize the charge separation between b jets and \mr{\bar b} jets.
 It is found to be fairly independent of the
centre-of-mass energy and the average value of 0.36 is used. 
The same \mr{\kappa} value is also used for
the b-depleted event sample.

The mean charge asymmetry 
${\langle Q_{\rm FB}\rangle\ = \langle Q^{\rm F}_{\rm jet} -Q^{\rm B}_{\rm jet}\rangle}$
is measured on both the b-enriched and b-depleted samples
as the average of the jet charge difference between the forward
and backward hemispheres, defined with respect to the thrust axis.
It is related to the quark forward-backward asymmetries ($A_{\rm FB}^{\rm q}$)
as follows

$${ {\langle Q_{\rm FB}\rangle\ =\
\frac{\sum \epsilon _{\rm q}\sigma _{\rm q} \delta _{\rm q} A_{\rm FB}^{\rm q} +
        \sum \epsilon _{\rm x}\sigma _{\rm x} \langle Q_{\rm FB}^{\rm x}\rangle}
     {\sum \epsilon _{\rm q}\sigma _{\rm q}  + \sum \epsilon _{\rm x} \sigma _{\rm x}} }}$$

\noindent where the index q indicates the quark flavours (u, d, s, c and b)  and
 the index ${\rm x}$ indicates the various background components
 (WW, ZZ and radiative \mr{q\bar q}).
In this expression 
${ \langle Q_{\rm FB}^{\rm x}\rangle}$  indicates
the background mean charge asymmetry ,
${\epsilon_{\rm q,x}}$ the selection efficiencies 
and ${\delta_{\rm q}}$ the charge separation  (defined as the mean of
 the ${Q_{\rm q} - Q_{\rm {\bar q}}}$ distribution).

 The asymmetry $A_{\rm FB}^{\rm b}$ is obtained from the b-enriched sample; it is
extracted from ${\langle Q_{\rm FB}^{\rm enr}\rangle}$, 
the charge asymmetry measured from the
data, using the previous formula.
The background mean charge asymmetry, the selection efficiencies and the
charge separations are taken from the simulation.
The non-b quark cross sections \mr{\sigma_q}\ and asymmetries $A_{\rm FB}^{\rm q}$
are computed with {\tt ZFITTER} for the signal definition
\sps \mr{>} 0.9, with \mr{|\cos\theta| < 0.9} for both quark and anti-quark.
 It is possible to reduce the
dependence of this measurement on the b efficiency
estimated from the simulation
 by replacing the product \mr{\epsilon_b\sigma_b} with ${N_{\rm data}^{\rm b}/{\cal L}}$,
  where ${N_{\rm data}^{\rm b}}$ is the number of b events in the data and
  \mr{{\cal L}} is the integrated luminosity.
 It follows:

$${{A_{\rm FB}^{\rm b}\ =\
\frac{N_{\rm data}^{\rm b} \langle Q_{\rm FB}^{\rm enr}\rangle
        - \sum^{\rm q\neq b} {\cal L} \epsilon _{\rm q}\sigma _{\rm q} \delta _{\rm q} 
         A_{\rm FB}^{\rm q}
        - \sum {\cal L} \epsilon _{\rm x}\sigma _{\rm x} \langle Q_{\rm FB}^{\rm x}\rangle}
     {(N_{\rm data}^{\rm b} - N_{\rm MC}^{\rm bkg})\delta _{\rm b}} }}$$

\noindent where ${N_{\rm MC}^{\rm bkg}}$ is the number of background
 events predicted by the simulation.
The measurement is corrected by a factor 1.03 to extrapolate
from the range \mr{|\cos \theta|<0.9} to
the nominal range \mr{|\cos \theta|<0.95}.
 The potentially large uncertainty originating from the
\cc\ contamination in the b-enriched sample is reduced
to a negligible level by a tight cut on \Pb{E}.

\begin{figure}[htbp]
\begin{center}
\includegraphics[width=15cm]{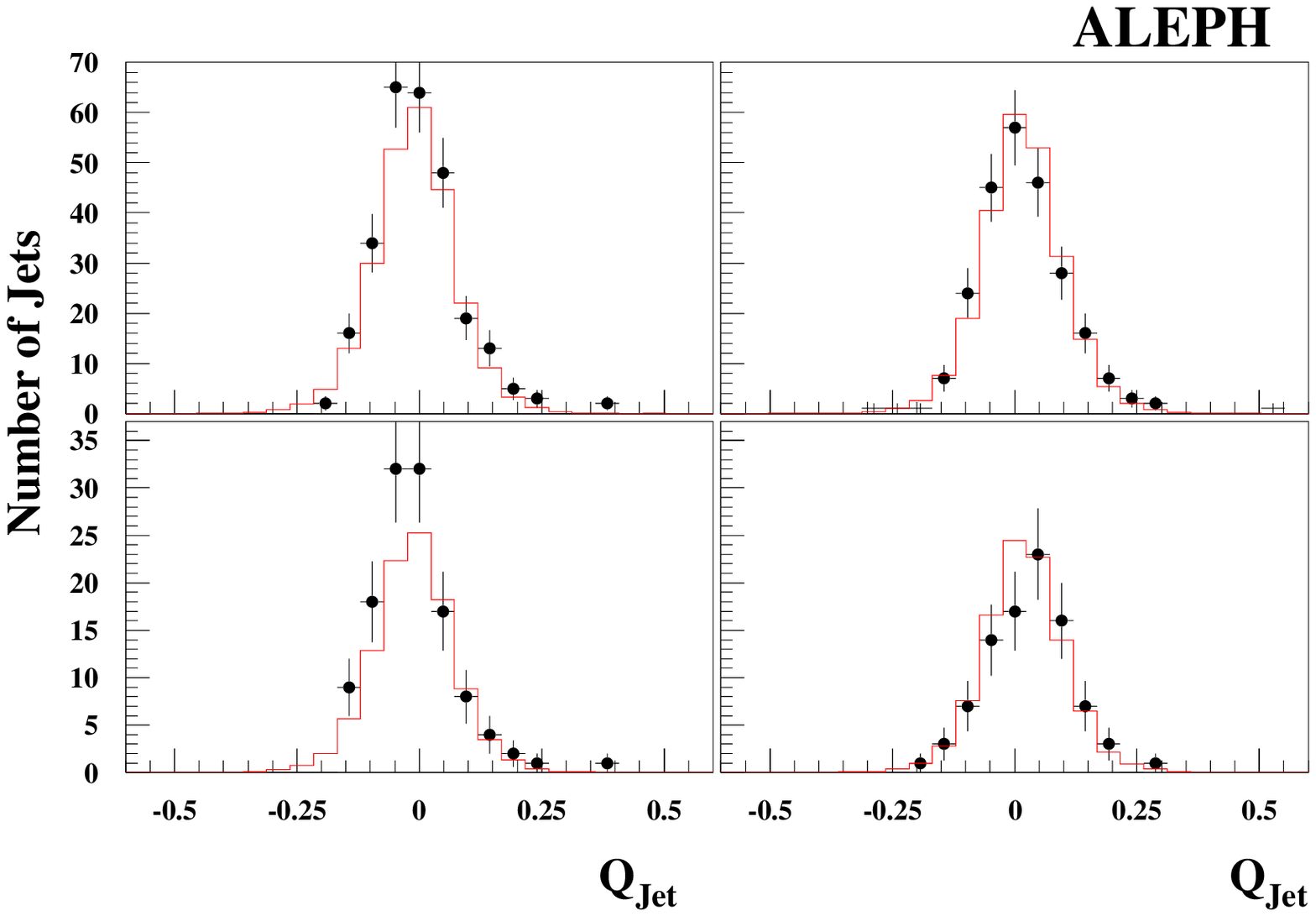}
\end{center}
\caption{\label{fig:afbb-qjet}{Distributions of the b-jet charge
from b-enriched event samples with a high-$p_T$ lepton
in the hemisphere opposite to the jet, for the data
(dots) and the simulation (histograms). The top (bottom) plots
are for $p_T>1$~GeV/$c$ ($p_T>1.7$~GeV/$c$), the left (right) plots
are for positive (negative) leptons.}}
\end{figure}

In order to evaluate the systematic uncertainty on the jet charge separation,
$\delta_q$ is measured with the data, using
semileptonic b decays for b quarks and
semileptonic WW events for light and c quarks.
 Semileptonic b decays are selected by requiring an electron or muon
with high transverse momentum in one jet. The charge of the
opposite b jet is then known. Because of the low event
statistics surviving this
selection, data taken at all energies must be combined. 
The difference between the jet charge distributions in data and simulation 
(Fig.~\ref{fig:afbb-qjet})
is propagated as systematic uncertainty to the $A_{\rm FB}^{\rm b}$ measurement,
representing the dominant systematic effect.
A similar procedure is applied to a selected sample 
of semileptonic W-pair events to measure the average lighter quarks charge separation.

 These and other sources of systematic uncertainties are summarized in
 Table~\ref{tab:afbb-syst}. The  $A_{\rm FB}^{\rm b}$ measurements
 are presented in Table~\ref{tab:afbb-meas}.

 Finally, the difference
 ${\Delta\ =\ \langle Q_{\rm FB}^{\rm depl}\rangle-\langle Q_{\rm FB}^{\rm MC}\rangle }$,
 measured with
 b-depleted samples, constrains
simultaneously $A_{\rm FB}^{\rm q}$ and \mr{\sigma_{q}} (q=u,d,s or c), 
providing additional sensitivity to physics beyond the
Standard Model. 
Assuming small deviations from the SM, the linearized equation
\[{\mathrm \Delta = \sum_{\rm q}
\frac{\partial\langle Q_{\rm FB}\rangle}{\partial \sigma_{\rm q}}\Delta\sigma _{\rm q} +
\sum_{\rm q}
\frac{\partial\langle Q_{\rm FB}\rangle}{\partial A^{\rm q}_{\rm FB}} 
\Delta A^{\rm q}_{\rm FB}}\]
\noindent is used to constrain
the deviations of $A_{\rm FB}^{\rm q}$ and \mr{\sigma_{q}} from the SM with the measured
values of
$\Delta$ at each centre-of-mass energy, as described in Ref.~\cite{aleph-183}. 
Examples of the coefficients of the above equations are shown in Table~\ref{tab:qfb-coef}.

The dominant systematic uncertainty originates from the
 jet charge, determined as explained
above. This and other sources are listed in Table~\ref{tab:qfb-syst}, while
the measurement results are reported in Table~\ref{tab:qfb-meas}.

\section{Interpretation in terms of new physics\label{sec:newphys}}

  New physics, if present, 
 could give rise to deviations of the measured
 cross sections and asymmetries from the Standard Model expectations.
 The results presented in the previous Sections indicate good agreement
 between the data and the SM predictions. As an example the global fit of the muon, tau
and hadronic exclusive cross sections and of the muon and tau asymmetries at the seven energies gives
$\chi^2/{\rm d.o.f.} = 29.79/35$. Stringent limits can be placed on scenarios beyond
 the Standard Model.

  Predictions of several models of new physics 
  are fitted to the
 data using binned maximum likelihoods, as explained in~\cite{aleph-183}.
  For this purpose, the measurements described in this
 paper are combined with those at lower energies reported
 in~\cite{aleph-183}.

Following the conclusions in Ref.~\cite{Lep2-WS}, theoretical
uncertainties of 0.26\%, 0.4\%, 0.4\%, 0.5\% and 2.0\%  are
assigned to the \qq , \mumu , \tautau , \ee\ (forward) and \ee\
(central) cross section predictions, respectively.

\subsection{Contact interactions}

 Four-fermion contact interactions, expected to occur for example
 if fermions are composite,  are characterized by a scale
 $\Lambda$, interpreted as the mass of a new heavy particle
 exchanged between the incoming
and outgoing fermions, and a coupling \emph{g} giving the strength
of the interaction. Conventionally, $g$ is assumed to satisfy
$g/\sqrt{4\pi}=1$. Following the notation of Ref.~\cite{th-ci}, the
effective Lagrangian for the four-fermion contact interaction in
the process \mr{e^+e^- \rightarrow f\bar f} is given by

$$ L^{\rm CI} = \frac{g^2\eta_{\rm sign}}{(1+\delta)\Lambda^2} 
\sum_{i,j={\rm L,R}} \eta_{ij}[\bar e_{i}\gamma^{\mu}e_{i}]
                        [\bar f_{j}\gamma_{\mu}f_{j}] $$

\noindent with
$\delta$~=~1 if $f=e$ and $\delta$~=~0 otherwise.
 The fields $e_{\rm L,R}\ (f_{\rm L,R})$
are left- or right-handed projections of electron (fermion) spinors,
and the coefficients $\eta_{ij}$ specify the relative contribution of
the different chirality combinations.  
 New physics can add
constructively or destructively to the SM Lagrangian, according to the
sign of $\eta_{\rm sign}$.
 Several models, defined in Table~\ref{tab:ci-models}, are considered in 
this analysis.

In the presence of
contact interactions, the differential cross section for \mr{e^+e^-
\rightarrow f\bar f} as a function of the polar angle $\theta$ of
the outgoing fermion with respect to the e$^-$ beam can be written
as
\[
\frac{{\mathrm d}\sigma}{{\mathrm d}\cos\theta} =
F_{\mathrm{SM}}(s,t)
  \left[1 + \eps   \frac{F_{\mathrm{IF}}^{\mathrm{Born}}(s,t)}
                        {F_{\mathrm{SM}}^{\mathrm{Born}}(s,t)}
          + \eps^2 \frac{F_{\mathrm{CI}}^{\mathrm{Born}}(s,t)}
                        {F_{\mathrm{SM}}^{\mathrm{Born}}(s,t)} \right]
\]
where $s,t$ are the Mandelstam variables and $\eps =
g^2\eta_{\mathrm{sign}}/(4\pi\lam^2)$. $F_{\mathrm{SM}}$ is
the Standard Model cross section. $F_{\mathrm{IF}}^{\mathrm{Born}}$
and $F_{\mathrm{CI}}^{\mathrm{Born}}$ are the contributions 
 from the interference between 
 the Standard Model and the contact interaction and from
the pure contact interaction, respectively. The above formula
is fitted to the data using a binned maximum likelihood function, 
as described in Ref.~\cite{aleph-183}. Limits are quoted at the 95\% C.L.
for $\Lambda^\pm$ corresponding to $\eta_{\rm sign}=\pm 1$.

\begin{figure}[htbp]
\begin{center}
\includegraphics[width=15cm]{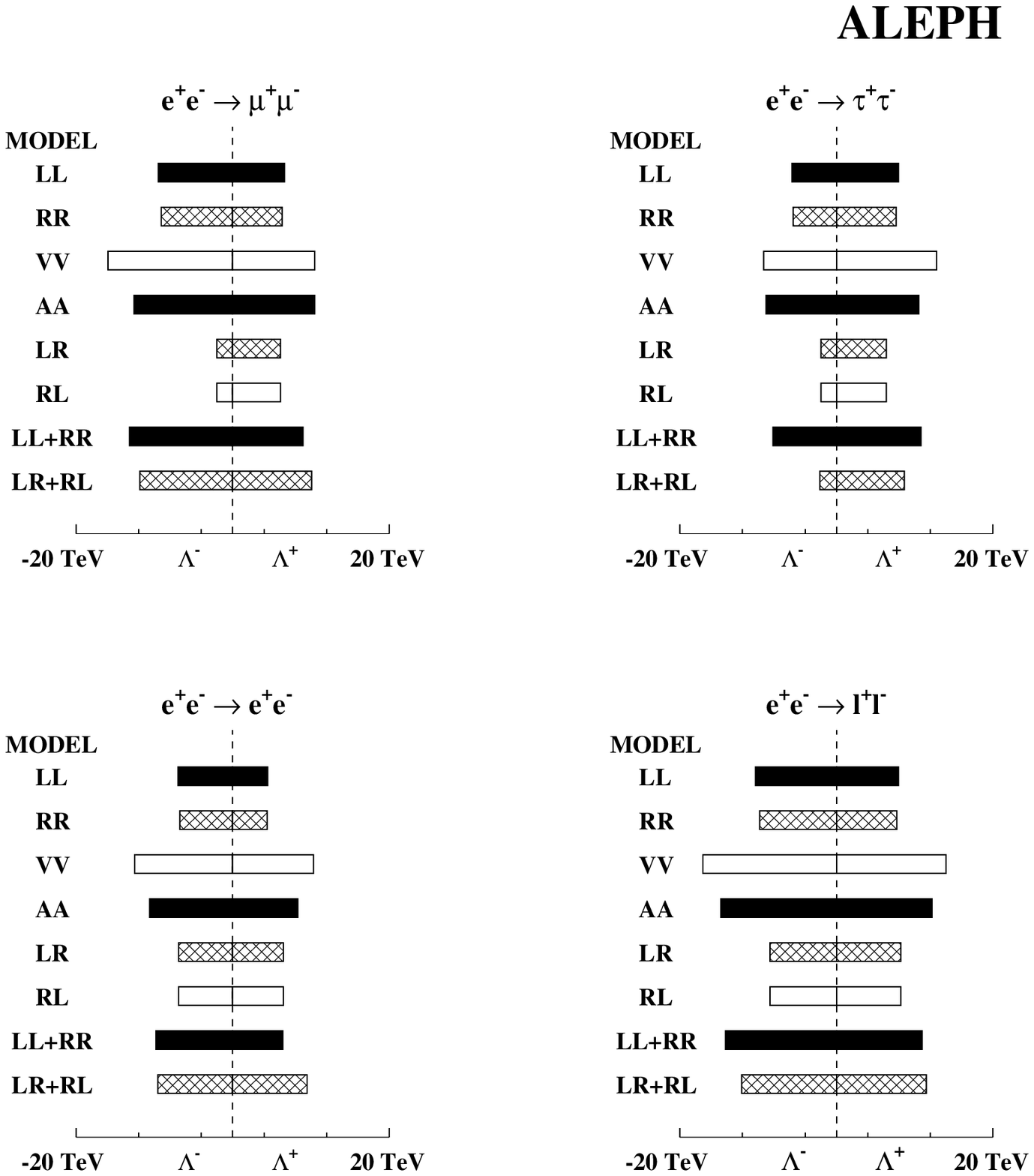}
\caption{\label{fig:cont_ll}{The $95$\%~C.L. excluded values of the scale of
contact interaction $\Lambda$ from di-lepton final states and
for various models. The results for the \mr {e^+e^-\to\ell^+\ell^-} 
process assume lepton universality of the contact interactions.}}
\end{center}
\end{figure}

 For leptonic final states, limits on the  scale $\Lambda$  are derived
 from the leptonic differential cross sections.
 The results are shown in Table~\ref{tab:contact_lepton} 
 and Fig.~\ref{fig:cont_ll}.

\begin{figure}[htbp]
\begin{center}
\includegraphics[width=15cm]{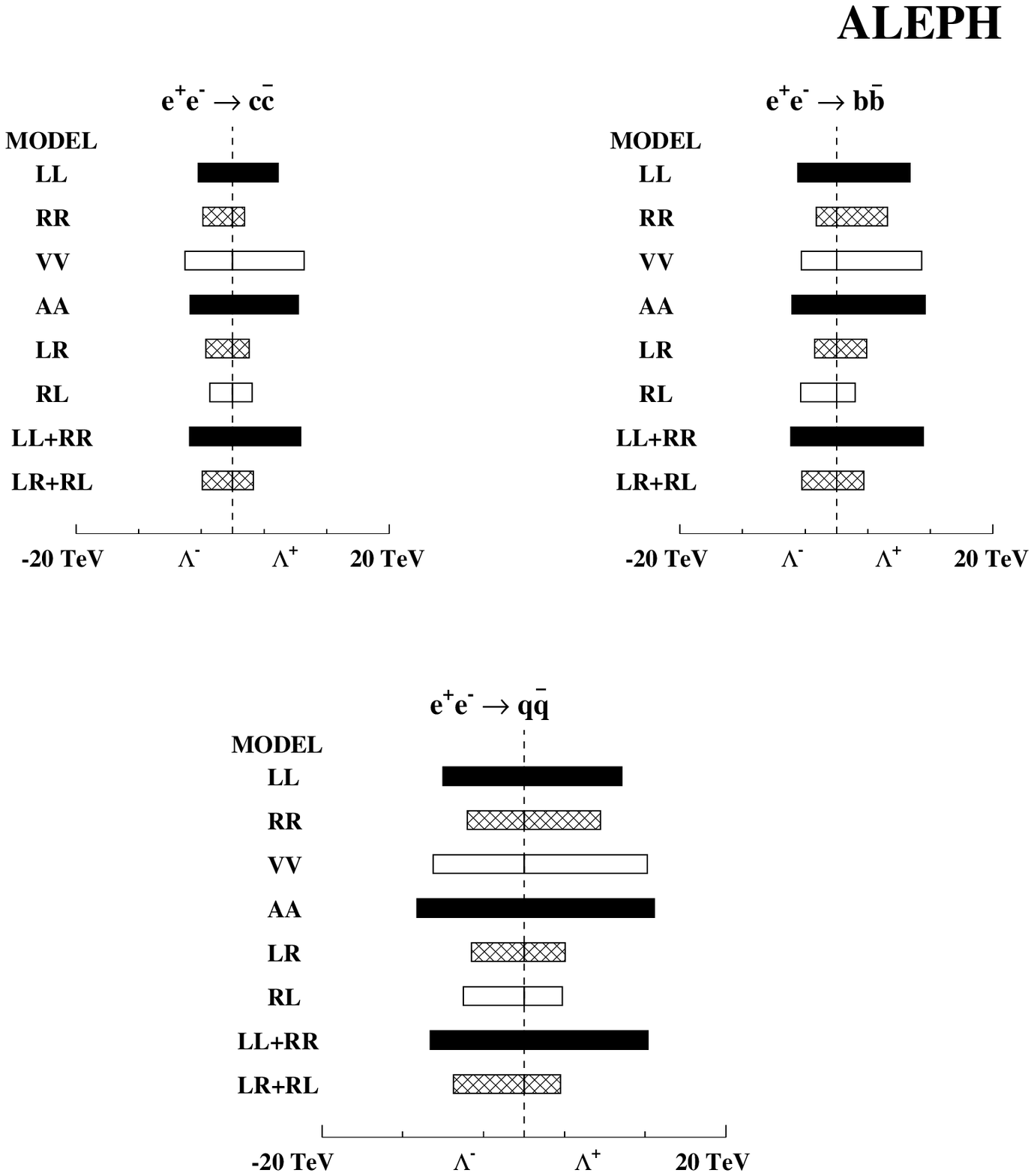}
\caption{\label{fig:cont_had}{The $95$\%~C.L. excluded values
 of the scale of contact interaction $\Lambda$ from hadronic final states.
 The results for \cc\ and \bb\ final states
 assume that the contact interactions affect only c and b quarks.}}
\end{center}
\end{figure}

 For generic hadronic final states, limits on $\Lambda$ are obtained from fits
to the hadronic cross sections assuming that the contact interaction
affects all flavours with equal strength.
 In addition, limits on models involving only couplings to
c or b quarks can be derived from the $R_{\rm c}$ and $Q_{\rm FB}^{\rm depl}$
or $R_{\rm b}$ and $A_{\rm FB}^{\rm b}$
 measurements respectively. The results are
 shown in Table~\ref{tab:ci-hadron} and Fig.~\ref{fig:cont_had}.
 Combining hadronic cross section measurements
 with observables in the charm sector
 improves the overall sensitivity, % and
%in this case, the correlation between both sets of measurements
%are neglected.
 whereas the gain is marginal for the bottom sector.

In summary, the ALEPH limits on the scale of contact interactions
$\Lambda$ are in the range 2-17~TeV, and most stringent
for the VV and AA models.
 Constraints on \mr{e^{+}e^{-}\ell^{+}\ell^{-}},
 \mr{e^{+}e^{-}b\bar{b}} and \mr{e^{+}e^{-}c\bar{c}} contact interactions
are of particular
interest because these couplings are not accessible
at \mr{p\bar{p}} and \mr{ep} colliders.
%Limits for models of
%$e^{+}e^{-}u\bar{u}$, $e^{+}e^{-}d\bar{d}$ contact interactions which violate parity
%(LL,RR,LR and RL) are severely constrained by atomic physics parity violation
%experiments and are the order of 15 \tevc ~\cite{exp-at-ci}.

\subsection{R-parity violating sneutrinos}

 Supersymmetric theories with R-parity violation have terms in the
 Lagrangian of the form ${\lambda_{ijk}{\rm L}_i{\rm L}_j\bar{\rm E}_k}$, where L
 denotes a left-handed lepton doublet superfield and \mr{\bar E}  a
 right-handed lepton singlet superfield~\cite{th-sneu}.
  The parameters \mr{\lambda} are Yukawa
 couplings and $i,j,k$ are generation indices. The couplings
 ${\lambda_{ijk}}$,
 assumed to be real in this analysis, are non-vanishing only for $i<j$.
  These terms allow for single production of sleptons in \mr {e^+e^-}
  collisions.

\begin{figure}[htbp]
\begin{center}
\includegraphics[width=15cm]{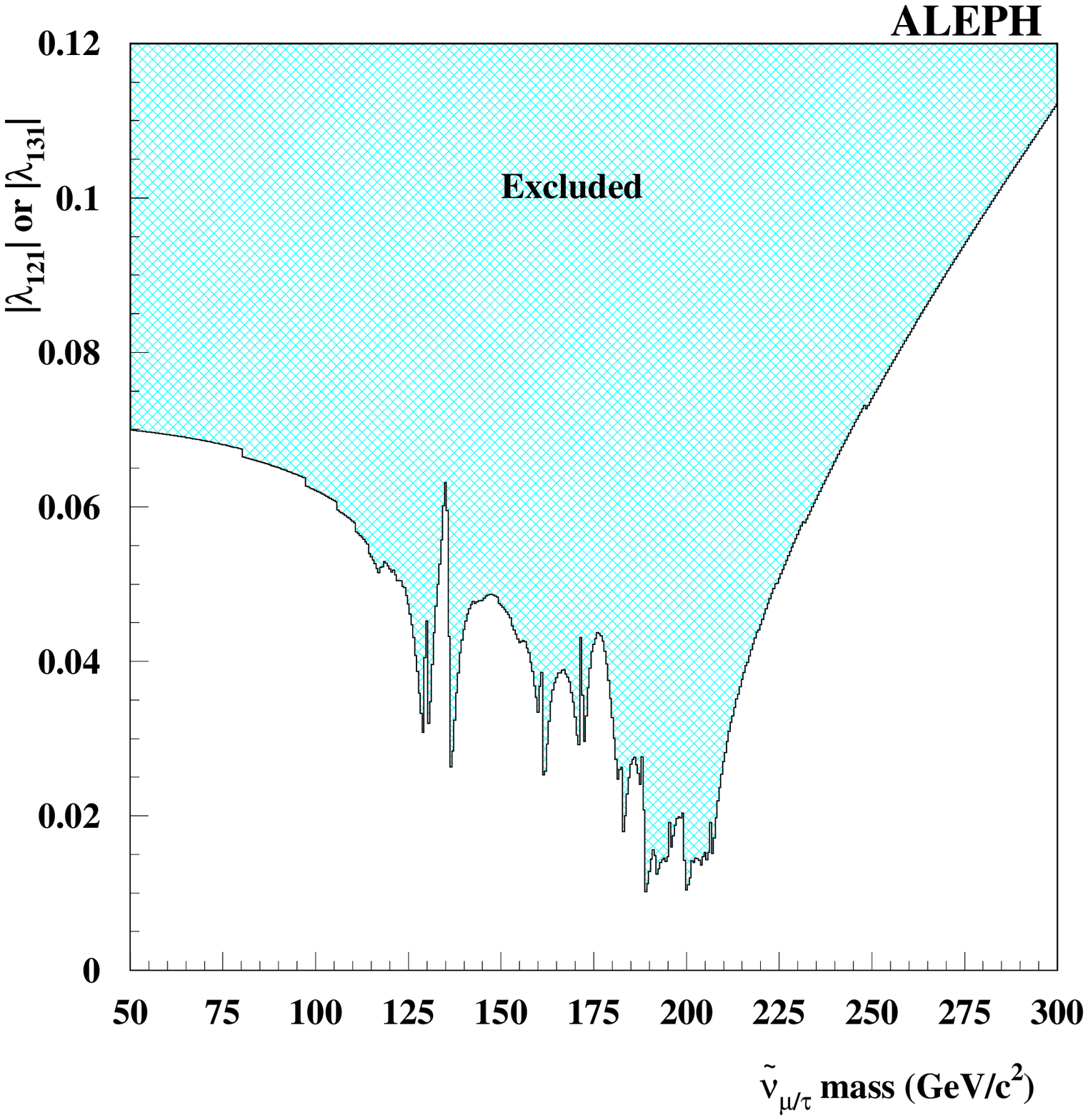}
\caption{\label{fig:snu-e}{The 95\%~C.L.  upper limits
 on the \mr{|{\lambda_{121}}|} coupling
versus the assumed \mr{\tilde{\nu}_{\mu}} mass and on
the \mr{|{\lambda_{131}}|} coupling versus the
assumed \mr{\tilde{\nu}_{\tau}}
mass, as obtained from the Bhabha cross section measurement.
The \mr{\tilde{\nu}_{\mu/\tau}} width is assumed to be 1~GeV/$c^2$.}}
\end{center}
\end{figure}

\begin{figure}[htbp]
\begin{center}
\includegraphics[width=14cm]{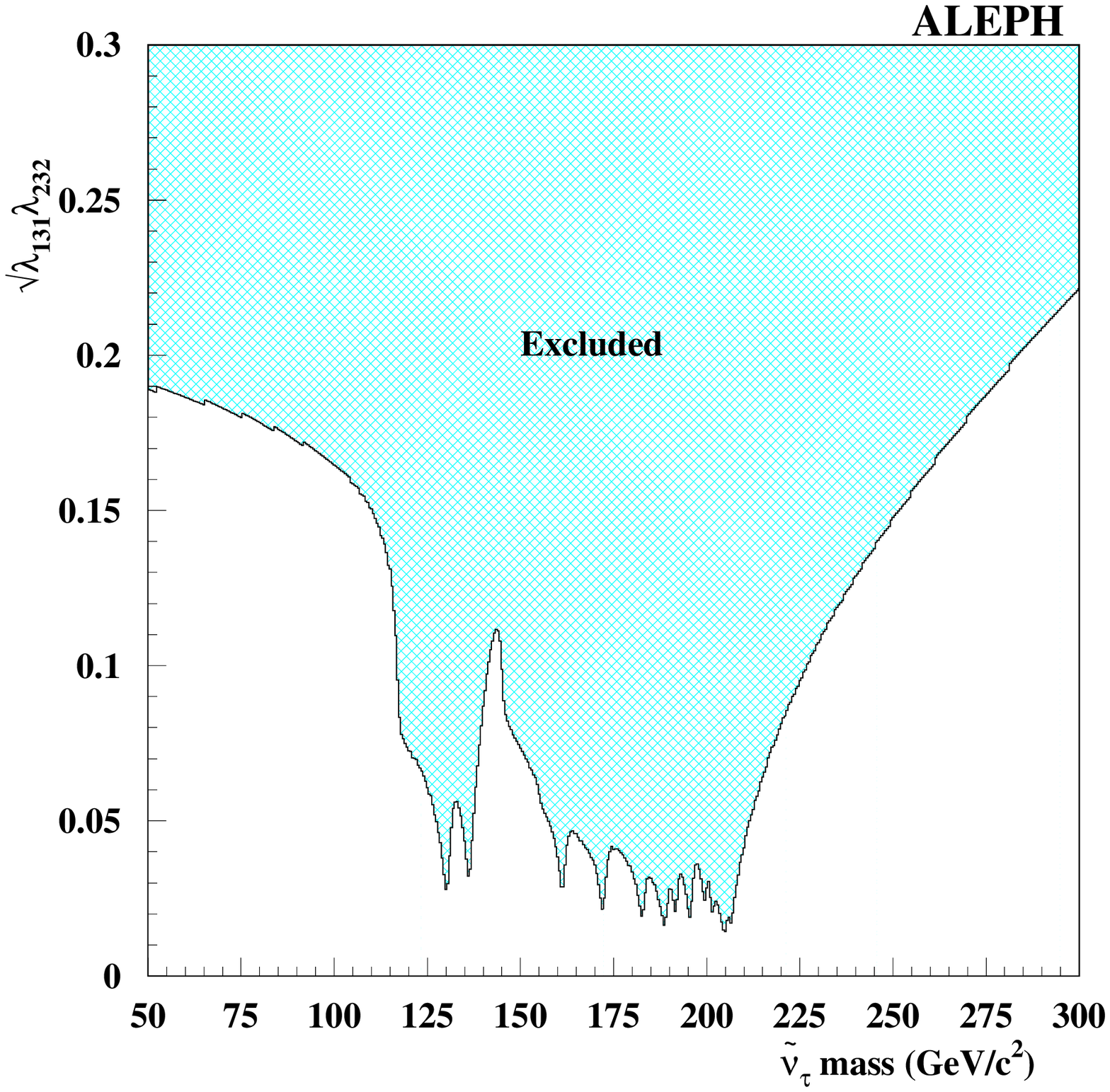}
\end{center}
\caption{\label{fig:snu-m}{The 95\%~C.L. upper limits
on \mr{\sqrt{|\lambda_{131}\lambda_{232}|}} versus the assumed
\mr{\tilde{\nu}_{\tau}} mass, as obtained from
 the \mr{\mu^+\mu^-} cross section measurement.
The \mr{\tilde{\nu}_{\tau}} width is assumed to be 1~GeV/$c^2$.}}
\end{figure}

\begin{figure}[htbp]
\begin{center}
\includegraphics[width=14cm]{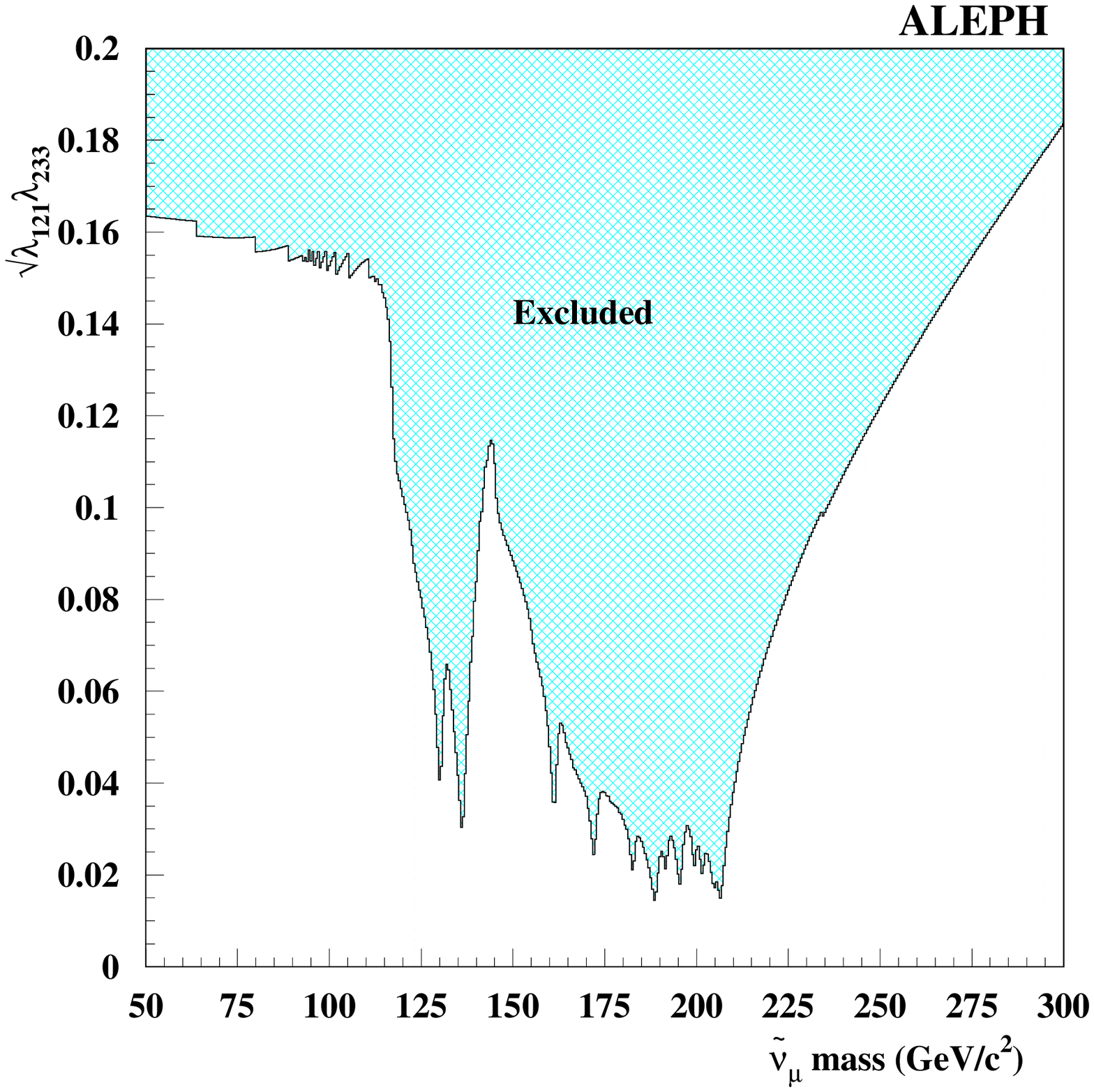}
\end{center}
\caption{\label{fig:snu-t}{The 95\%~C.L. upper limits
 on \mr{\sqrt{|\lambda_{121}\lambda_{233}}|}
versus the assumed \mr{\tilde{\nu}_{\mu}} mass,
 as obtained from the \mr{\tau^+\tau^-} cross section measurement.
The \mr{\tilde{\nu}_{\mu}} width is assumed to be 1~GeV/$c^2$.}}
\end{figure}

  At LEP, R-parity violating sneutrinos could be exchanged in the
 $s$ or $t$ channel, leading to deviations of di-lepton production
 from the SM expectations. Table~\ref{tab:sneu-models} shows the
 most interesting cases.
 Sneutrino exchange in the $s$ channel gives rise to a resonant state,
 assumed here to have a width of 1~GeV/$c^2$~\cite{th-sneu}.
 Limits on couplings are extracted as explained in Ref.~\cite{aleph-183}
 using leptonic differential cross section measurements.
  Figures~\ref{fig:snu-e}-\ref{fig:snu-t} show the resulting
  constraints for processes
 involving sneutrino exchange.

\subsection{Leptoquarks and R-Parity violating squarks}

 At LEP, the $t$ channel exchange of a leptoquark can modify the \qq\
cross section and jet charge asymmetry. In scenarios where
leptoquarks couple to the first-generation  leptons and to the
second- or third-generation quarks, more stringent limits can be
placed by using in addition the relevant heavy-flavour observables
$R_{\rm b}$, $R_{\rm c}$ and $A_{\rm FB}^{\rm b}$. Comparisons of the measurements with
the predictions given in
Ref.~\cite{th-lq} allow upper limits to be set on the leptoquark
coupling  as a function of its mass.

 Leptoquarks are classified according to the spin, weak
isospin I and hypercharge. Scalar and vector leptoquarks are
denoted by symbols \mr{S_I} and \mr{V_I}, and different
hypercharge states are indicated by a tilde. In addition, ``L" or ``R"
specifies if the leptoquark couples to the right- or
left-handed leptons exclusively.
 The \mr{\tilde S_{\frac{1}{2}}(L)} and \mr{S_0(L)} leptoquarks are
equivalent to up-type anti-squarks and down-type squarks,
 respectively, in supersymmetric theories with an R-parity breaking
term ${\lambda^\prime_{1jk}{\rm L}_1{\rm Q}_j\bar{\rm D}_k}$ ($j,k$=1,2,3). Limits
on leptoquark couplings are then equivalent to limits on
${\lambda_{1jk}}$.

\begin{figure}[htbp]
\begin{center}
\includegraphics[width=15cm]{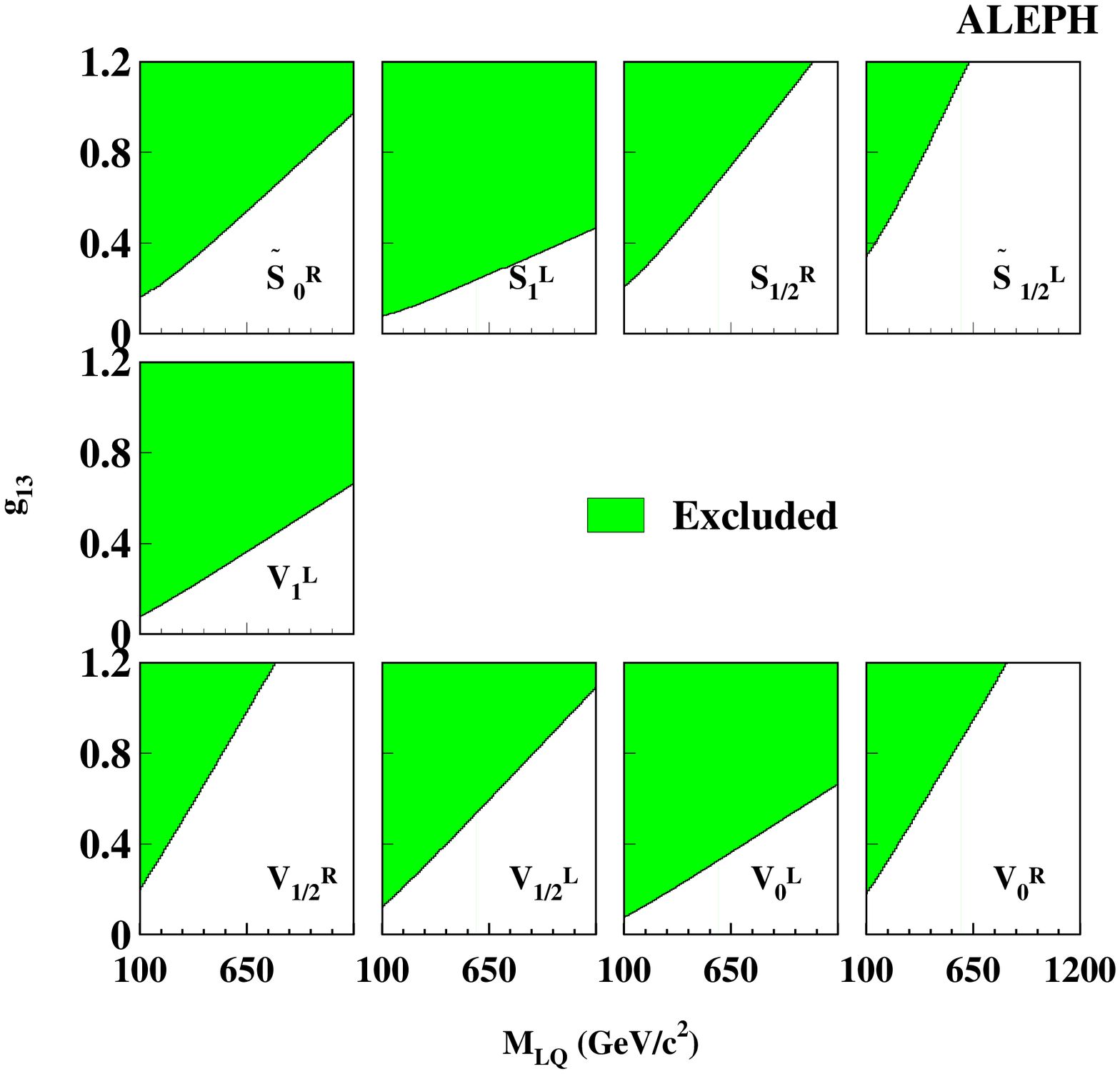}
\end{center}
\caption{\label{fig:lq-lim}{The 95\%~C.L. upper limits on the
coupling versus mass for leptoquarks coupling
to the third quark generation.}}
\end{figure}

 Table~\ref{tab:lq-limits} gives, for various
leptoquark type, the 95\% C.L. lower limit on the mass
$M_{\rm LQ}$ assuming the leptoquark couples with strength
$g^2=4\pi$$\alpha_{em}^{2}$.
 Figure~\ref{fig:lq-lim} shows the exclusion contour in the plane
 coupling-mass for leptoquarks coupling to the third quark
generation.
%The inclusion of the forward-backward charge
%asymmetry \mr{Q_{FB}} measurement improves the sensitivity.

% Constraints on leptoquarks coupling to the 1st quark generation are
%complementary to
% HERA~\cite{exp-lq-h1} and Tevatron limits
%~\cite{exp-lq-tevatron}.

\input{ze}

\subsection{Limits on TeV-scale gravity}

 A solution to the hierarchy problem has been proposed
in Ref.~\cite{th-grav}, where gravity is characterized by a
fundamental scale $M_{\rm D}$ which could be as low as 1~TeV,
 provided that
 space has \mr{\delta} extra dimensions compactified
 to a size $R$. The effective Gravitational constant is
 then given by ${g_{\rm N}^{-1} = 8\pi R^\delta M_{\rm D}^{2+\delta}}$.
 Hence, gravity can become strong
at small distances, leading for example to deviations
 of the \mr{e^+e^-\rightarrow\ e^+e^-} differential cross section
from the SM  expectation. These deviations are
parametrized by a cut-off \mr{\Lambda_T}~\cite{th-grav-param} of
the same order of magnitude as $M_{\rm D}$.

\begin{figure}[htbp]
\begin{center}
\includegraphics[width=15cm]{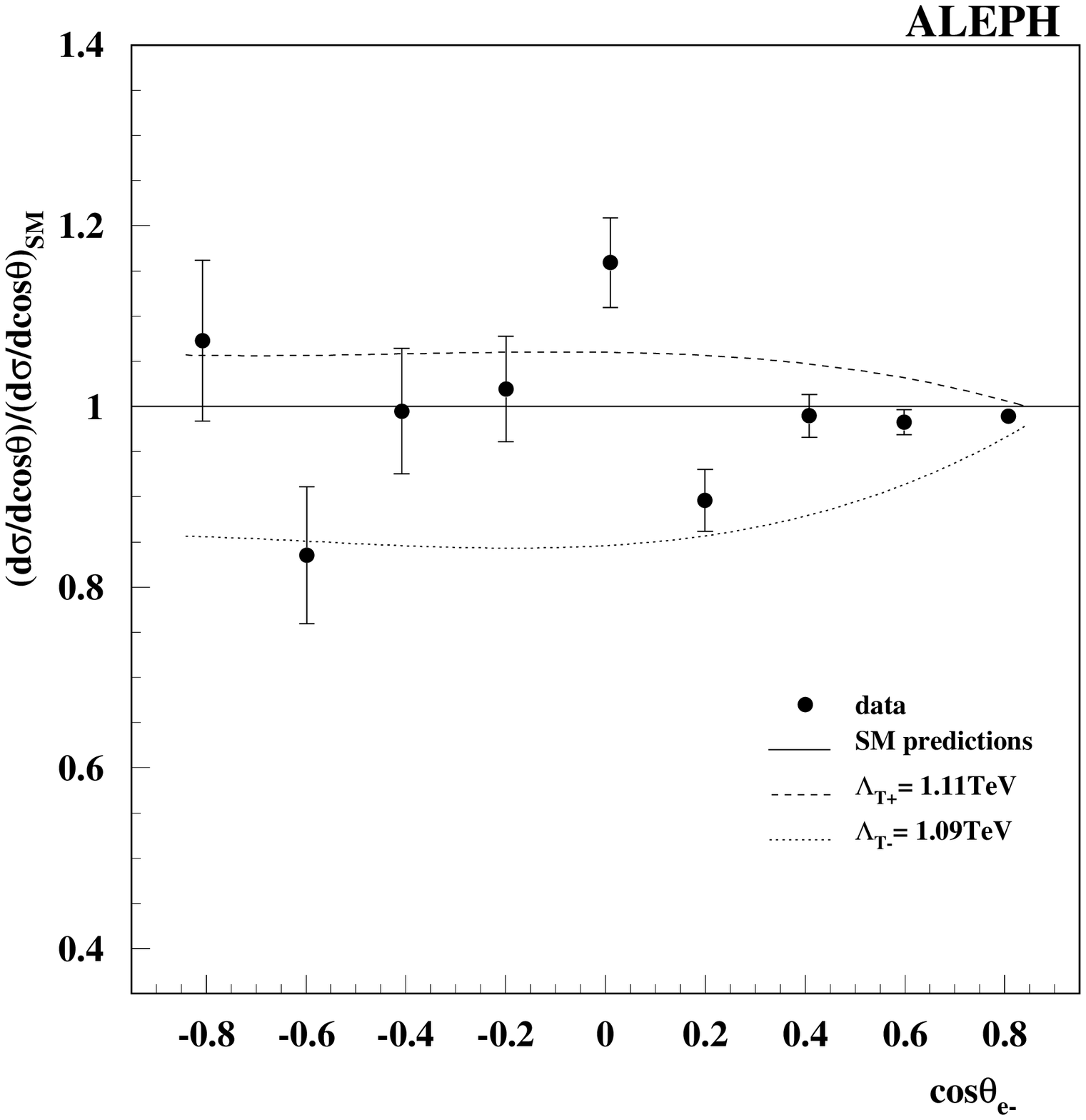}
\end{center}
\caption{\label{fig:grav}{Ratio between the measured cross sections
 and the SM
cross sections for Bhabha scattering, as a function of polar angle,
 obtained using data collected at $\sqrt{s}=189-209$~GeV.
 The dashed and dotted curves indicate the expected ratios in the
 presence of TeV-scale gravity, for  two values of the cut-off
 parameters \mr{\Lambda_T^{\pm}}.}}
\end{figure}

  Figure~\ref{fig:grav} shows the  \mr{e^+e^-\ \rightarrow\ e^+e^-}
 differential cross sections measured with
  data collected at $\sqrt{s}$=189-209~GeV,
  normalized to the SM prediction, together with the expected
  deviations from
  TeV-scale gravity models. Using all data,
   a lower limit of 1.1~TeV (1.2~TeV) is obtained on
   \mr{\Lambda_T^-} (\mr{\Lambda_T^+}), for
   destructive (constructive) interference with the SM prediction.
In computing the limits the luminosity measurement was assumed unaffected and the theoretical
errors of 0.5\% (2.0\%) assigned to the forward (central) \mr{e^+e^-} cross sections were 
taken as uncorrelated.

\section{Conclusions\label{sec:conclu}}

   Several measurements of di-fermion final states
  using data collected by ALEPH at $\sqrt{s}$=189-209~GeV
 have been presented. In the leptonic sector, total
 and 
 differential cross sections, as well 
 as muon and tau forward-backward asymmetries, have
 been derived. In the hadronic sector,  cross sections,
  forward-backward charge asymmetries for light and c quarks,
  b quark forward-backward asymmetries, 
  and the $R_{\rm b}$ and $R_{\rm c}$ ratios have been measured. 
Similar measurements have been performed by the DELPHI~\cite{exp-delphi}, 
L3~\cite{exp-l3}  and OPAL~\cite{exp-opal-ci} Collaborations.

 The results are consistent with the Standard Model expectations and
 have been used to place constraints on several
  scenarios of new physics:
 four-fermion contact interactions, R-parity violating
 sneutrinos and squarks, leptoquarks, additional Z bosons
 and TeV-scale gravity. These constraints improve on previous ALEPH
 limits, and are similar to those obtained by the other LEP Collaborations.

Additional interpretations in terms of new physics, using measurements presented in this paper,
can be found in Refs.~\cite{strumia-1,strumia-2}.

%The limits on the energy scale $\Lambda$ of four-fermion contact
%interactions are in the range 2-17 TeV depending on the model.
%For R-parity violating sneutrinos, the values of the Yukawa coupling $\lambda >
%0.12$ can be excluded for \mr{m_{\tilde{\nu}}} in the range 130 - 210 \gevcc for
%processes involving sneutrino exchange in the s-channel.
%Limits on the LQ mass for leptoquarks coupling to the first, second or third
%generationof quark are extracted.
%The limits for TeV scale gravity reach 1.2 TeV in the sensitive channel
%\mr{e^+e^-\ \rightarrow\ e^+e^-}.

\section*{Acknowledgments}
We thank our colleagues from the CERN accelerator divisions for
the successful running of LEP at high energy. We are
indebted to the engineers and technicians in all our institutions
for their contribution to the good performance of
ALEPH. Those of us from non-member states thank CERN for its
hospitality.

\clearpage
\newpage

%% Tables
\input{tab-lum_cor.tex}
%% Hadrons
\input{tab-had.tex}
%% Leptons
\clearpage
\input{tab-leptons.tex}
\clearpage
%% HF
\input{tab-had-events.tex}
\input{tab-bmod.tex}
\input{tab-cmod.tex}
%\input{tab-rb-syst.tex}
\input{tab-rb-syst-comb.tex}
\input{tab-rb-meas.tex}
%\input{tab-rc-syst.tex}
\input{tab-rc-syst-comb.tex}
\input{tab-rc-meas.tex}
%\input{tab-afbb-syst.tex}
\input{tab-afbb-syst-comb.tex}
\input{tab-afbb-meas.tex}
\input{tab-coef.tex}
%\input{tab-qfb-syst.tex}
\input{tab-qfb-syst-comb.tex}
\input{tab-qfb-meas.tex}

\input{tab-ci-models}
\input{tab-ci-lepton}
\input{tab-ci-hadron}
\input{tab-sneu-models}

\input{tab-lq-limits}

\input{tab-zprime}

\clearpage
\newpage
\begin{figure}[htbp]
\begin{center}
\includegraphics[width=12cm]{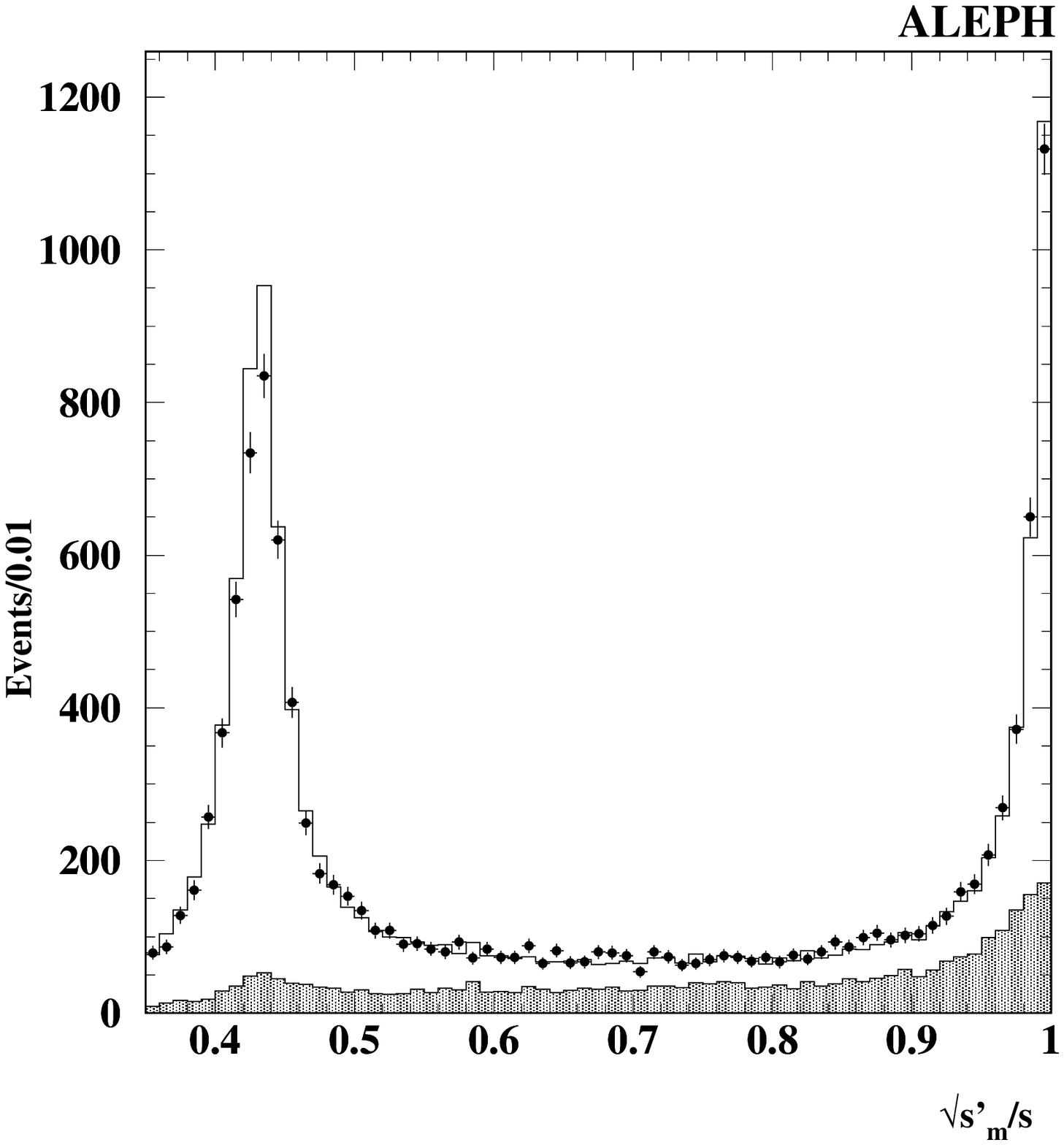}
 \end{center}
% \vspace{-6cm}
 {\lblcaption{fig:sprim_207}{The $\sqrt{s^{\prime}_m/s}$ distribution
 for hadronic events collected at $\sqrt{s}=$207~GeV.
  The data (dots) are compared to the
 Monte Carlo expectation (histogram). The shaded area shows the background
 contribution.}}
 \end{figure}

\clearpage
\newpage
\begin{figure}[htbp]
\begin{center}
%\begin{flushleft}
%\begin{sloppypar}
\includegraphics[width=10cm]{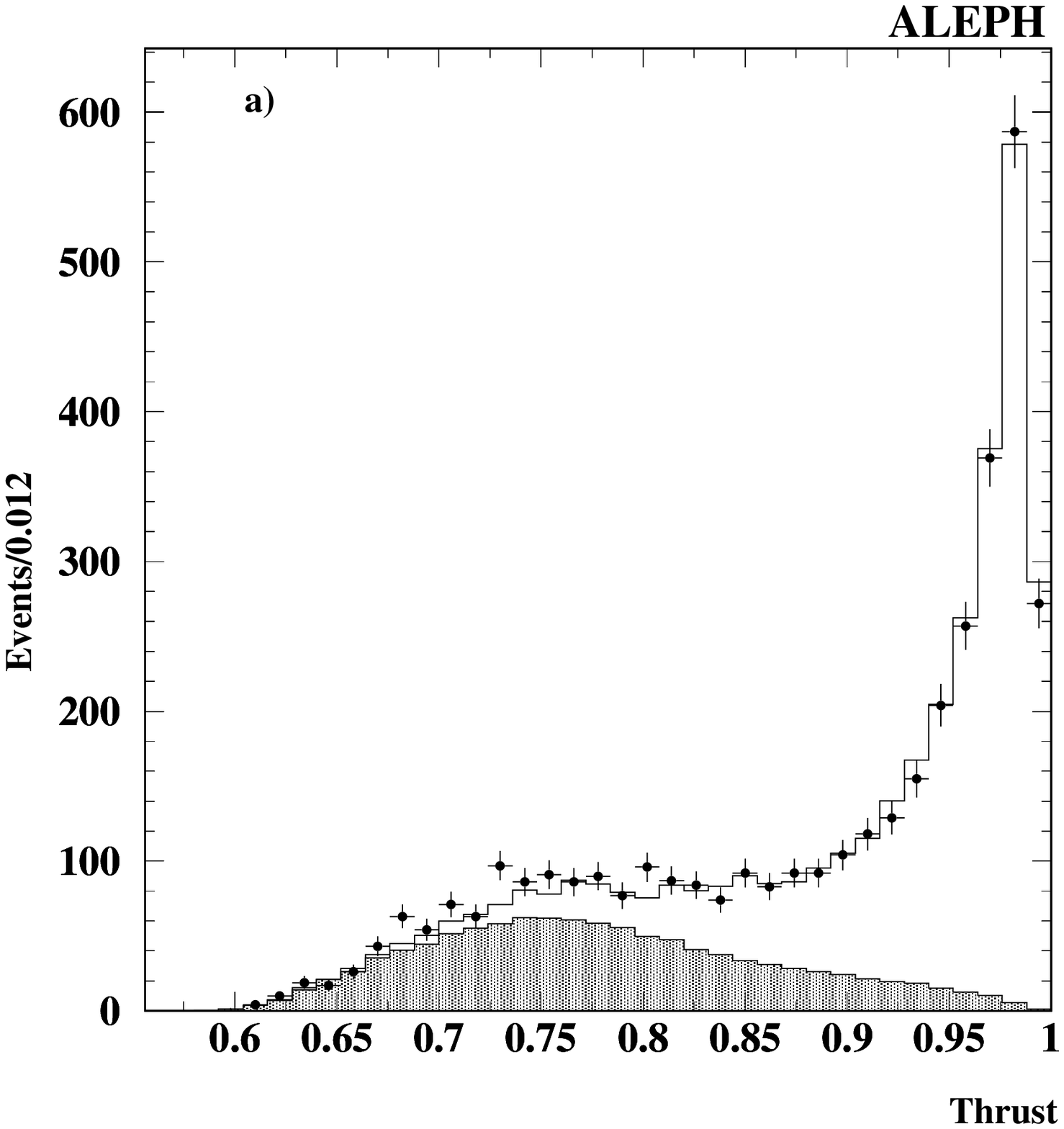}
                        \hspace{0.1cm}
\includegraphics[width=10cm]{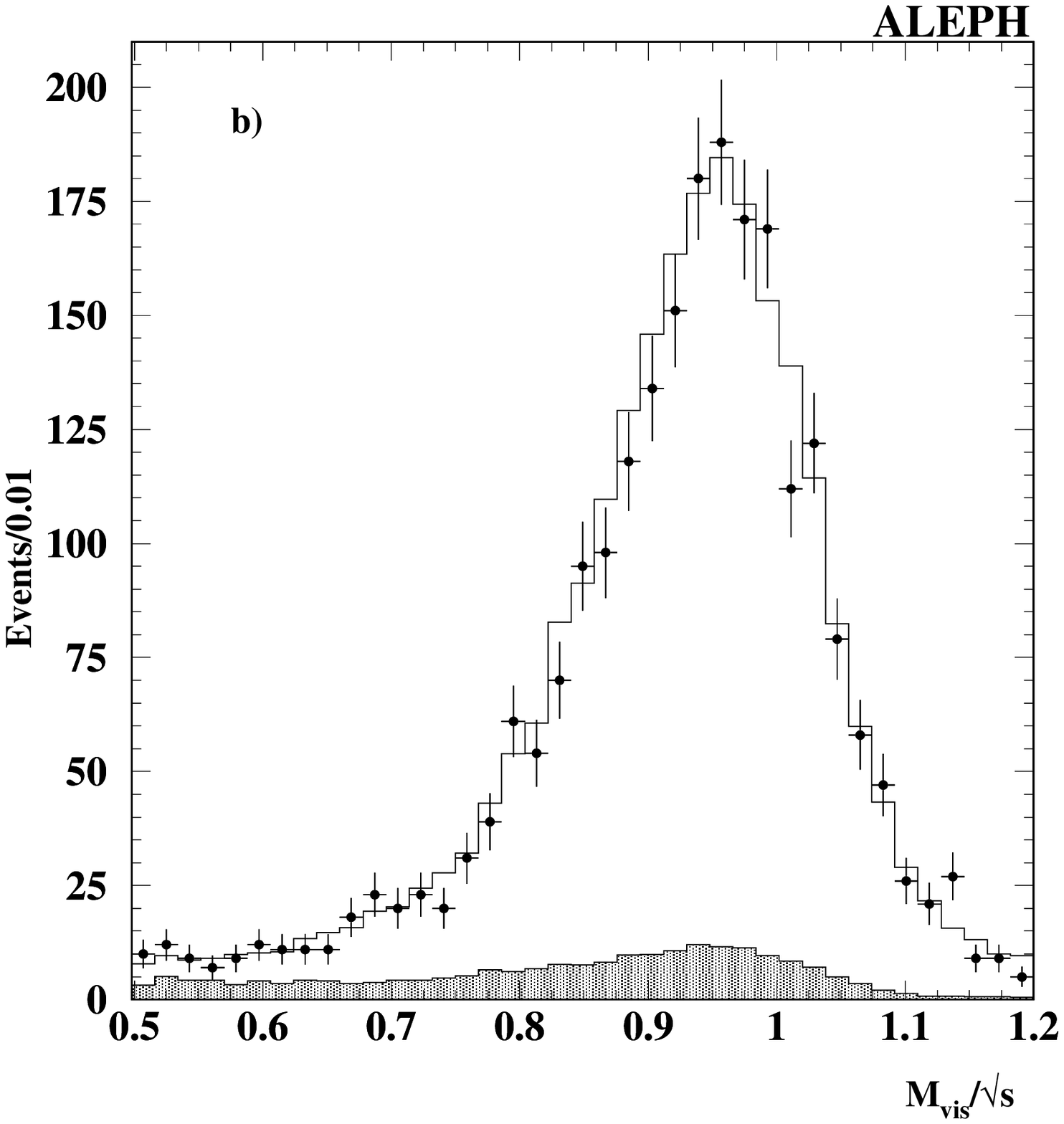}
                        \hspace{0.1cm}
%                            \end{sloppypar}
%                           \end{flushleft}
                           \end{center}
                       { \lblcaption{fig:mvis_207}{For exclusive hadronic final states,
                at $\sqrt{s}=207\,{\rm GeV}$, 
                the distributions of the event thrust (a) and of the visible mass
             (normalized to the collision energy)
            for events with thrust $>$ 0.85 (b). The data
            (dots) are compared to the
                        Monte Carlo expectation (histograms).
                 The shaded areas show the background
                       contribution.}}
\end{figure}

\clearpage
\newpage
\begin{figure}[htbp]
\begin{center}
 \includegraphics[width=15cm]{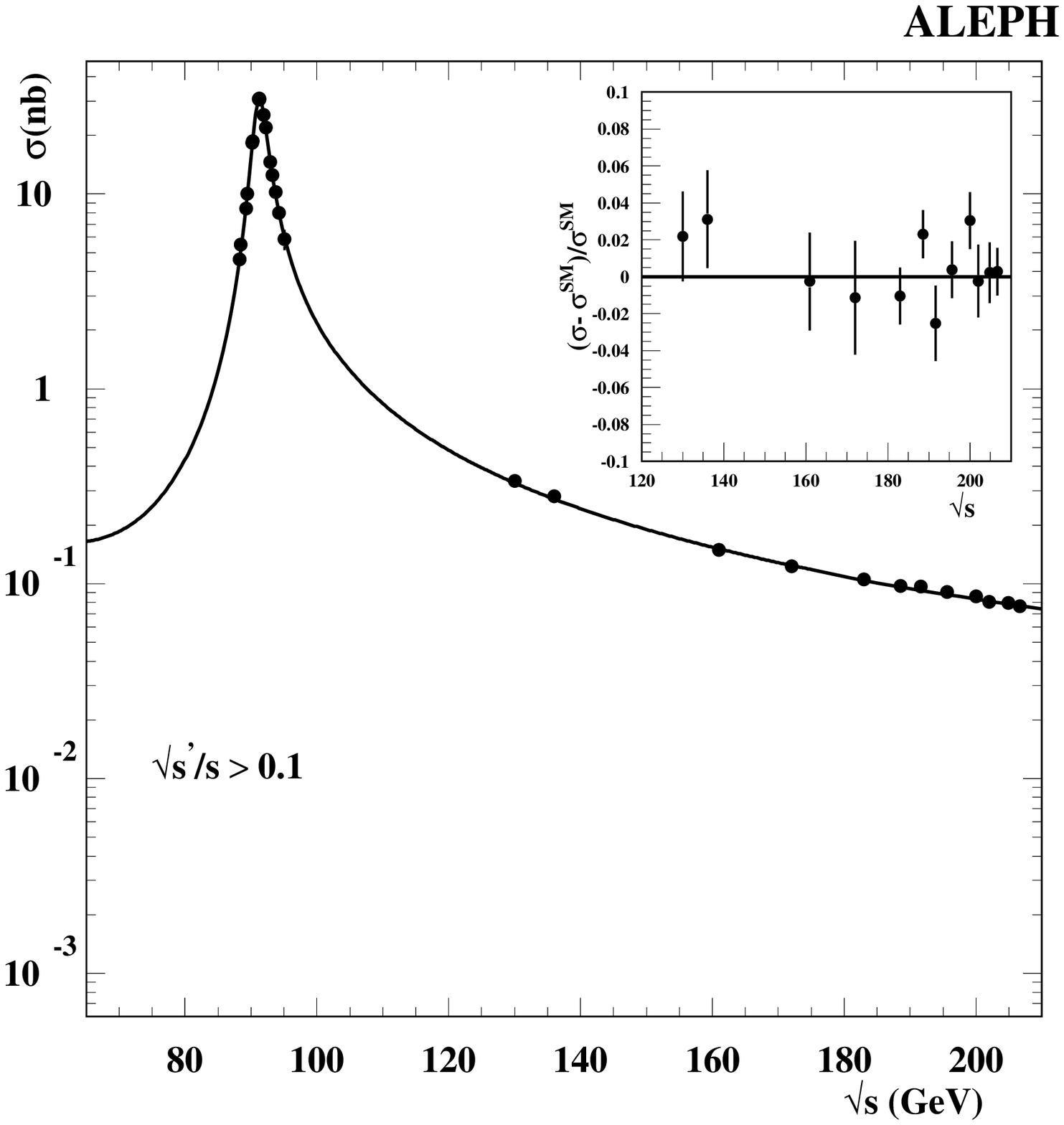}
                        \hspace{0.1cm}
                           \end{center}
                        {\lblcaption{fig:cross_had_01}{Measured inclusive
            hadronic cross section (dots), as a function of the centre-of-mass
            energy. The full curve indicates the {\tt ZFITTER} prediction.
                        The insert shows the difference between the measurements
            and the Standard Model predictions, normalized to the predicted cross
            sections. Measurements at centre-of-mass energies below 189~GeV are
from Ref.~\cite{aleph-lep1} and \cite{aleph-183}.}}
\end{figure}

\clearpage
\newpage
\begin{figure}[htbp]
\begin{center}
                        \includegraphics[width=14cm]{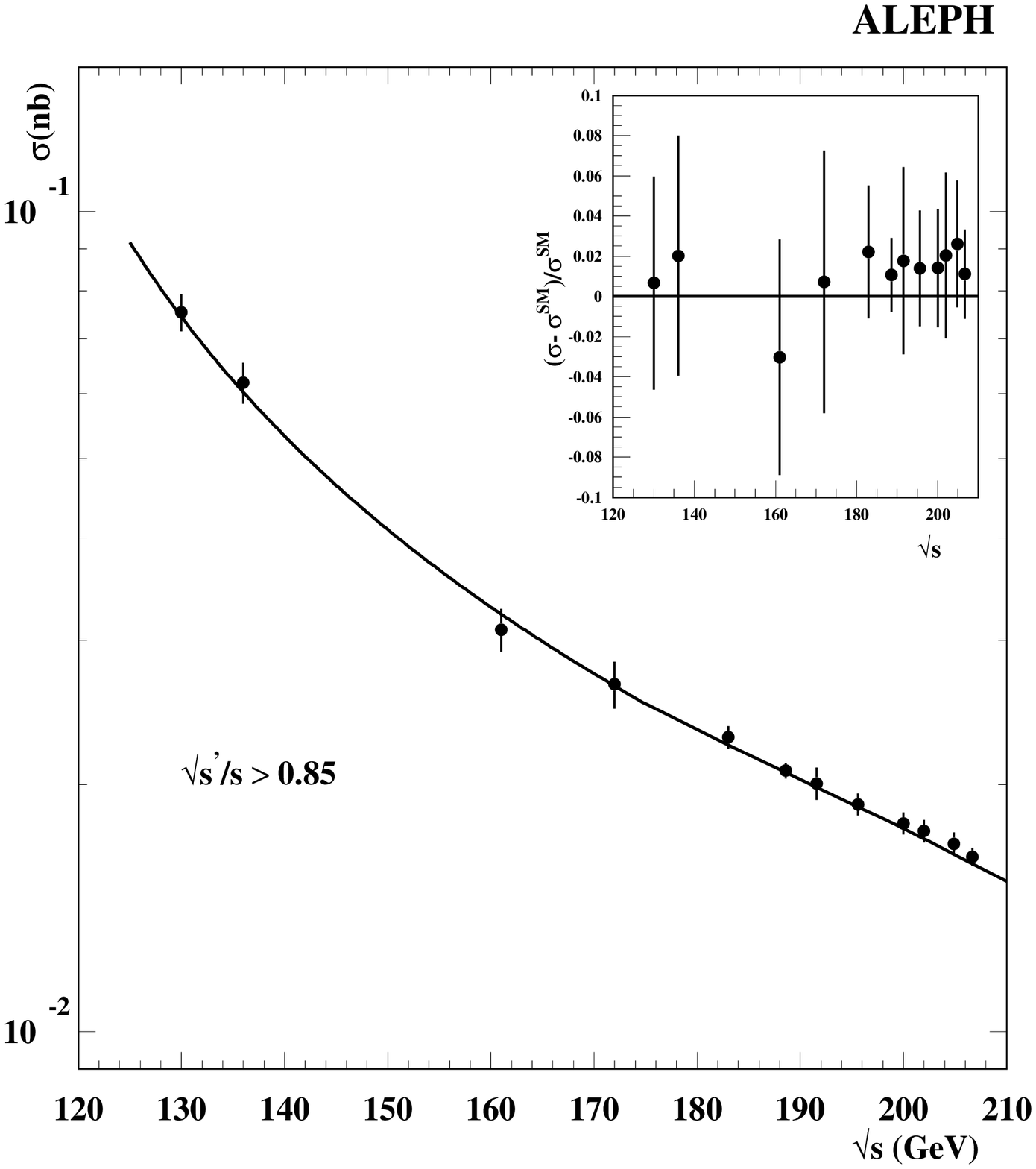}
                      \end{center}
                        {\lblcaption{fig:cross_had_085}{Measured exclusive
            hadronic cross section (dots), as a function of the centre-of-mass
            energy. The full curve indicates the {\tt ZFITTER} prediction.
                    The insert shows the difference between the measurements
            and the Standard Model predictions, normalized to the predicted cross
            sections. The deviation of the seven highest energy points with respect to the Standard
Model prediction, with correlations taken into account, corresponds to 
$\chi^2/{\rm d.o.f.} =1.8$. Measurements at centre-of-mass energies below 189~GeV are
from Ref.~\cite{aleph-183}, corrected for the different definition of the exclusive
processes.}}
 \end{figure}

\clearpage
\newpage
\begin{figure}[htbp]
\begin{center}
\includegraphics[width=15cm]{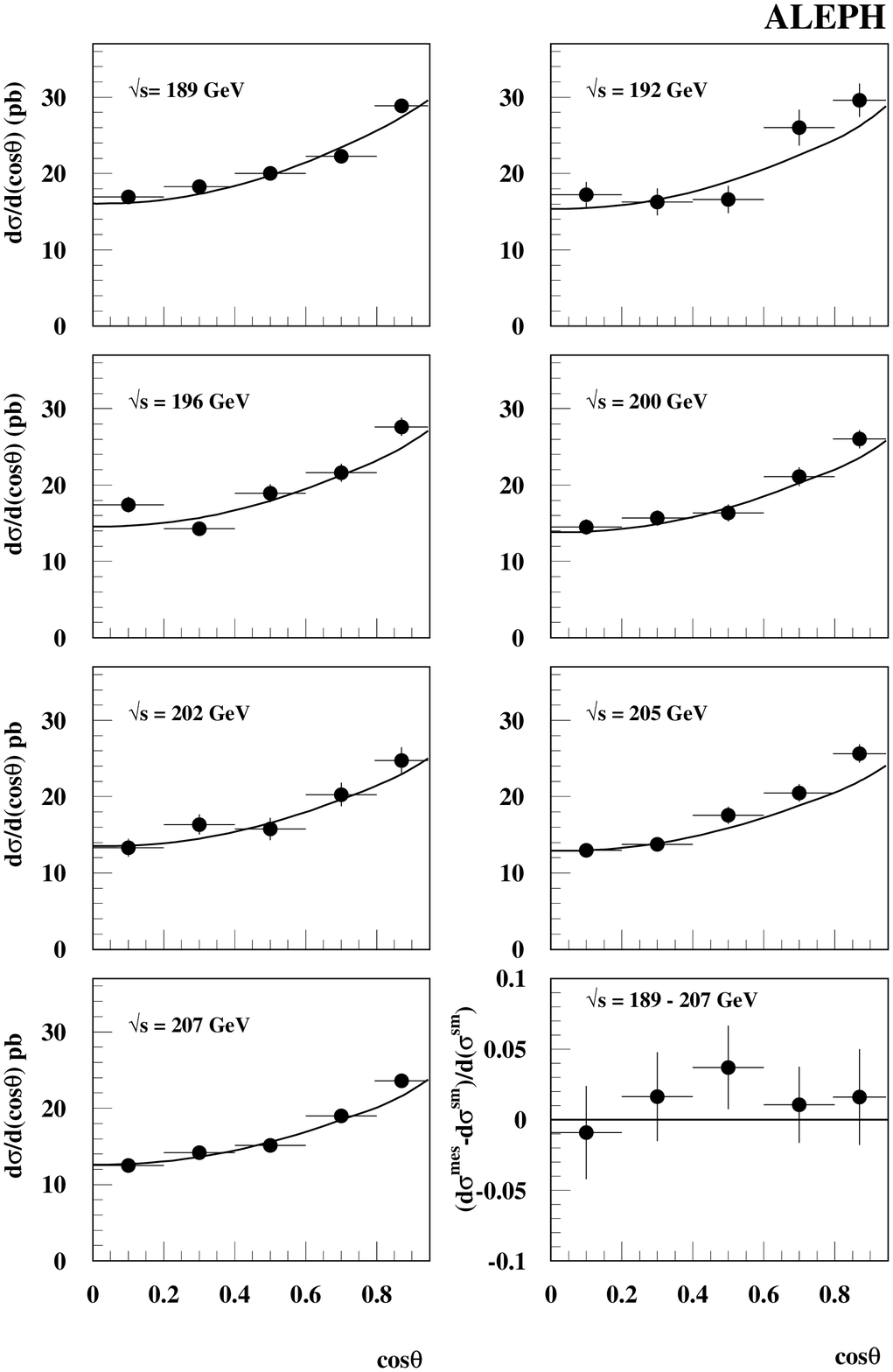}
   \end{center}
                         {\lblcaption{fig:diff-cross-had}{Measured differential
             cross sections for \qq\ exclusive production (closed symbols),
              as a function of polar
             angle and for several centre-of-mass energies.
      The full curves indicate the {\tt ZFITTER} predictions.
      The right bottom plot shows the luminosity weighted sum of the 
differences between the measurements and the Standard Model predictions,
normalized to the luminosity weighted sum of the predicted cross sections, for all
energy points together.}}
                       \end{figure}

\clearpage
\newpage
\begin{figure}[htbp]
\begin{center}
\includegraphics[width=14cm]{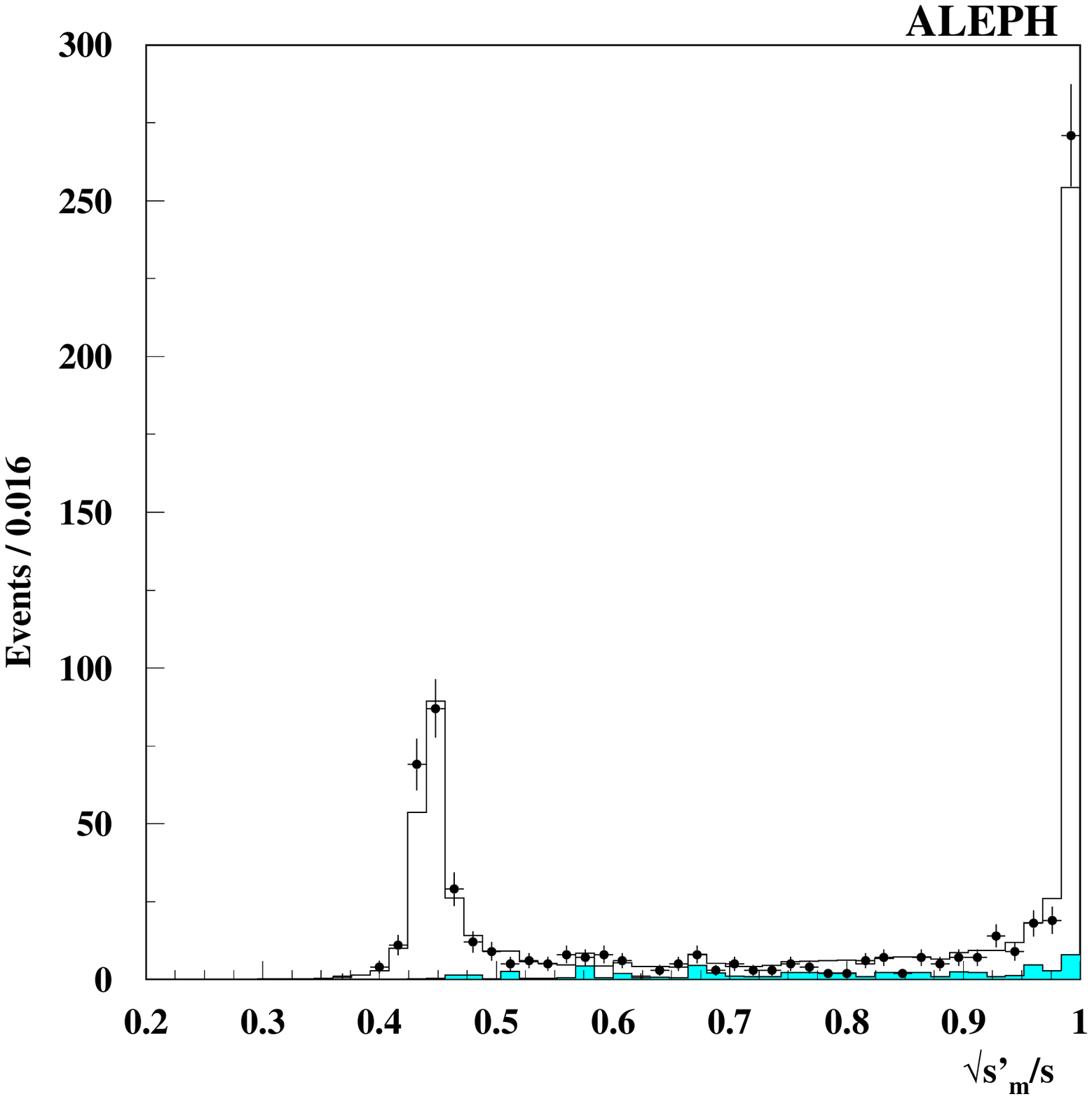}
 \end{center}
% \vspace{-6cm}
 {\lblcaption{fig:spsmu_207}{The $\sqrt{s^{\prime}_m/s}$ distribution
 for muon-pair events collected at $\sqrt{s}=$207~GeV.
  The data (dots) are compared to the
 Monte Carlo expectation (histogram). The filled area shows the background
 contribution.}}
 \end{figure}

\clearpage
\newpage
\begin{figure}[htbp]
\begin{center}
\includegraphics[width=15cm]{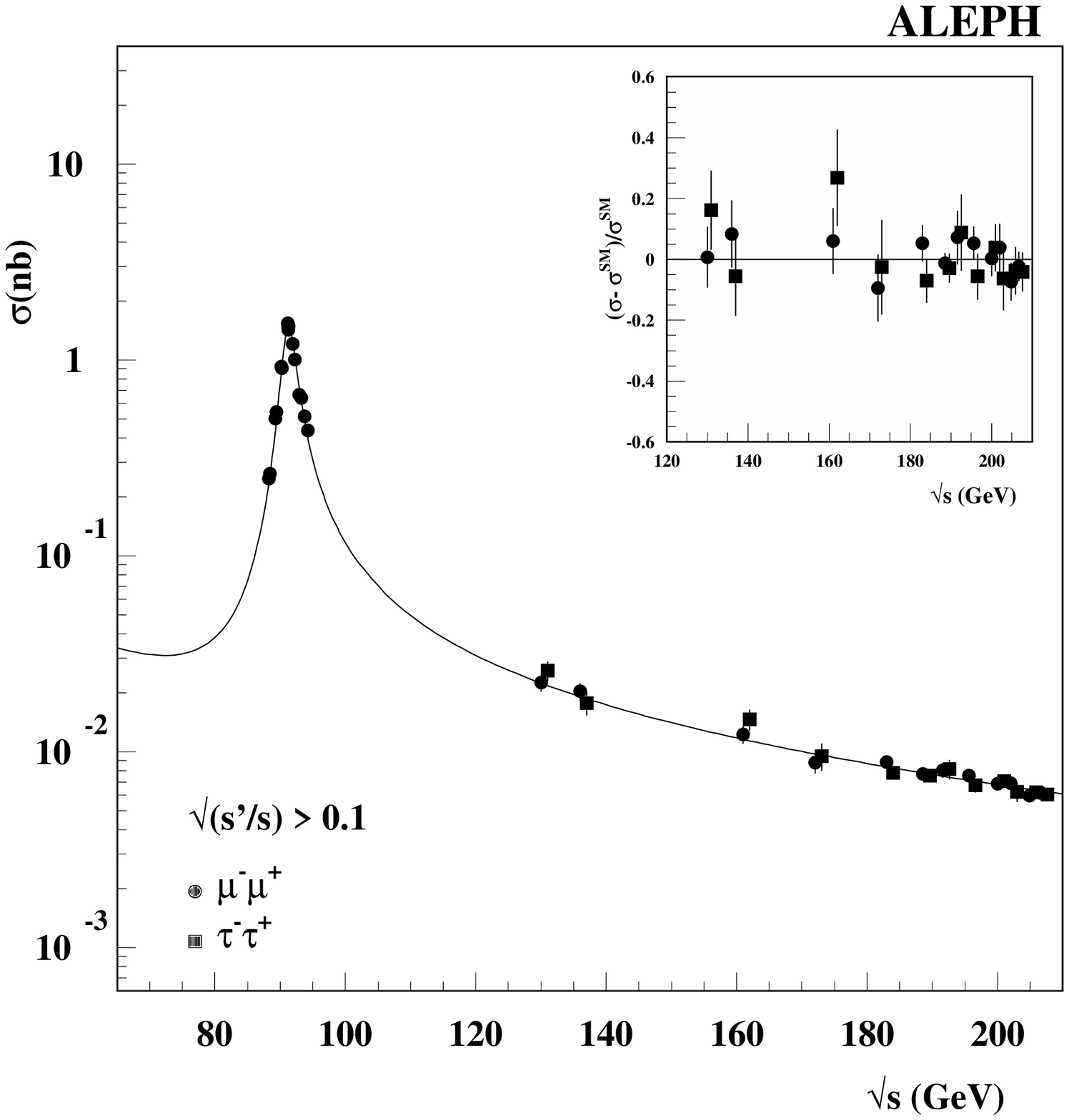}
\end{center}
\caption{\label{fig:lep_line_01}{Measured inclusive cross
sections for muon-pair (dots) and tau-pair (squares) production,
as a function of the centre-of-mass
energy (points are shifted for visibility). The full curve indicates the
{\tt ZFITTER} prediction (for the \mr{\mu^+\mu^-} channel). The insert shows the difference
between the measurements and the Standard Model predictions, normalized to
the predicted cross sections.  Measurements at centre-of-mass energies below 189~GeV are
from Ref.~\cite{aleph-lep1} and \cite{aleph-183}.}}
\end{figure}

\clearpage
\newpage
\begin{figure}[htbp]
\begin{center}
\includegraphics[width=14cm]{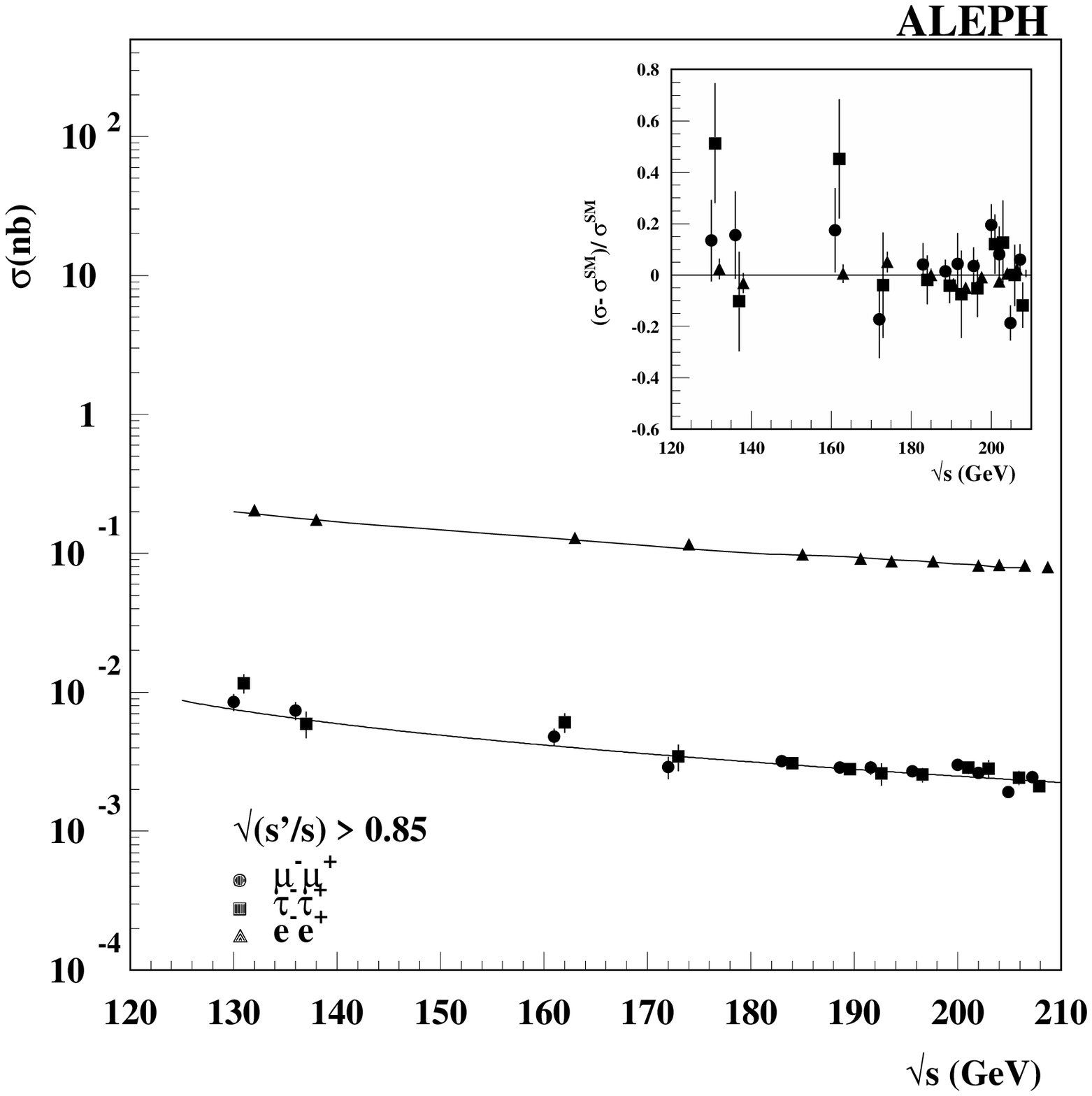}
\end{center}
\caption{\label{fig:lep_line_085}{Measured exclusive
             cross sections for di-lepton production ($|\cos\theta^*|<0.9$ range
             for the \mr {e^+e^-} channel),
             as a function of the centre-of-mass
            energy (points are shifted for visibility). 
            The full curvess indicate the {\tt BHWIDE} prediction
            for the \mr{e^+e^-} channel and the {\tt ZFITTER} prediction for
            the \mr{\mu^+\mu^-} channel.
            The insert shows the difference between the measurements
            and the Standard Model predictions, normalized to the predicted 
            cross sections.  Measurements at centre-of-mass energies below 189~GeV are
from Ref.~\cite{aleph-183}, corrected for the different definition of the exclusive 
processes.}}
\end{figure}

\clearpage
\newpage
\begin{figure}[htbp]
\begin{center}
\includegraphics[width=15cm]{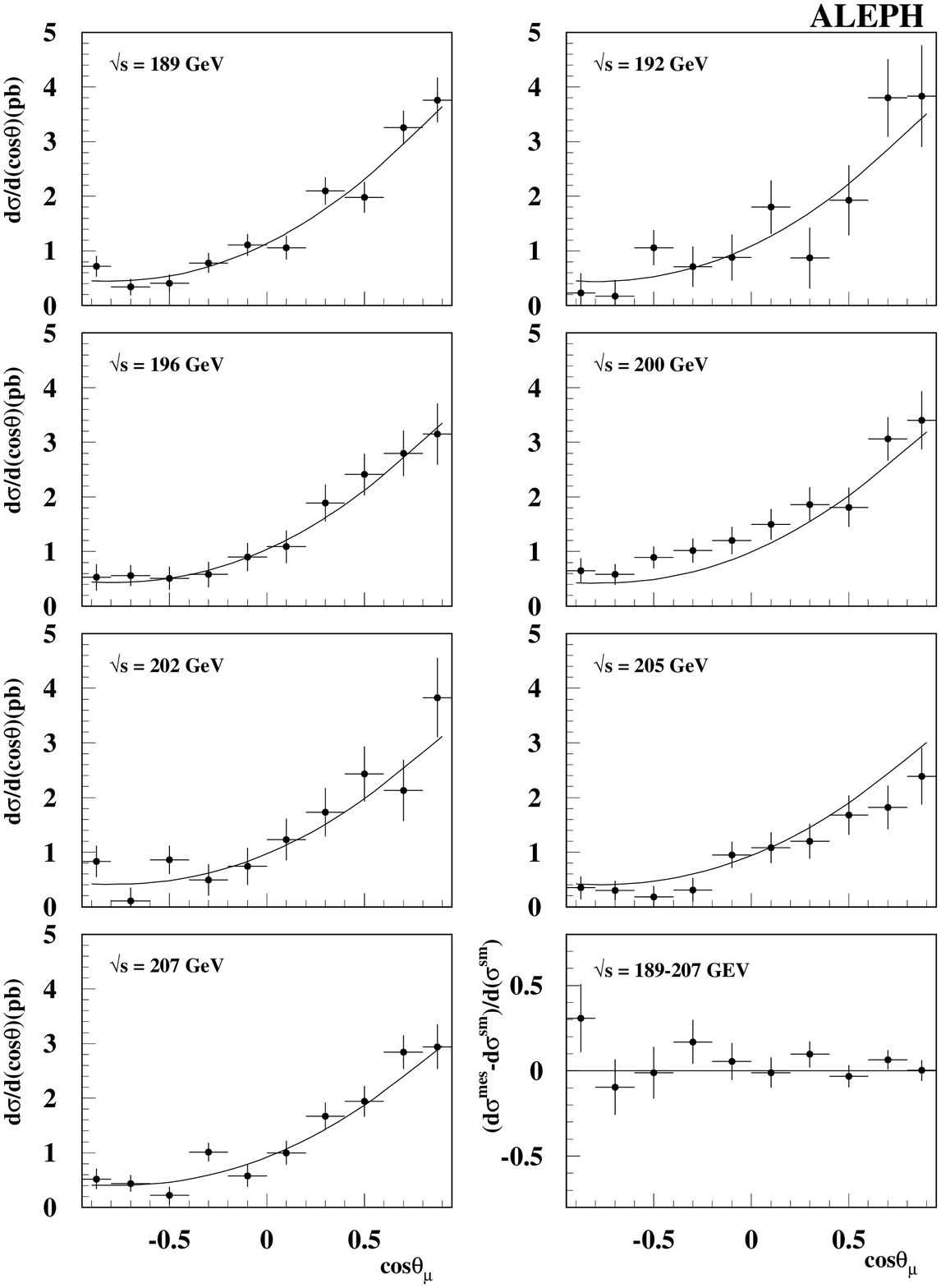}
\end{center}
\caption{\label{fig:xdiff_m}{Measured differential
             cross sections for \mr {\mu^+\mu^-} production (dots),
             as a function of polar
             angle and for several centre-of-mass energies.
      The full curves indicate the {\tt ZFITTER} predictions.
      The right bottom plot shows the luminosity weighted sum of the 
differences between the measurements and the Standard Model predictions,
 normalized to the luminosity weighted sum of the predicted cross sections, for all
energy points together.}}
\end{figure}

\clearpage
\newpage
\begin{figure}[htbp]
\begin{center}
\includegraphics[width=15cm]{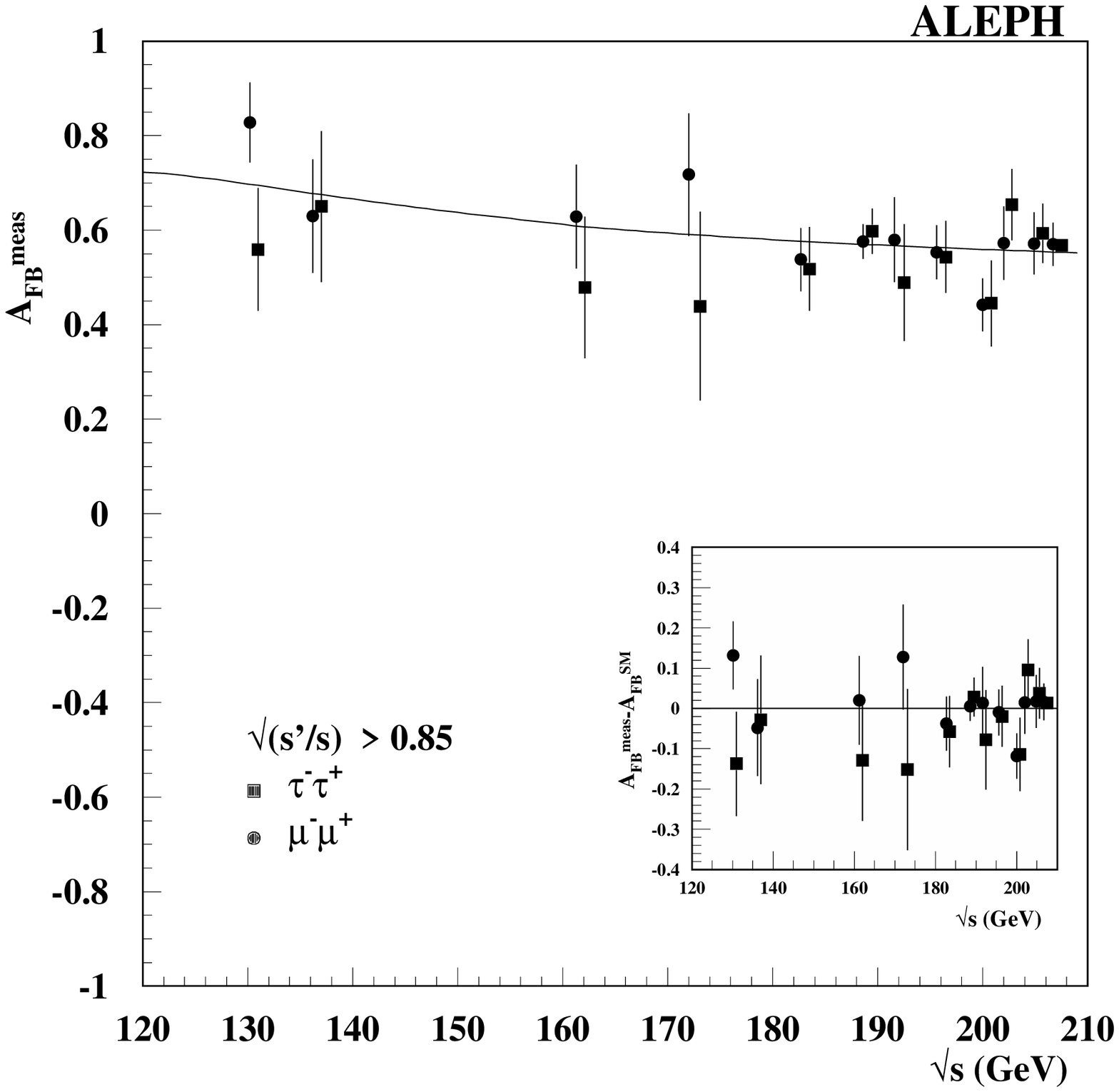}
\end{center}
\caption{\label{fig:asym}{Measured forward-backward asymmetries for muon-pair
and tau-pair production, over the $|\cos\theta|<0.95$ range, 
as a function of the centre-of-mass energy.
 The curve indicates the {\tt ZFITTER} prediction.
 The insert shows the difference between the measurements and the Standard Model 
  predictions.
  Measurements at centre-of-mass energies below 189~GeV are from Ref.~\cite{aleph-183},
corrected for the different definition of the exclusive processes.}}
\end{figure}

\clearpage
\newpage
\begin{figure}[htbp]
\begin{center}
\includegraphics[width=14cm]{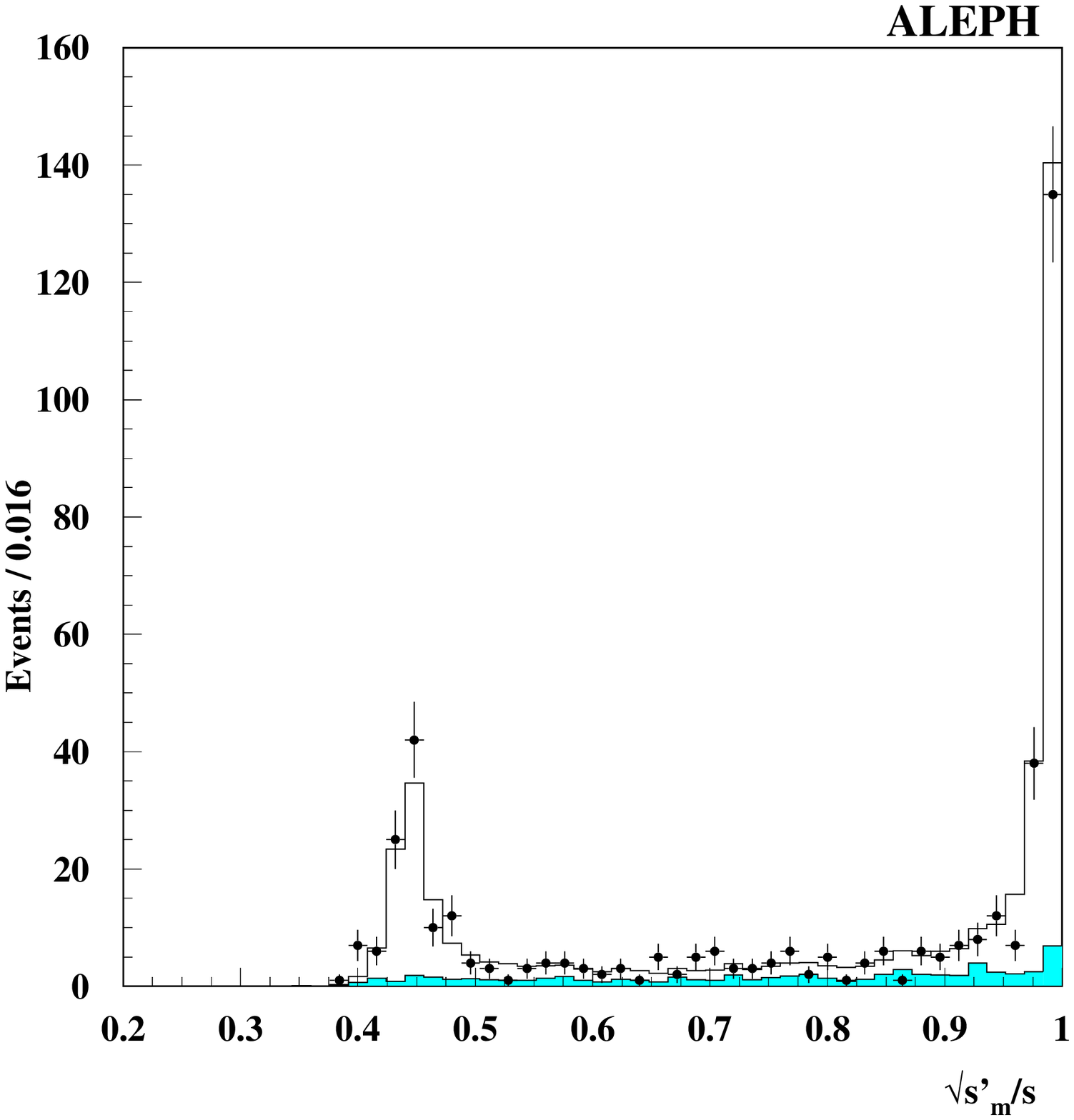}
 \end{center}
% \vspace{-6cm}
 {\lblcaption{fig:spstau_207}{The $\sqrt{s^{\prime}_m/s}$ distribution
 for tau-pair events collected at $\sqrt{s}=$207~GeV.
  The data (dots) are compared to the
 Monte Carlo expectation (histogram). The filled area shows the background
 contribution.}}
 \end{figure}

\clearpage
\newpage
\begin{figure}[htbp]
\begin{center}
\includegraphics[width=15cm]{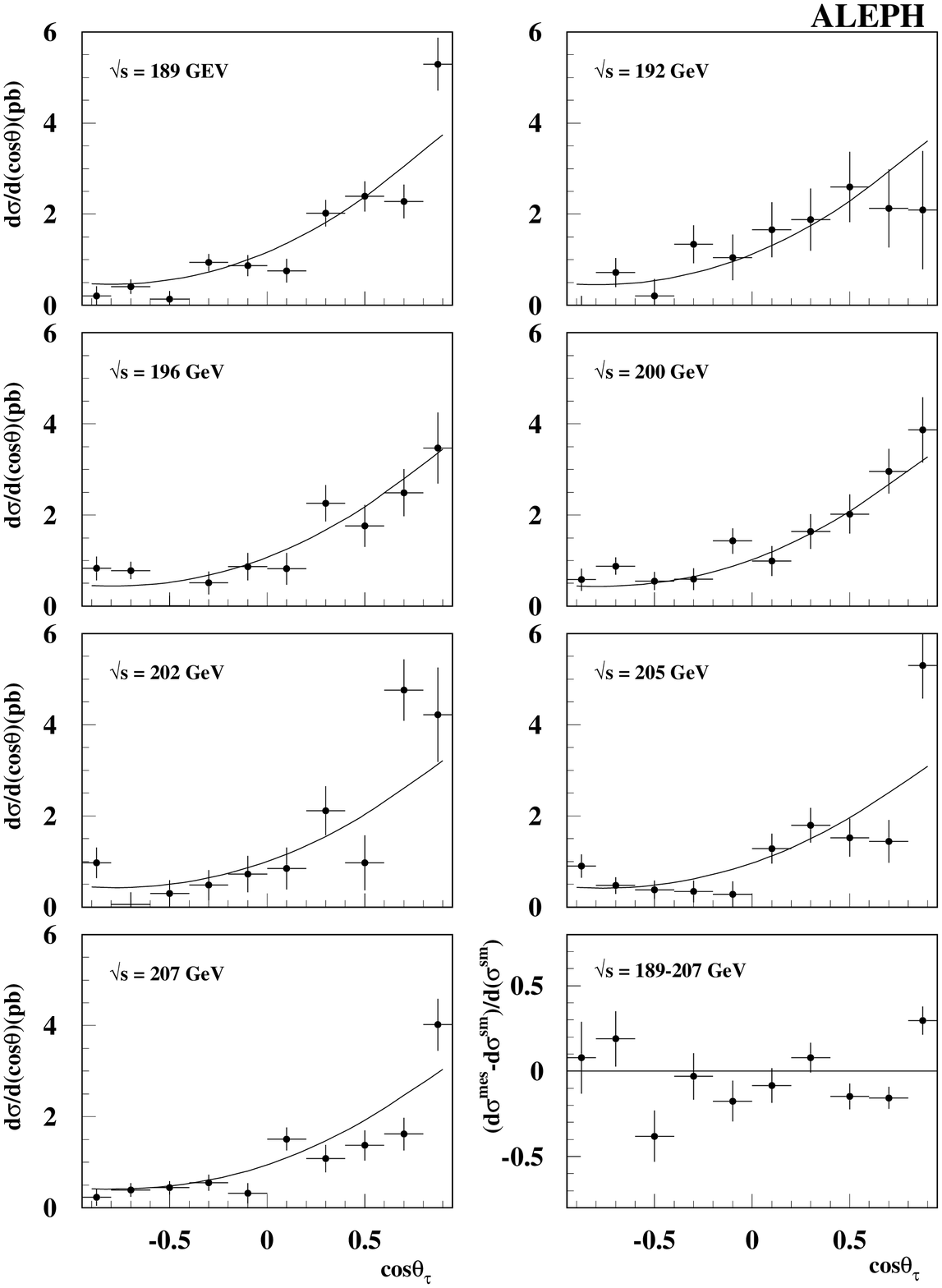}
\end{center}
\caption{\label{fig:xdiff_t}{Measured differential
             cross sections for \mr {\tau^+\tau^-} production (dots),
              as a function of polar
             angle and for several centre-of-mass energies.
      The full curves indicate the {\tt ZFITTER} predictions.
      The right bottom plot shows the luminosity weighted sum of the
differences between the measurements and the Standard Model predictions,
 normalized to the luminosity weighted sum of the predicted cross sections,  
for all energy points together. }}
\end{figure}

\clearpage
\newpage
\begin{figure}[htbp]
\begin{center}
\includegraphics[width=14cm]{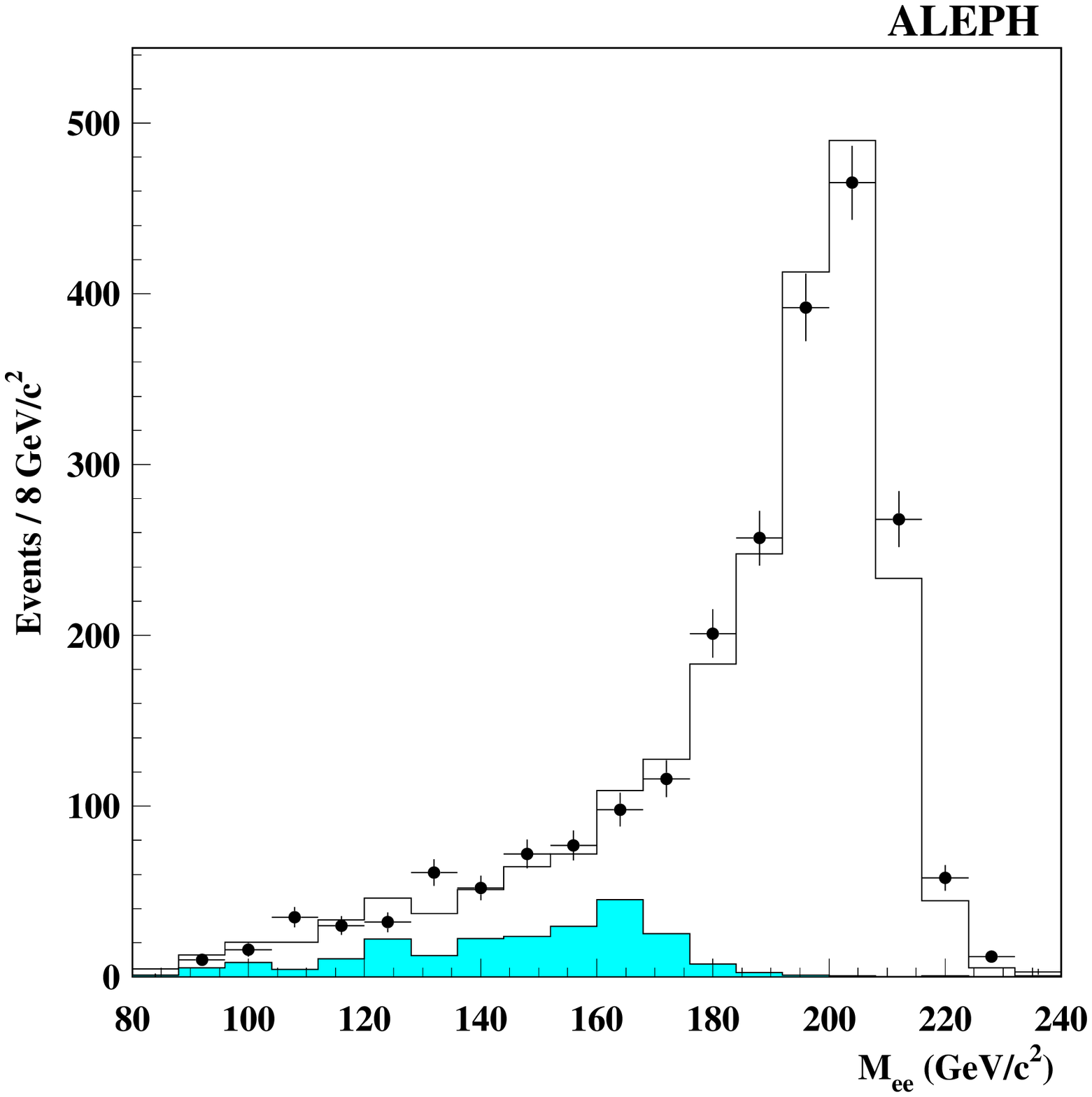}
 \end{center}
% \vspace{-6cm}
 {\lblcaption{fig:mee_207}{The $M_{{\rm e}^+{\rm e}^-}$ distribution
 for electron-pair events collected at $\sqrt{s}=$207~GeV in the angular range
$-0.9<\cos\theta <0.7$.
  The data (dots) are compared to the
 Monte Carlo expectation (histogram). The filled area shows the background
 contribution.}}
 \end{figure}

\clearpage
\newpage
\begin{figure}[htbp]
\begin{center}
\includegraphics[width=15cm]{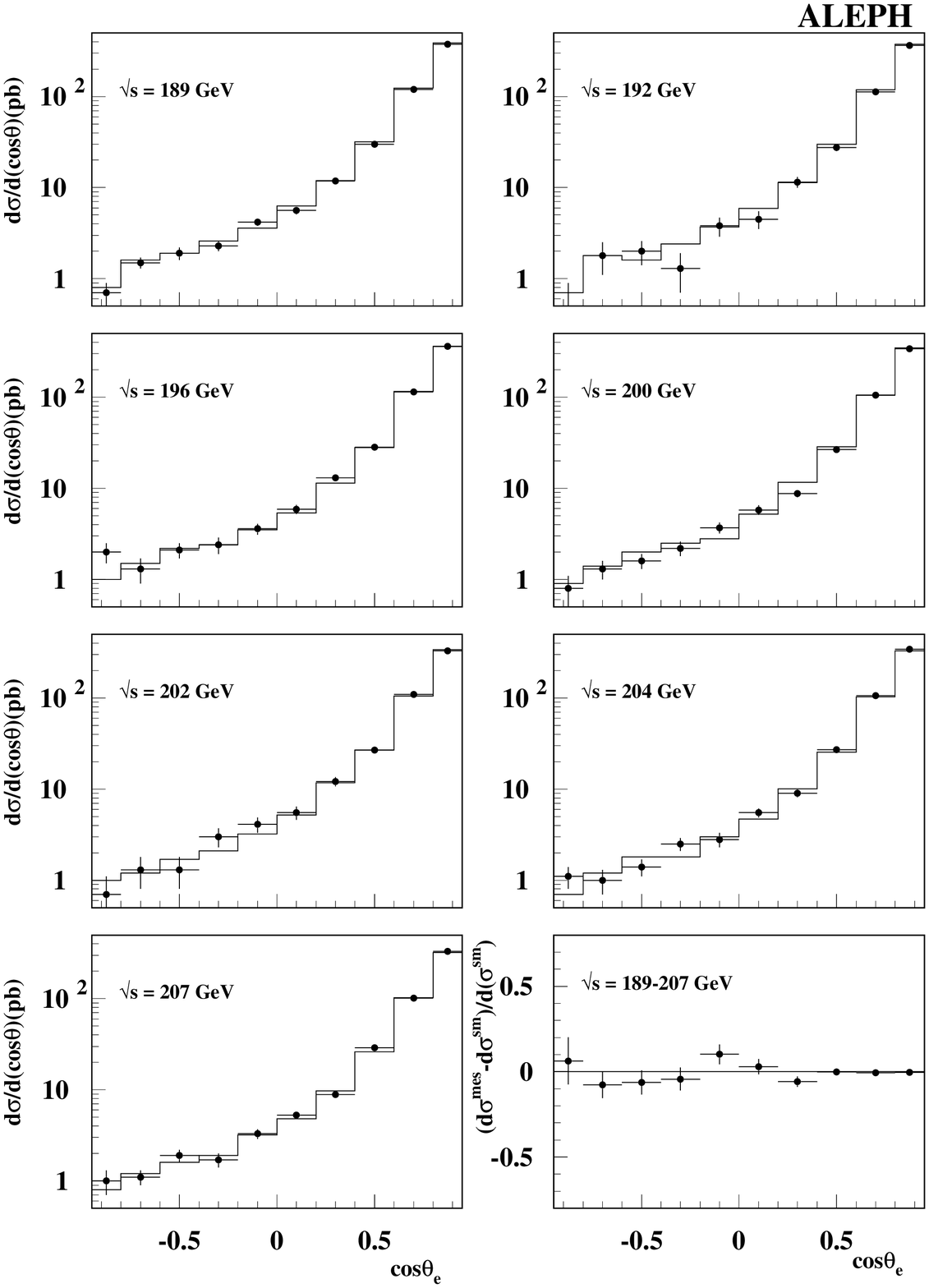}
\end{center}
\caption{\label{fig:xdiff_e}{Measured differential
             cross sections for \mr {e^+e^-} production (dots),
              as a function of polar
             angle and for several centre-of-mass energies.
      The full curves indicate the {\tt BHWIDE} predictions.
      The right bottom plot shows the luminosity weighted sum of the
differences between the measurements and the Standard Model predictions,
 normalized to the luminosity weighted sum of the predicted cross sections, for all
energy points together.}}
\end{figure}

\clearpage
\newpage

\clearpage
\newpage

%\begin{figure}[htbp]
%\begin{center}
%\includegraphics[width=14cm]{afb-lept-2.eps}
%\includegraphics[width=14cm]{afb-lept-3.eps}
%\end{center}
%\caption{\label{fig:afbb-ratio} {Normalized difference between
%data and Monte Carlo for the mean (dots) and the RMS (open symbols)
%values of the b-jet charge, as a function of the $p_T$ cut
%on the positive (top) or negative (bottom) lepton. Data collected
%at all LEP2 energies are combined.}}
%\end{figure}

\clearpage
\newpage

\clearpage
\newpage

\clearpage
\newpage

\clearpage

\clearpage
\newpage

\renewcommand{\thesection}{\Alph{section}}
\setcounter{section}{0}
\section*{Appendix : {\tt ZFITTER} steering flags and input parameters \label{annexe-zf}}

 The main flags used in the {\tt ZFITTER} Monte Carlo are listed below:

\begin{itemize}

    \item{General flags.}
    As advised in Ref.~\cite{Lep2-WS}, CONV=2 is used to properly take into account
    the angular dependence of the electroweak box diagrams; INTF=2
    is used to include the contribution from initial and
    final state interferences; BOXD=2 is selected to take into account box
    contributions.

    \item{Hadron flags.} FINR=0 describes $\sqrt{s^\prime}$ as the mass of the propagator
    excluding FSR.
    \item{Lepton flags.} FINR=1 describes $\sqrt{s^\prime}$ as the
    invariant mass of the outgoing lepton system.
\end{itemize}

 The input parameters required by {\tt ZFITTER} have been set as follows:

\begin{itemize}
    \item{ $m_{\rm {Z}}\ =\ 91.1875 \ {\rm GeV}/c^2 $}
    \item{ $m_t\ =\ 174.3 \ {\rm GeV}/c^2 $}
    \item{ $m_H\ =\ 150 \ {\rm GeV}/c^2 $}
    \item{ $1/\alpha_{QED}^{\rm Z}\ = 128.896 $}
    \item{ $\alpha_S = 0.118 $ }
 %   \item{ $\Delta\alpha^{(5)}_{\rm had}=0.02761 $}

\end{itemize}

\end{document}

%% file: authb.tex
%------------------------------------------------------------------------
% authob12pt.tex
% authors' list for papers at LEP 1.5 and 2 energies
%-----------------------------------------------------------------------
\pagestyle{empty}
\newpage
\small
%
% remember the old settings
%
\newlength{\saveparskip}
\newlength{\savetextheight}
\newlength{\savetopmargin}
\newlength{\savetextwidth}
\newlength{\saveoddsidemargin}
\newlength{\savetopsep}
\setlength{\saveparskip}{\parskip}
\setlength{\savetextheight}{\textheight}
\setlength{\savetopmargin}{\topmargin}
\setlength{\savetextwidth}{\textwidth}
\setlength{\saveoddsidemargin}{\oddsidemargin}
\setlength{\savetopsep}{\topsep}
%
% text dimensions for the author list
%
\setlength{\parskip}{0.0cm}
\setlength{\textheight}{25.0cm}
\setlength{\topmargin}{-1.5cm}
\setlength{\textwidth}{16 cm}
\setlength{\oddsidemargin}{-0.0cm}
\setlength{\topsep}{1mm}
\pretolerance=10000
%%\begin{document}
%\centerline{EUROPEAN ORGANIZATION FOR NUCLEAR RESEARCH}
%\centerline{EUROPEAN LABORATORY FOR PARTICLE PHYSICS (CERN)}
%\vspace{1cm}
%\begin{flushright}CERN-EP-2000-
%\12 May 2003 - last update
%\end{flushright}
\centerline{\large\bf The ALEPH Collaboration}
\footnotesize
\vspace{0.5cm}
{\raggedbottom
\begin{sloppypar}
\samepage\noindent
S.~Schael
\nopagebreak
\begin{center}
\parbox{15.5cm}{\sl\samepage
Physikalisches Institut der RWTH-Aachen, D-52056 Aachen, Germany}
\end{center}\end{sloppypar}
\vspace{2mm}
\begin{sloppypar}
\noindent
R.~Barate,
R.~Bruneli\`ere,
I.~De~Bonis,
D.~Decamp,
C.~Goy,
S.~J\'ez\'equel,
J.-P.~Lees,
F.~Martin,
E.~Merle,
\mbox{M.-N.~Minard},
B.~Pietrzyk,
B.~Trocm\'e
\nopagebreak
\begin{center}
\parbox{15.5cm}{\sl\samepage
Laboratoire de Physique des Particules (LAPP), IN$^{2}$P$^{3}$-CNRS,
F-74019 Annecy-le-Vieux Cedex, France}
\end{center}\end{sloppypar}
\vspace{2mm}
\begin{sloppypar}
\noindent
S.~Bravo,
M.P.~Casado,
M.~Chmeissani,
J.M.~Crespo,
E.~Fernandez,
M.~Fernandez-Bosman,
Ll.~Garrido,$^{15}$
M.~Martinez,
A.~Pacheco,
H.~Ruiz
\nopagebreak
\begin{center}
\parbox{15.5cm}{\sl\samepage
Institut de F\'{i}sica d'Altes Energies, Universitat Aut\`{o}noma
de Barcelona, E-08193 Bellaterra (Barcelona), Spain$^{7}$}
\end{center}\end{sloppypar}
\vspace{2mm}
\begin{sloppypar}
\noindent
A.~Colaleo,
D.~Creanza,
N.~De~Filippis,
M.~de~Palma,
G.~Iaselli,
G.~Maggi,
M.~Maggi,
S.~Nuzzo,
A.~Ranieri,
G.~Raso,$^{24}$
F.~Ruggieri,
G.~Selvaggi,
L.~Silvestris,
P.~Tempesta,
A.~Tricomi,$^{3}$
G.~Zito
\nopagebreak
\begin{center}
\parbox{15.5cm}{\sl\samepage
Dipartimento di Fisica, INFN Sezione di Bari, I-70126 Bari, Italy}
\end{center}\end{sloppypar}
\vspace{2mm}
\begin{sloppypar}
\noindent
X.~Huang,
J.~Lin,
Q. Ouyang,
T.~Wang,
Y.~Xie,
R.~Xu,
S.~Xue,
J.~Zhang,
L.~Zhang,
W.~Zhao
\nopagebreak
\begin{center}
\parbox{15.5cm}{\sl\samepage
Institute of High Energy Physics, Academia Sinica, Beijing, The People's
Republic of China$^{8}$}
\end{center}\end{sloppypar}
\vspace{2mm}
\begin{sloppypar}
\noindent
D.~Abbaneo,
T.~Barklow,$^{26}$
O.~Buchm\"uller,$^{26}$
M.~Cattaneo,
B.~Clerbaux,$^{23}$
H.~Drevermann,
R.W.~Forty,
M.~Frank,
F.~Gianotti,
J.B.~Hansen,
J.~Harvey,
D.E.~Hutchcroft,$^{30}$,
P.~Janot,
B.~Jost,
M.~Kado,$^{2}$
P.~Mato,
A.~Moutoussi,
F.~Ranjard,
L.~Rolandi,
D.~Schlatter,
F.~Teubert,
A.~Valassi,
I.~Videau
\nopagebreak
\begin{center}
\parbox{15.5cm}{\sl\samepage
European Laboratory for Particle Physics (CERN), CH-1211 Geneva 23,
Switzerland}
\end{center}\end{sloppypar}
\vspace{2mm}
\begin{sloppypar}
\noindent
F.~Badaud,
S.~Dessagne,
A.~Falvard,$^{20}$
D.~Fayolle,
P.~Gay,
J.~Jousset,
B.~Michel,
S.~Monteil,
D.~Pallin,
J.M.~Pascolo,
P.~Perret
\nopagebreak
\begin{center}
\parbox{15.5cm}{\sl\samepage
Laboratoire de Physique Corpusculaire, Universit\'e Blaise Pascal,
IN$^{2}$P$^{3}$-CNRS, Clermont-Ferrand, F-63177 Aubi\`{e}re, France}
\end{center}\end{sloppypar}
\vspace{2mm}
\begin{sloppypar}
\noindent
J.D.~Hansen,
J.R.~Hansen,
P.H.~Hansen,
A.C.~Kraan,
B.S.~Nilsson
\nopagebreak
\begin{center}
\parbox{15.5cm}{\sl\samepage
Niels Bohr Institute, 2100 Copenhagen, DK-Denmark$^{9}$}
\end{center}\end{sloppypar}
\vspace{2mm}
\begin{sloppypar}
\noindent
A.~Kyriakis,
C.~Markou,
E.~Simopoulou,
A.~Vayaki,
K.~Zachariadou
\nopagebreak
\begin{center}
\parbox{15.5cm}{\sl\samepage
Nuclear Research Center Demokritos (NRCD), GR-15310 Attiki, Greece}
\end{center}\end{sloppypar}
\vspace{2mm}
\begin{sloppypar}
\noindent
A.~Blondel,$^{12}$
\mbox{J.-C.~Brient},
F.~Machefert,
A.~Roug\'{e},
H.~Videau
\nopagebreak
\begin{center}
\parbox{15.5cm}{\sl\samepage
Laoratoire Leprince-Ringuet, Ecole
Polytechnique, IN$^{2}$P$^{3}$-CNRS, \mbox{F-91128} Palaiseau Cedex, France}
\end{center}\end{sloppypar}
\vspace{2mm}
\begin{sloppypar}
\noindent
V.~Ciulli,
E.~Focardi,
G.~Parrini
\nopagebreak
\begin{center}
\parbox{15.5cm}{\sl\samepage
Dipartimento di Fisica, Universit\`a di Firenze, INFN Sezione di Firenze,
I-50125 Firenze, Italy}
\end{center}\end{sloppypar}
\vspace{2mm}
\begin{sloppypar}
\noindent
A.~Antonelli,
M.~Antonelli,
G.~Bencivenni,
F.~Bossi,
G.~Capon,
F.~Cerutti,
V.~Chiarella,
P.~Laurelli,
G.~Mannocchi,$^{5}$
G.P.~Murtas,
L.~Passalacqua
\nopagebreak
\begin{center}
\parbox{15.5cm}{\sl\samepage
Laboratori Nazionali dell'INFN (LNF-INFN), I-00044 Frascati, Italy}
\end{center}\end{sloppypar}
\vspace{2mm}
%\pagebreak
\begin{sloppypar}
\noindent
J.~Kennedy,
J.G.~Lynch,
P.~Negus,
V.~O'Shea,
A.S.~Thompson
\nopagebreak
\begin{center}
\parbox{15.5cm}{\sl\samepage
Department of Physics and Astronomy, University of Glasgow, Glasgow G12
8QQ,United Kingdom$^{10}$}
\end{center}\end{sloppypar}
\vspace{2mm}
%\pagebreak
\begin{sloppypar}
\noindent
S.~Wasserbaech
\nopagebreak
\begin{center}
\parbox{15.5cm}{\sl\samepage
Utah Valley State College, Orem, UT 84058, U.S.A.}
\end{center}\end{sloppypar}
\vspace{2mm}
%\pagebreak
\begin{sloppypar}
\noindent
R.~Cavanaugh,$^{4}$
S.~Dhamotharan,$^{21}$
C.~Geweniger,
P.~Hanke,
V.~Hepp,
E.E.~Kluge,
A.~Putzer,
H.~Stenzel,
K.~Tittel,
M.~Wunsch$^{19}$
\nopagebreak
\begin{center}
\parbox{15.5cm}{\sl\samepage
Kirchhoff-Institut f\"ur Physik, Universit\"at Heidelberg, D-69120
Heidelberg, Germany$^{16}$}
\end{center}\end{sloppypar}
\vspace{2mm}
\begin{sloppypar}
\noindent
R.~Beuselinck,
W.~Cameron,
G.~Davies,
P.J.~Dornan,
M.~Girone,$^{1}$
R.D.~Hill,
N.~Marinelli,
J.~Nowell,
S.A.~Rutherford,
J.K.~Sedgbeer,
J.C.~Thompson,$^{14}$
R.~White
\nopagebreak
\begin{center}
\parbox{15.5cm}{\sl\samepage
Department of Physics, Imperial College, London SW7 2BZ,
United Kingdom$^{10}$}
\end{center}\end{sloppypar}
\vspace{2mm}
\begin{sloppypar}
\noindent
V.M.~Ghete,
P.~Girtler,
E.~Kneringer,
D.~Kuhn,
G.~Rudolph
\nopagebreak
\begin{center}
\parbox{15.5cm}{\sl\samepage
Institut f\"ur Experimentalphysik, Universit\"at Innsbruck, A-6020
Innsbruck, Austria$^{18}$}
\end{center}\end{sloppypar}
\vspace{2mm}
\begin{sloppypar}
\noindent
E.~Bouhova-Thacker,
C.K.~Bowdery,
D.P.~Clarke,
G.~Ellis,
A.J.~Finch,
F.~Foster,
G.~Hughes,
R.W.L.~Jones,
M.R.~Pearson,
N.A.~Robertson,
M.~Smizanska
\nopagebreak
\begin{center}
\parbox{15.5cm}{\sl\samepage
Department of Physics, University of Lancaster, Lancaster LA1 4YB,
United Kingdom$^{10}$}
\end{center}\end{sloppypar}
\vspace{2mm}
\begin{sloppypar}
\noindent
O.~van~der~Aa,
C.~Delaere,$^{28}$
G.Leibenguth,$^{31}$
V.~Lemaitre$^{29}$
\nopagebreak
\begin{center}
\parbox{15.5cm}{\sl\samepage
Institut de Physique Nucl\'eaire, D\'epartement de Physique, Universit\'e Catholique de Louvain, 1348 Louvain-la-Neuve, Belgium}
\end{center}\end{sloppypar}
\vspace{2mm}
\begin{sloppypar}
\noindent
U.~Blumenschein,
F.~H\"olldorfer,
K.~Jakobs,
F.~Kayser,
A.-S.~M\"uller,
G.~Quast$^{32}$
B.~Renk,
H.-G.~Sander,
S.~Schmeling,
H.~Wachsmuth,
C.~Zeitnitz,
T.~Ziegler
\nopagebreak
\begin{center}
\parbox{15.5cm}{\sl\samepage
Institut f\"ur Physik, Universit\"at Mainz, D-55099 Mainz, Germany$^{16}$}
\end{center}\end{sloppypar}
\vspace{2mm}
\begin{sloppypar}
\noindent
A.~Bonissent,
P.~Coyle,
C.~Curtil,
A.~Ealet,
D.~Fouchez,
P.~Payre,
A.~Tilquin
\nopagebreak
\begin{center}
\parbox{15.5cm}{\sl\samepage
Centre de Physique des Particules de Marseille, Univ M\'editerran\'ee,
IN$^{2}$P$^{3}$-CNRS, F-13288 Marseille, France}
\end{center}\end{sloppypar}
\vspace{2mm}
\begin{sloppypar}
\noindent
F.~Ragusa
\nopagebreak
\begin{center}
\parbox{15.5cm}{\sl\samepage
Dipartimento di Fisica, Universit\`a di Milano e INFN Sezione di
Milano, I-20133 Milano, Italy.}
\end{center}\end{sloppypar}
\vspace{2mm}
\begin{sloppypar}
\noindent
A.~David,
H.~Dietl,$^{33}$
G.~Ganis,$^{27}$
K.~H\"uttmann,
G.~L\"utjens,
W.~M\"anner$^{33}$,
\mbox{H.-G.~Moser},
R.~Settles,
M.~Villegas,
G.~Wolf
\nopagebreak
\begin{center}
\parbox{15.5cm}{\sl\samepage
Max-Planck-Institut f\"ur Physik, Werner-Heisenberg-Institut,
D-80805 M\"unchen, Germany\footnotemark[16]}
\end{center}\end{sloppypar}
\vspace{2mm}
\begin{sloppypar}
\noindent
J.~Boucrot,
O.~Callot,
M.~Davier,
L.~Duflot,
\mbox{J.-F.~Grivaz},
Ph.~Heusse,
A.~Jacholkowska,$^{6}$
L.~Serin,
\mbox{J.-J.~Veillet}
\nopagebreak
\begin{center}
\parbox{15.5cm}{\sl\samepage
Laboratoire de l'Acc\'el\'erateur Lin\'eaire, Universit\'e de Paris-Sud,
IN$^{2}$P$^{3}$-CNRS, F-91898 Orsay Cedex, France}
\end{center}\end{sloppypar}
\vspace{2mm}
\begin{sloppypar}
\noindent
%\samepage
P.~Azzurri, 
G.~Bagliesi,
T.~Boccali,
L.~Fo\`a,
A.~Giammanco,
A.~Giassi,
F.~Ligabue,
A.~Messineo,
F.~Palla,
G.~Sanguinetti,
A.~Sciab\`a,
G.~Sguazzoni,
P.~Spagnolo,
R.~Tenchini,
A.~Venturi,
P.G.~Verdini
\samepage
\begin{center}
\parbox{15.5cm}{\sl\samepage
Dipartimento di Fisica dell'Universit\`a, INFN Sezione di Pisa,
e Scuola Normale Superiore, I-56010 Pisa, Italy}
\end{center}\end{sloppypar}
\vspace{2mm}
\begin{sloppypar}
\noindent
O.~Awunor,
G.A.~Blair,
G.~Cowan,
A.~Garcia-Bellido,
M.G.~Green,
T.~Medcalf,$^{25}$
A.~Misiejuk,
J.A.~Strong,$^{25}$
P.~Teixeira-Dias
\nopagebreak
\begin{center}
\parbox{15.5cm}{\sl\samepage
Department of Physics, Royal Holloway \& Bedford New College,
University of London, Egham, Surrey TW20 OEX, United Kingdom$^{10}$}
\end{center}\end{sloppypar}
\vspace{2mm}
\begin{sloppypar}
\noindent
R.W.~Clifft,
T.R.~Edgecock,
P.R.~Norton,
I.R.~Tomalin,
J.J.~Ward
\nopagebreak
\begin{center}
\parbox{15.5cm}{\sl\samepage
Particle Physics Dept., Rutherford Appleton Laboratory,
Chilton, Didcot, Oxon OX11 OQX, United Kingdom$^{10}$}
\end{center}\end{sloppypar}
\vspace{2mm}
%\pagebreak
\begin{sloppypar}
\noindent
\mbox{B.~Bloch-Devaux},
D.~Boumediene,
P.~Colas,
B.~Fabbro,
E.~Lan\c{c}on,
\mbox{M.-C.~Lemaire},
E.~Locci,
P.~Perez,
J.~Rander,
B.~Tuchming,
B.~Vallage
\nopagebreak
\begin{center}
\parbox{15.5cm}{\sl\samepage
CEA, DAPNIA/Service de Physique des Particules,
CE-Saclay, F-91191 Gif-sur-Yvette Cedex, France$^{17}$}
\end{center}\end{sloppypar}
%\nopagebreak
\vspace{2mm}
\begin{sloppypar}
\noindent
A.M.~Litke,
G.~Taylor
\nopagebreak
\begin{center}
\parbox{15.5cm}{\sl\samepage
Institute for Particle Physics, University of California at
Santa Cruz, Santa Cruz, CA 95064, USA$^{22}$}
\end{center}\end{sloppypar}
%\pagebreak
\vspace{2mm}
\begin{sloppypar}
\noindent
C.N.~Booth,
S.~Cartwright,
F.~Combley,$^{25}$
P.N.~Hodgson,
M.~Lehto,
L.F.~Thompson
\nopagebreak
\begin{center}
\parbox{15.5cm}{\sl\samepage
Department of Physics, University of Sheffield, Sheffield S3 7RH,
United Kingdom$^{10}$}
\end{center}\end{sloppypar}
\vspace{2mm}
\begin{sloppypar}
\noindent
A.~B\"ohrer,
S.~Brandt,
C.~Grupen,
J.~Hess,
A.~Ngac,
G.~Prange
\nopagebreak
\begin{center}
\parbox{15.5cm}{\sl\samepage
Fachbereich Physik, Universit\"at Siegen, D-57068 Siegen, Germany$^{16}$}
\end{center}\end{sloppypar}
\vspace{2mm}
\begin{sloppypar}
\noindent
C.~Borean,
G.~Giannini
\nopagebreak
\begin{center}
\parbox{15.5cm}{\sl\samepage
Dipartimento di Fisica, Universit\`a di Trieste e INFN Sezione di Trieste,
I-34127 Trieste, Italy}
\end{center}\end{sloppypar}
\vspace{2mm}
\begin{sloppypar}
\noindent
H.~He,
J.~Putz,
J.~Rothberg
\nopagebreak
\begin{center}
\parbox{15.5cm}{\sl\samepage
Experimental Elementary Particle Physics, University of Washington, Seattle,
WA 98195 U.S.A.}
\end{center}\end{sloppypar}
\vspace{2mm}
\begin{sloppypar}
\noindent
S.R.~Armstrong,
K.~Berkelman,
K.~Cranmer,
D.P.S.~Ferguson,
Y.~Gao,$^{13}$
S.~Gonz\'{a}lez,
O.J.~Hayes,
H.~Hu,
S.~Jin,
J.~Kile,
P.A.~McNamara III,
J.~Nielsen,
Y.B.~Pan,
\mbox{J.H.~von~Wimmersperg-Toeller}, 
W.~Wiedenmann,
J.~Wu,
Sau~Lan~Wu,
X.~Wu,
G.~Zobernig
\nopagebreak
\begin{center}
\parbox{15.5cm}{\sl\samepage
Department of Physics, University of Wisconsin, Madison, WI 53706,
USA$^{11}$}
\end{center}\end{sloppypar}
\vspace{2mm}
\begin{sloppypar}
\noindent
G.~Dissertori
\nopagebreak
\begin{center}
\parbox{15.5cm}{\sl\samepage
Institute for Particle Physics, ETH H\"onggerberg, 8093 Z\"urich,
Switzerland.}
\end{center}\end{sloppypar}
}
\footnotetext[1]{Also at CERN, 1211 Geneva 23, Switzerland.}
\footnotetext[2]{Now at Fermilab, PO Box 500, MS 352, Batavia, IL 60510, USA}
\footnotetext[3]{Also at Dipartimento di Fisica di Catania and INFN Sezione di
 Catania, 95129 Catania, Italy.}
\footnotetext[4]{Now at University of Florida, Department of Physics, Gainesville, Florida 32611-8440, USA}
\footnotetext[5]{Also IFSI sezione di Torino, INAF, Italy.}
\footnotetext[6]{Also at Groupe d'Astroparticules de Montpellier, Universit\'{e} de Montpellier II, 34095, Montpellier, France.}
\footnotetext[7]{Supported by CICYT, Spain.}
\footnotetext[8]{Supported by the National Science Foundation of China.}
\footnotetext[9]{Supported by the Danish Natural Science Research Council.}
\footnotetext[10]{Supported by the UK Particle Physics and Astronomy Research
Council.}
\footnotetext[11]{Supported by the US Department of Energy, grant
DE-FG0295-ER40896.}
\footnotetext[12]{Now at Departement de Physique Corpusculaire, Universit\'e de
Gen\`eve, 1211 Gen\`eve 4, Switzerland.}
\footnotetext[13]{Also at Department of Physics, Tsinghua University, Beijing, The People's Republic of China.}
\footnotetext[14]{Supported by the Leverhulme Trust.}
\footnotetext[15]{Permanent address: Universitat de Barcelona, 08208 Barcelona,
Spain.}
\footnotetext[16]{Supported by Bundesministerium f\"ur Bildung
und Forschung, Germany.}
\footnotetext[17]{Supported by the Direction des Sciences de la
Mati\`ere, C.E.A.}
\footnotetext[18]{Supported by the Austrian Ministry for Science and Transport.}
\footnotetext[19]{Now at SAP AG, 69185 Walldorf, Germany}
\footnotetext[20]{Now at Groupe d' Astroparticules de Montpellier, Universit\'e de Montpellier II, 34095 Montpellier, France.}
\footnotetext[21]{Now at BNP Paribas, 60325 Frankfurt am Mainz, Germany}
\footnotetext[22]{Supported by the US Department of Energy,
grant DE-FG03-92ER40689.}
\footnotetext[23]{Now at Institut Inter-universitaire des hautes Energies (IIHE), CP 230, Universit\'{e} Libre de Bruxelles, 1050 Bruxelles, Belgique}
\footnotetext[24]{Now at Dipartimento di Fisica e Tecnologie Relative, Universit\`a di Palermo, Palermo, Italy.}
\footnotetext[25]{Deceased.}
\footnotetext[26]{Now at SLAC, Stanford, CA 94309, U.S.A}
\footnotetext[27]{Now at CERN, 1211 Geneva 23, Switzerland}
\footnotetext[28]{Research Fellow of the Belgium FNRS}
\footnotetext[29]{Research Associate of the Belgium FNRS} 
\footnotetext[30]{Now at Liverpool University, Liverpool L69 7ZE, United Kingdom} 
\footnotetext[31]{Supported by the Federal Office for Scientific, Technical and Cultural Affairs through
the Interuniversity Attraction Pole P5/27} 
\footnotetext[32]{Now at Institut f\"ur Experimentelle Kernphysik Universit\"at Karlsruhe, Germany}
\footnotetext[33]{Now at Henryk Niewodnicznski Institute of Nuclear Physics, Polish Academy of Sciences, Cracow, Poland}   
\setlength{\parskip}{\saveparskip}
\setlength{\textheight}{\savetextheight}
\setlength{\topmargin}{\savetopmargin}
\setlength{\textwidth}{\savetextwidth}
\setlength{\oddsidemargin}{\saveoddsidemargin}
\setlength{\topsep}{\savetopsep}
\normalsize
\newpage
\pagestyle{plain}
\setcounter{page}{1}

%% file: hadron.tex
\section{Hadronic final states \label{sec:hadronic}}

The selection of hadronic final states is described in Ref.~\cite{aleph-183};
events with high charged-track multiplicity are required.

  For inclusive processes, the cross sections
 are determined, after background subtraction, using a global
 efficiency correction.
   Backgrounds and selection efficiencies, which are both obtained
  from Monte Carlo studies, are listed in
  Table~\ref{tab:effqq} as a function of centre-of-mass energy.
   The main background arises from W pair and Z
 pair production. The contribution from $\gamma\gamma$ interactions
 is suppressed by requiring the event visible mass to be larger
 than 50~GeV/c$^2$.
  The measured cross sections are presented
 in Table~\ref{tab:crossqq}, together with the
  {\tt ZFITTER} predictions over the same acceptance as the experimental
  measurements. 

For the exclusive cross sections the events are divided into two hemispheres
(hereafter called jets) with respect to the thrust axis, determined after
removing the ISR photons.
  The quantity $\sqrt{s^{\prime}_{m}/s}$ is measured
 from the reconstructed jet directions and 
a cut $\sqrt{s^{\prime}_{m}/s}>0.85$ is applied. 
The $\sqrt{s^{\prime}_{m}/s}$ distribution  for the data collected at
$\sqrt{s}$=207~GeV is displayed in Fig.~\ref{fig:sprim_207}, 
together with the expected background.
 In the exclusive region, the latter is dominated by:

\begin{itemize}
\item W-pair production. For these events, the thrust ($T$) distribution
 extends to lower values than for \qq\ events,
 as shown in Fig.~\ref{fig:mvis_207}a. A cut $T>0.85$
 rejects approximately $80$\% of this background.
\item Fermion-pair events where, due to photon
radiation by both colliding electrons, the measured
$\sqrt{s^{\prime}_{m}/s}$ from the jets directions is above 0.85.
 This background is reduced
by requiring that the event visible mass, calculated excluding
ISR photons with energies above 10~GeV, is greater than $70$\%
of the centre-of-mass energy. The residual background is called
 ``radiative background".
 Figure~\ref{fig:mvis_207}b shows the
visible mass distribution for events
with $\sqrt{s^{\prime}/s}\;>\;0.85$ and thrust value exceeding
0.85. The systematic uncertainty on this radiative background accounts for the
small discrepancy visible in Fig.~\ref{fig:sprim_207}.

\end{itemize}

 The contribution from four-fermion processes other than
 WW production is found to be small. It is taken into
account by including an additional $0.1$\% systematic uncertainty on the
exclusive cross section measurements. Other
 systematic uncertainties
arise from the knowledge of the calorimeter calibration
 and of the detector response to the hadronization process. These
 uncertainties are taken as fully correlated between years. 
The evaluation of the detector response uncertainties includes the calorimeter
effects described in Ref.~\cite{aleph-wmass}, which were shown to have 
negligible impact on this measurement.

 The efficiencies for the exclusive process
and the background contributions are summarized
in Table~\ref{tab:effqq} and the measured cross sections are presented
in Table~\ref{tab:crossqq}.\

The systematic uncertainties for the inclusive and exclusive processes are listed
in Table~\ref{tab:error}. Figures~\ref{fig:cross_had_01} and~\ref{fig:cross_had_085}
 show the measured inclusive and exclusive \qq\ cross sections
  as a function of energy.
  The exclusive differential cross sections as a function of the thrust 
axis polar angle are shown in Fig.~\ref{fig:diff-cross-had}
  (in this case the selection efficiencies have been determined
 in angular bins).

%% file: dileptons_cor.tex
\section{Leptonic final states \label{sec:leptonic}}
For the ${\rm e}^+{\rm e}^- \rightarrow \mu^{+}\mu^{-}$ and 
${\rm e}^+{\rm e}^- \rightarrow\tau^{+}\tau^{-}$ 
channels, cross section measurements are
provided for the inclusive and exclusive processes as defined
in Section \ref{sec:signal}.
 The inclusive cross sections are determined after
background subtraction and a global efficiency correction, while
the exclusive cross sections are computed as the sum of the measured
cross sections in bins of $\cos\theta$. Asymmetries are extracted by
a counting method from the $\cos\theta^*$ distributions, where
$\theta^*$ is the scattering angle between the incoming e$^-$ and
the outgoing $\ell^-$ in the $\ell^+\ell^-$ rest frame.
 The asymmetry $A_{\rm FB}$ is defined as:

$$ A_{\rm FB} = \frac{N_{\rm F}-N_{\rm B}}{N_{\rm F}+N_{\rm B}}$$

\noindent where $N_{\rm F}$ and $N_{\rm B}$ are the numbers of events with the negative
 lepton in the forward and backward regions, respectively.
 Acceptance corrections, as well as corrections
 for asymmetric distributions of the main backgrounds,
  are determined with Monte Carlo samples.

 For the ${\rm e}^+{\rm e}^-\rightarrow {\rm e}^+{\rm e}^-$ channel,
  because of the dominant contribution from the
$t$-channel photon exchange, the cross section is provided only
for $\sqrt{s^\prime/s}>0.85$ over two angular ranges: $-0.9<
\cos\theta^*<0.9$ and
$-0.9<\cos\theta^*<0.7$.

For all leptonic channels, the background contamination,
 estimated from simulation,
stems from $\gamma\gamma$ processes, four-fermion final states
${\rm W^{+}W^{-}}$, ${\rm ZZ}$, ${\rm Ze^{+}e^{-}}$
and production of other di-fermion species.
As for the hadronic final state, for the exclusive selection only,
events reconstructed with $\sqrt{s_m^\prime/s}>0.85$ but with a
$\ell^+\ell^-$ invariant mass below $0.85\sqrt{s}$ are called
radiative event background.
%
%
%%%%%%%%%%%%%%%%%%%%%%%%
\subsection{The $\mu^{+}\mu^{-}$ channel}
The selection of  muon pairs is described in Ref.~\cite{aleph-183}.
 For the inclusive selection, the main background  comes from
$\gamma\gamma \rightarrow \mu^{+}\mu^{-}$ and is largely reduced  by
requiring that the invariant mass of the muon pair exceeds
$ 60 \  {\rm GeV}/c^{2}$. For the exclusive selection the background from radiative events
is removed by asking that the invariant mass of the muon pair
exceeds $0.8\sqrt{s}$.
The $\sqrt{s^{\prime}_{m}/s}$ distribution  for the data collected at
$\sqrt{s}$=207~GeV is displayed in Fig.~\ref{fig:spsmu_207}.

 The $\mu^{+}\mu^{-}$ selection efficiencies, evaluated  using the
{\tt KK} Monte Carlo, are listed in Table~\ref{tab:eff_back_mm}.
 The main systematic uncertainty is due to the simulation
of the muon identification efficiency and is estimated from the
difference between data and simulation for the
muon identification efficiency in muon-pair
events recorded at the Z peak.

The background contamination is also given in Table~\ref{tab:eff_back_mm}.
For the inclusive selection, a major contribution to the systematic
uncertainty on the estimated background comes from
the normalization of the $\gamma\gamma\rightarrow \mu^{+}\mu^{-}$ process,
 and is determined by comparing data
 and Monte Carlo in the  $\mu^{+}\mu^{-}$ mass range
$15 \ {\rm GeV}/c^{2} < M_{\mu^{+}\mu^{-}} < 50 \ {\rm GeV}/c^{2}$.
 Other systematic uncertainties on the inclusive background
arise from the knowledge of the $\textrm{$\tau^{+}\tau^{-}$}$,
${\rm W^{+}W^{-}}$, $\textrm{ZZ}$ and
${\rm Ze^{+}e^{-}}$ cross sections,
 and are at the level of $3$\%, $1$\%, $5$\% and $10$\%, respectively.
 For the exclusive selection, the dominant background
 systematic uncertainty comes from radiative events, and is 
 estimated from the difference between the data and the
 Monte Carlo prediction in the  region
 $ 60< M_{\mu^{+}\mu^{-}} < 150$~GeV/$c^{2}$.

    The measured cross sections are presented in Table~\ref{tab:cross_mm}
   and in Figs.~\ref{fig:lep_line_01} and~\ref{fig:lep_line_085}, and
   compared to the SM prediction from  {\tt ZFITTER}.
 The dominant
  contributions to the systematic uncertainties on the cross sections
 (Table~\ref{tab:syst_cross_mm}) come from the limited
 statistics of the Monte Carlo samples
  and from the knowledge of the integrated luminosity,
 of the muon identification efficiency and of the background contamination.

 The  differential cross sections are shown in
Table~\ref{tab:cross_diff_mm} and Fig.~\ref{fig:xdiff_m}, while
 the asymmetry results are presented in Table~\ref{tab:assy_mm} and in
Fig.~\ref{fig:asym}. The dominant
systematic uncertainty on the asymmetry comes from the statistical error
on the Monte Carlo based corrections to the $\mu^{+}\mu^{-}$ acceptance.
%
%
%%%%%%%%%%%%%%%%%%%%%%%%%%%%%%%%%%%%%%%%%%%%%%%

\subsection{ The $\tau^+\tau^-$ channel}

   As described in Ref.~\cite{aleph-183},
  the selection of tau pairs requires two collimated jets with low
 charged-track multiplicity. Each event is divided into two hemispheres
 and is accepted if at least one hemisphere contains a tau candidate
 decaying into either a muon, or charged hadrons, or charged hadrons plus one or
 more $\pi^0$.

  The dominant backgrounds are reduced in the following way.
  Criteria against the Bhabha process
 are applied to events containing two
 high-momentum charged tracks.
      An additional cut on 
    the polar angle of both tau-jet candidates 
    is introduced ($|\cos\theta|<0.92$), in order to
    accept only tracks for which the ionisation estimator $dE/dx$, used to 
reject electron candidates,
     is accurately determined.
 WW events are rejected by requiring the acoplanarity angle between the
two tau candidates to be smaller than 250~mrad.
 Di-muon events are removed by demanding that one of the two
 hemispheres does not
 contain a muon. Finally,
the tau-pair  visible invariant mass is required to exceed $25\,{\rm
GeV}/c^{2}$ in order to reduce the $\gamma\gamma \to
\ell^{+}\ell^{-}$ contamination.
The $\sqrt{s^{\prime}_{m}/s}$ distribution  for the data collected at
$\sqrt{s}$=207~GeV is displayed in Fig.~\ref{fig:spstau_207}.

 The resulting
 selection efficiencies and the total background contamination 
are listed in Table~\ref{tab:eff_back_tt}. 
  For the inclusive selection, the systematic uncertainty on the
  dominant $\gamma\gamma \to \ell^{+}\ell^{-}$ background is estimated
 by comparing data and Monte Carlo in the  $\tau^{+}\tau^{-}$ mass range
$15 \ {\rm GeV}/c^{2}<M_{\tau^{+}\tau^{-}}<50 \ {\rm GeV}/c^{2}$.
  Bhabha and $\textrm{WW}$  cross section uncertainties amount
to $3$\% and $1$\% respectively.
The systematic uncertainty for the exclusive selection is
dominated by the limited knowledge of the radiative background cross
section. The uncertainty on the latter is determined as the relative
difference between the simulated and the observed numbers of
$\tau^+\tau^-$ events selected with a value of $\sqrt{s^\prime_m/s}$ 
between 0.5 and  0.8,
assumed to be identical for values in excess of 0.85.

   The measured cross sections are presented in Table~\ref{tab:cross_tt} and
  Figs.~\ref{fig:lep_line_01} and \ref{fig:lep_line_085}, together with
  the SM  prediction.
   The systematic uncertainties on these measurements
  are listed in Table~\ref{tab:syst_cross_tt}.
   Table~\ref{tab:cross_diff_tt} and Fig.~\ref{fig:xdiff_t}  show
  the  differential cross sections, while the
  asymmetry results are given in Table~\ref{tab:assy_tt} and in
  Fig.~\ref{fig:asym}. Asymmetric contributions
  from the main backgrounds (Bhabha and radiative events) are significant,
% For each background contamination,
%the bias on $A_{FB}$ is $ \delta A_{FB} = C_{bkg} A_{bkg}$
%where $C_{bkg}$ and $A_{bkg}$ are respectively the percentage and the asymmetry of
%the background events.
  and the statistical error on the estimated Bhabha asymmetry
  yields the dominant
 systematic uncertainty on the $\tau^{+}\tau^{-}$ asymmetry.
\subsection{The ${\rm e}^+{\rm e}^-$ channel}

The selection of electron pairs~\cite{aleph-183} requires that the two
most energetic tracks with opposite sign in the event
satisfy the following conditions:

\begin{displaymath}
\sum_{i=1}^2 p_i > 0.30 \sqrt{s} \ \ \ \ \ {\rm and} \ \ \ \
\sum_{i=1}^2 E_i > 0.40 \sqrt{s} \ \ \ \ {\rm and} \ \ \ \
\sum_{i=1}^2 p_i + \mathcal{E}_i > 1.0 \sqrt{s}
\end{displaymath}

\noindent where $p_{i}$, $E_i$ and $\mathcal{E}_i$ are the track momentum,
the ECAL energy deposition associated to the track, and
the total calorimeter energy associated to the track (including
the ECAL and HCAL energies as well as the energy from a radiated photon),
 respectively.
The previous cuts reduce significantly the contamination
from tau and muon pairs. In addition,  events
 with both tracks identified as muons are discarded.
 Finally, the  requirement on  the invariant mass of the ${\rm e}^+{\rm e}^-$ pair 
candidate $(M_{{\rm e}^+{\rm e}^-} > 80 \ {\rm GeV}/c^2)$
 suppresses most of the residual radiative background.
The $M_{{\rm e}^+{\rm e}^-}$ distribution  for the data collected at
$\sqrt{s}$=207~GeV is displayed in Fig.~\ref{fig:mee_207}.

 The resulting
 selection efficiencies and the total background contamination are listed in 
Table~\ref{tab:eff_back_ee}.
 The background is dominated by radiative events whose normalization  is extracted
from fits to the $M_{{\rm e}^+{\rm e}^-}$  and $(\Sigma p + \Sigma \mathcal{E})$ 
 experimental distributions using the
expected shapes for the ${\rm e}^+{\rm e}^-$ signal and radiative background. 
 For both selections, $-0.9< \cos\theta^*<0.9$ and
$-0.9<\cos\theta^*<0.7$, the statistical uncertainty on the fit result contributes
 the dominant systematic uncertainty on the background estimation.

  The cross section measurements are compared to the SM prediction
 from the {\tt BHWIDE} generator in Table~\ref{tab:cross_ee} and Fig.~\ref{fig:lep_line_085}.
  The main contributions to the systematic uncertainties
  are listed in Table~\ref{tab:syst_cross_ee}. Finally,
  Table~\ref{tab:cross_diff_ee} and Fig.~\ref{fig:xdiff_e} show the measured
  differential cross section.

%% file: ze.tex
\subsection{Extra Z bosons \label{sec:zprime}}

 Several extensions to the Standard
 Model~\cite{bib-Zprime} predict the existence of at least one 
 additional neutral gauge boson Z$^{\prime}$.   
  Two classes of models are considered here: $E_6$ models
  and Left-Right (LR) models. 
  In $E_6$ models, the  Z$^{\prime}$ properties depend on the
 breaking pattern of the gauge symmetry, governed by the parameter
 $\theta_6$. Limits on the Z$^{\prime}$ mass are derived 
 here for the choices
 $\theta_6=0,\,\pi/2,\, \pm\arctan\sqrt{5/3}$, known as
 the $\chi,\,\psi,\,\eta$ and $I$ models.
 In LR models, right-handed extensions to the Standard Model gauge 
 group are introduced. The Z$^{\prime}$ couplings to fermions depend
 on the parameter $\alpha_{\rm LR}$, which is set here to
 the value $\alpha_{\rm LR}=1.53$ (as predicted in LR symmetric models). 
  More details can be found in~\cite{aleph-183}. 

 Limits on the Z$^{\prime}$ mass are derived using the method
 described in Ref.~\cite{aleph-183}.   
  The theoretical predictions for the 
  two-fermion exclusive cross-sections and asymmetries 
  are obtained from {\tt ZFITTER} 6.10 used together with {\tt ZEFIT}~\cite{bib-ZEFIT} and they
are compared to the corresponding measurements presented above.
  
The most conservative $m_{\rm Z^\prime}$ lower limits, 
with respect to the Z/Z$^\prime$ mixing angle, are presented in Table~\ref{tab-zprimlim}.
Constraints on extra gauge boson have been also set at the Tevatron~\cite{CDF,CDF2,D0}.

%% file: tab-lum_cor.tex
%Luminosity
\begin{table}[hbtp]
\caption{\small\label{tab:lum}{ Luminosity weighted centre-of-mass energies
 and integrated luminosities for the data samples presented in this paper.
  The total (statistical and systematic combined) uncertainties
 on the integrated luminosities  are given. The last column contains the
data sample names used in this paper.}} \vspace*{0.5cm}
  \begin{center}
    \begin{tabular}{|c|c|c|c|}
      \hline
      Year &  $E_{\rm cm}$ (GeV) & Luminosity (\mr{pb^{-1}}) & \\
      \hline
      1998   & 188.63  & 174.2 \plmo 0.8 & 189 \\
      1999   & 191.58  &  28.9 \plmo 0.1 & 192 \\
             & 195.52  &  79.9 \plmo 0.4 & 196 \\
             & 199.52  &  86.3 \plmo 0.4 & 200 \\
             & 201.62  &  41.9 \plmo 0.2 & 202 \\
     2000    & 204.86  &  81.60\plmo 0.4 & 205 \\
             & 206.53  & 133.6 \plmo 0.6 & 207 \\
      \hline
    \end{tabular}
  \end{center}
\end{table}

%% file: tab-had.tex
\begin{table}
\caption{\small \label{tab:effqq} {Selection efficiencies and
background fractions for the $\rm{q\overline q}$ channel  for
the inclusive and exclusive processes, as a function of the
centre-of-mass energy. The statistical uncertainties are also given.}}
\begin{center}
\begin{tabular}{|c|c|c|c|}
\hline
 $\sqrt{s^\prime/s} $ & $E_{\rm cm}$ &  Efficiency   & Background  \\
     {\rm cut}                    & (GeV)              & (\%)             & (\%) \\
\hline
    0.1  &     189     & $83.9\pm 0.2  $  &    $ 20.2\pm 0.1  $      \\
         &     192     & $83.9\pm 0.3  $  &    $ 22.2\pm 0.1  $      \\
         &     196     & $83.3\pm 0.3  $  &    $ 22.2\pm 0.1  $      \\
         &     200     & $82.5\pm 0.3  $  &    $ 22.1\pm 0.1  $      \\
         &     202     & $82.3\pm 0.3  $  &    $ 22.3\pm 0.1  $      \\
         &     205     & $81.4\pm 0.3  $  &    $ 23.3\pm 0.1  $      \\
         &     207     & $81.1\pm 0.3  $  &    $ 23.7\pm 0.1  $      \\
 \hline
  0.85                  &     189     & $81.9\pm 0.3  $  &    $ 5.82\pm 0.07  $      \\
  $|\cos\theta| < 0.95$ &     192     & $82.2\pm 0.3  $  &    $ 6.10\pm 0.08  $      \\
                        &     196     & $82.3\pm 0.3  $  &    $ 6.34\pm 0.08  $      \\
                        &     200     & $82.5\pm 0.3  $  &    $ 6.56\pm 0.08  $      \\
                        &     202     & $83.3\pm 0.3  $  &    $ 6.93\pm 0.08  $      \\
                        &     205     & $82.0\pm 0.3  $  &    $ 8.47\pm 0.09  $      \\
                        &     207     & $81.9\pm 0.3  $  &    $ 8.69\pm 0.09  $      \\
\hline
\end{tabular}
 \end{center}
\end{table}

\begin{table}
\caption{\small  \label{tab:crossqq} {Measured $\rm{q\overline q}$
cross sections for the inclusive and exclusive processes, as a
function of the centre-of-mass energy, with their statistic and
systematic uncertainties. 
 The corresponding Standard Model predictions from {\tt ZFITTER}
 are given in the last column.}}\vspace*{.5cm}
\begin{center}
\begin{tabular}{| c | c |  c |  c | c |}

\hline
 $\sqrt{s^\prime/s}  $   &  ${E_{\rm cm}}$ & Number & $\sigma_{\rm{q\overline q}}$        & SM prediction \\
         {\rm cut}        & (GeV)             & of events & (pb)             & (pb)  \\
\hline
      0.1                & 189     & 18617 & $101.65\pm 0.83\pm 0.83$  & 99.35 \\
                         & 192     & 2898  & $93.00\pm 1.95\pm 0.73$   & 95.41 \\
                         & 196     & 7776  & $90.89\pm 1.17\pm 0.66$   & 90.55 \\
                         & 200     & 8102  & $88.65\pm 1.12\pm 0.63$   & 86.02 \\
                         & 202     & 3710  & $83.59\pm 1.56\pm 0.63$   & 83.78 \\
                         & 205     & 6989  & $80.71\pm 1.12\pm 0.46$   & 80.53 \\
                         & 207     & 11183 & $79.16\pm 0.85\pm 0.43$   & 78.94 \\
\hline
       0.85              & 189     & 3153  & $20.80\pm 0.38\pm 0.17$   & 20.58 \\
  $|\cos\theta| < 0.95$  & 192     & 508   & $20.07\pm 0.92\pm 0.16$   & 19.72 \\
                         & 196     & 1329  & $18.93\pm 0.54\pm 0.16$   & 18.67 \\
                         & 200     & 1367  & $17.94\pm 0.51\pm 0.16$   & 17.69 \\
                         & 202     & 658   & $17.56\pm 0.71\pm 0.15$   & 17.21 \\
                         & 205     & 1238  & $16.94\pm 0.52\pm 0.15$   & 16.51 \\
                         & 207     & 1958  & $16.34\pm 0.38\pm 0.14$   & 16.16 \\
\hline
\end{tabular}
\end{center}

\end{table}
\begin{table}
\caption{\small \label{tab:error}
{Contributions to the systematic uncertainties (in \%)
 on the measured $\rm{q\overline q}$ cross sections, averaged among the centre-of-mass energies.}}
\begin{center}
\begin{tabular}{|l|cc|}
\hline

  Source   & \multicolumn{2}{c|}{$\sqrt{s^\prime/s}$ cut} \\
                       & 0.1  & 0.85 \\

\hline

             MC statistics                     & 0.30    & 0.30 \\
             Energy scale                      & 0.36    & 0.30 \\
             Detector response                 & 0.38    & 0.60 \\
  \rm ${\gamma\gamma\to\rm{q\overline q}}$     & 0.05    &      \\
             WW                                & 0.19    & 0.05 \\
             Radiative background              &         & 0.21 \\
             Other 4-f backgrounds             &         & 0.03 \\
             Luminosity                        & 0.45    & 0.45 \\
\cline{1-3}
             Total                             & 0.78    & 0.89 \\

\hline
\end{tabular}

\end{center}
\end{table}

%had_table_2
%\begin{table}
%\caption{\small  \label{tab:cross-corr} { Cross sections for all
%recorded energies, together with their statistical and systematic
%errors. The exclusive processes  are defined in the angular range
%$|\cos\theta| < 0.95 $. The errors are splitted into 6 terms
%corresponding to : statistical errors, correlated errors between
%years, channels, between experiments, between channels and years
%and uncorrelated errors.}} \vspace*{.5cm}
%\begin{center}
%\begin{tabular}{ c   c   c   c   c   c   c   c }
%\hline
%  $\mathrm{E_{cm}}$ &  Statistical & Correlated & Correlated & Correlated & Correlated &uncorrelated \\
%  GeV &  error & between & between & between & between & \\
%   & & years & channels & experiments & everything & \\
%   & pb & pb & pb & pb & pb & pb \\
%  \hline
%         189  &      0.380  &  0.156 &  0.108 & 0.021 & 0.052 &  0.021\\
%         192  &      0.920 &  0.151 & 0.155 & 0.020 & 0.050 & 0.040\\
%         196  &      0.540 &  0.144 & 0.115 &  0.015 &  0.047 &  0.038\\
%         200  &      0.510 &  0.138 & 0.113 &  0.014 & 0.045 &  0.036 \\
%         202  &      0.710 &  0.137 & 0.133 &  0.012 & 0.044 & 0.035\\
%         205  &      0.520  & 0.129 & 0.100 & 0.012 &  0.042 &  0.034\\
%         207  &      0.380  & 0.124 & 0.087 &  0.011 &  0.041 &  0.033\\
%\hline
%\end{tabular}
%\end{center}

%\end{table}

%% file: tab-leptons.tex
% Mu efficiency
\begin{table}
\centering
  \caption{\small\label{tab:eff_back_mm}{Selection efficiencies and background fractions for
  the \mr {\mu^+\mu^-} channel for the inclusive and exclusive processes. The statistical uncertainties
  are also given.}}
  \vspace*{0.5cm}
  \begin{center}
    \begin{tabular}{|c|c|c|c|}
      \hline
      $\sqrt{s^\prime/s}$ & ${E_{\rm cm}}$ & Efficiency  & Background  \\
         {\rm cut}  &     \mr{(GeV)}    &   (\mr{\%})   &   (\mr{\%})   \\

      \hline
          0.1          &       189         & $74.7\pm 0.2$  & $8.2\pm0.5$  \\
                       &       192         & $75.1\pm 0.2$  & $7.6\pm0.6$   \\
                       &       196         & $74.6\pm 0.1$  & $9.0\pm0.7$   \\
                       &       200         & $74.1\pm 0.1$  & $10.4\pm0.8$  \\
                       &       202         & $74.0\pm 0.2$  & $10.0\pm0.8$  \\
                       &       205         & $73.5\pm 0.1$  & $11.1\pm0.9$  \\
                       &       207         & $73.1\pm 0.1$  & $11.8\pm1.0$  \\
      \hline
          0.85         &       189         & $95.8\pm 0.1$  & $1.8\pm0.1$  \\
 $|\cos\theta| < 0.95$ &       192         & $96.1\pm 0.1$  & $1.9\pm0.1$  \\
                       &       196         & $96.1\pm 0.1$  & $1.9\pm0.1$  \\
                       &       200         & $96.1\pm 0.1$  & $1.9\pm0.1$  \\
                       &       202         & $96.2\pm 0.1$  & $1.9\pm0.1$  \\
                       &       205         & $96.1\pm 0.1$  & $2.1\pm0.1$  \\
                       &       207         & $95.9\pm 0.12$  & $1.9\pm0.1$  \\

      \hline
    \end{tabular}
  \end{center}
\end{table}

%
%
%Mu x-sec
\begin{table}
\centering
  \caption{\small\label{tab:cross_mm}{Measured \mr {\mu^+\mu^-} cross sections
   for the inclusive and exclusive processes. 
  The numbers of selected events and
  the predicted SM cross sections from {\tt ZFITTER} are also given.} }
  \vspace*{0.5cm}
  \begin{center}
    \begin{tabular}{|c|c|c|c|c|}
      \hline
      $\sqrt{s^\prime/s} $    &  ${E_{\rm cm}}$ & Number  & $\sigma_{\rm {\mu^+\mu^-}}$   & \mr{SM} \mr{prediction} \\
       {\rm cut}         &  \mr{(GeV)}  & of events &    \mr{(pb)}        &         \mr{(pb)}           \\

      \hline
            0.1              &   189    & 1090  & $7.68\pm 0.26\pm0.06$  & $7.78$ \\
                             &   192    & 189   & $8.04\pm 0.66\pm0.07$  & $7.50$ \\
                             &   196    & 493   & $7.53\pm 0.39\pm0.07$  & $7.15$ \\
                             &   200    & 489   & $6.85\pm 0.36\pm0.07$  & $6.83$ \\
                             &   202    & 238   & $6.92\pm 0.52\pm0.07$  & $6.66$ \\
                             &   205    & 402   & $5.96\pm 0.35\pm0.06$  & $6.43$ \\
                             &   207    & 683   & $6.16\pm 0.28\pm0.07$  & $6.31$ \\
      \hline
           0.85              &   189    & 489   & $2.88\pm 0.13\pm0.02$  & $2.83$ \\
 $|\cos\theta| < 0.95$       &   192    & 81    & $2.86\pm 0.33\pm0.02$  & $2.73$ \\
                             &   196    & 211   & $2.70\pm 0.19\pm0.02$  & $2.61$ \\
                             &   200    & 252   & $2.99\pm 0.20\pm0.02$  & $2.50$ \\
                             &   202    & 107   & $2.64\pm 0.26\pm0.02$  & $2.44$ \\
                             &   205    & 154   & $1.92\pm 0.16\pm0.02$  & $2.36$ \\
                             &   207    & 321   & $2.46\pm 0.14\pm0.02$  & $2.32$ \\
      \hline
    \end{tabular}
  \end{center}
\end{table}
\begin{table}
\centering
  \caption{\small\label{tab:syst_cross_mm}{Contributions to the systematic uncertainties (in \%)
   on the measured \mr {\mu^+\mu^-} cross sections, averaged among the centre-of-mass energies.}}
  \vspace*{0.5cm}
  \begin{center}
    \begin{tabular}{|l|cc|}
      \hline
       Source      &        \multicolumn{2}{c|}{$\sqrt{s^\prime/s}$ cut} \\
                   &            0.1 & 0.85 \\
      \hline
             MC statistics              & 0.76 & 0.17 \\
             Muon identification        & 0.20 & 0.20 \\
             Background contamination   & 0.20 & 0.53 \\
             Luminosity                 & 0.45 & 0.45 \\
\cline{1-3}
             Total                      & 0.93 & 0.75 \\

\hline
    \end{tabular}

  \end{center}
\end{table}
\begin{table}
\centering
  \caption{\small\label{tab:cross_diff_mm}{Measured differential cross section (pb) 
   for the \mr {\mu^+\mu^-}
  channel for $\sqrt{s^\prime/s}>0.85$, as a function of the 
  polar angle of the negative muon with respect to the incoming electron. Statistical and systematic uncertainties are
combined.} }
  \vspace*{0.5cm}
  \begin{center}
    \begin{tabular}{|c|ccccccc|}
      \hline
                                                      &\multicolumn{7}{c|} {${E_{\rm cm}}$ (GeV) }  \\
    $\cos\theta$               & 189          & 192          & 196          & 200          & 202    & 205       & 207 \\
                         &              & &&&&& \\

      \hline
$[-0.95,-0.80]$                    &$0.7\pm0.1$ &$0.2\pm0.3$ &$0.5\pm0.2$ &$0.7\pm0.2$ &$0.8\pm0.3$ &$0.3\pm0.2$ &$0.5\pm0.2$\\
$[-0.80,-0.60]$                    &$0.3\pm0.1$ &$0.2\pm0.3$ &$0.6\pm0.2$ &$0.6\pm0.3$ &$0.1\pm0.2$ &$0.3\pm0.2$ &$0.4\pm0.1$\\
$[-0.60,-0.40]$                    &$0.4\pm0.1$ &$1.1\pm0.3$ &$0.5\pm0.2$ &$0.9\pm0.2$ &$0.9\pm0.2$ &$0.2\pm0.2$ &$0.2\pm0.1$\\
$[-0.40,-0.20]$                    &$0.8\pm0.2$ &$0.7\pm0.4$ &$0.6\pm0.2$ &$1.0\pm0.2$ &$0.5\pm0.3$ &$0.3\pm0.2$ &$1.0\pm0.2$\\
$[-0.20,\,\,\,\,\,0.00]$           &$1.1\pm0.2$ &$0.9\pm0.4$ &$0.9\pm0.2$ &$1.2\pm0.2$ &$0.7\pm0.3$ &$1.0\pm0.2$ &$0.6\pm0.2$\\
$[\,\,\,\,\,0.00, \,\,\,\,\,0.20]$ &$1.1\pm0.2$ &$1.8\pm0.5$ &$1.1\pm0.3$ &$1.5\pm0.3$ &$1.2\pm0.4$ &$1.1\pm0.3$ &$1.0\pm0.2$\\
$[\,\,\,\,\,0.20,\,\,\,\,\,0.40]$  &$2.1\pm0.2$ &$0.9\pm0.6$ &$1.9\pm0.3$ &$1.9\pm0.3$ &$1.7\pm0.4$ &$1.2\pm0.3$ &$1.7\pm0.2$\\
$[\,\,\,\,\,0.40,\,\,\,\,\,0.60]$  &$2.0\pm0.3$ &$1.9\pm0.6$ &$2.4\pm0.4$ &$1.8\pm0.3$ &$2.4\pm0.5$ &$1.7\pm0.3$ &$1.9\pm0.3$\\
$[\,\,\,\,\,0.60,\,\,\,\,\,0.80]$  &$3.3\pm0.3$ &$3.8\pm0.7$ &$2.8\pm0.4$ &$3.1\pm0.4$ &$2.1\pm0.6$ &$1.8\pm0.4$ &$2.9\pm0.3$\\
$[\,\,\,\,\,0.80,\,\,\,\,\,0.95]$  &$3.8\pm0.4$ &$3.8\pm0.9$ &$3.2\pm0.5$ &$3.4\pm0.5$ &$3.8\pm0.7$ &$2.4\pm0.5$ &$2.9\pm0.4$\\
      \hline
    \end{tabular}
  \end{center}
\end{table}
\begin{table}
\centering
  \caption{\small\label{tab:assy_mm}{ Measured \mr {\mu^+\mu^-} forward-backward asymmetry
   for $\sqrt{s^\prime/s}>0.85$ and in the range $|\cos\theta|<0.95$.
  The SM prediction from {\tt ZFITTER} is given in the last column.} }
  \vspace*{0.5cm}
  \begin{center}
    \begin{tabular}{|c|c|c|}
      \hline
            ${E_{\rm cm}}$ (GeV)     & $ A_{\rm FB}^{\rm {\mu^+\mu^-}}$   & \mr{SM} \mr{prediction} \\

      \hline
               189           & $0.576\pm 0.036\pm0.009$      & $0.570$ \\
               192           & $0.580\pm 0.090\pm0.009$      & $0.567$ \\
               196           & $0.553\pm 0.057\pm0.006$      & $0.563$ \\
               200           & $0.442\pm 0.056\pm0.006$      & $0.560$ \\
               202           & $0.573\pm 0.078\pm0.010$      & $0.558$ \\
               205           & $0.572\pm 0.066\pm0.010$      & $0.555$ \\
               207           & $0.570\pm 0.046\pm0.007$      & $0.554$ \\
      \hline
    \end{tabular}
  \end{center}
\end{table}
%%%%%%%%%%%%%%%%%%%%%%%
%%%%%%%%%%%%%%%%%%%%%%%
%%%%%%%%%%%%%%%%%%%%%%%
\begin{table}
\centering
  \caption{\small\label{tab:eff_back_tt}{ Selection efficiencies and background
  fractions for
  the \mr {\tau^+\tau^-} channel for the inclusive and exclusive processes. 
  The statistical uncertainties are also given.}}
  \vspace*{0.5cm}
  \begin{center}
    \begin{tabular}{|c|c|c|c|}
      \hline
      $\sqrt{s^\prime/s}$ & ${E_{\rm cm}}$ & Efficiency  & Background  \\
       {\rm cut}     &     \mr{(GeV)}    &   (\mr{\%})   &   (\mr{\%})   \\

      \hline
          0.1          &       189         & $42.6\pm 0.2$  &  $12.4\pm0.5$ \\
                       &       192         & $42.0\pm 0.2$  &  $12.9\pm0.4$ \\
                       &       196         & $41.7\pm 0.2$  &  $14.4\pm0.5$ \\
                       &       200         & $41.5\pm 0.2$  &  $13.9\pm0.5$ \\
                       &       202         & $41.4\pm 0.2$  &  $15.9\pm0.6$ \\
                       &       205         & $41.2\pm 0.2$  &  $15.5\pm0.5$ \\
                       &       207         & $41.0\pm 0.2$  &  $17.7\pm0.6$ \\
      \hline
          0.85         &       189         & $63.9\pm 0.3$   & $14.7\pm0.6$  \\
$|\cos\theta| < 0.95$  &       192         & $63.8\pm 0.3$   & $15.4\pm0.6$  \\
                       &       196         & $63.4\pm 0.3$   & $15.2\pm0.7$  \\
                       &       200         & $63.5\pm 0.3$   & $13.2\pm0.6$  \\
                       &       202         & $63.5\pm 0.3$   & $14.5\pm0.7$  \\
                       &       205         & $63.2\pm 0.3$   & $16.2\pm0.7$  \\
                       &       207         & $62.8\pm 0.3$   & $18.3\pm0.9$ \\

      \hline
    \end{tabular}
  \end{center}
\end{table}
\begin{table}

\centering
  \caption{\small\label{tab:cross_tt}{
  Measured \mr {\tau^+\tau^-} cross sections
   for the inclusive and exclusive processes.
  The numbers of selected events and
  the predicted SM cross sections from {\tt ZFITTER} are also given.}}
  \vspace*{0.5cm}
  \begin{center}
    \begin{tabular}{|c|c|c|c|c|}
      \hline
      $\sqrt{s^\prime/s}$    &  ${E_{\rm cm}}$ & Number & $\sigma_{\rm {\tau^+\tau^-}}$   & \mr{SM} \mr{prediction} \\
         {\rm cut}            &  \mr{(GeV)}  & of events &    (\mr{pb})                &         (\mr{pb})           \\

      \hline
            0.1              &  189    & 642  & $7.56\pm 0.36\pm0.12$  & $7.77$ \\
                             &  192    & 114  & $8.16\pm 0.93\pm0.15$  & $7.49$ \\
                             &  196    & 263  & $6.75\pm 0.52\pm0.14$  & $7.2$ \\
                             &  200    & 295  & $7.09\pm 0.51\pm0.11$  & $6.82$ \\
                             &  202    & 129  & $6.24\pm 0.70\pm0.09$  & $6.66$ \\
                             &  205    & 246  & $6.19\pm 0.50\pm0.07$  & $6.43$ \\
                             &  207    & 402  & $6.05\pm 0.39\pm0.09$  & $6.31$ \\
\hline
            0.85             &  189    & 356  & $2.79\pm 0.20\pm0.05$  & $2.91$ \\
  $|\cos\theta| < 0.95$      &  192    & 59   & $2.60\pm 0.47\pm0.07$  & $2.81$ \\
                             &  196    & 158  & $2.55\pm 0.29\pm0.07$  & $2.69$ \\
                             &  200    & 184  & $2.88\pm 0.29\pm0.07$  & $2.57$ \\
                             &  202    & 85   & $2.83\pm 0.41\pm0.04$  & $2.51$ \\
                             &  205    & 149  & $2.43\pm 0.29\pm0.04$  & $2.43$ \\
                             &  207    & 220  & $2.10\pm 0.21\pm0.04$  & $2.38$ \\
      \hline
    \end{tabular}
  \end{center}
\end{table}
\begin{table}
\centering
  \caption{\small\label{tab:syst_cross_tt}{
   Contributions to the systematic uncertainties (in \%)
   on the measured \mr {\tau^+\tau^-} cross sections, averaged among the centre-of-mass energies.}}
  \vspace*{0.5cm}
  \begin{center}
    \begin{tabular}{|l|cc|}
      \hline
        Source         & \multicolumn{2}{c|}{$\sqrt{s^\prime/s}$ cut} \\
                       & 0.1 & 0.85  \\
      \hline
            MC statistics             & 0.65  & 0.79 \\
            Detector response         & 1.37  & 1.61 \\
            Background contamination  & 0.36  & 0.29 \\
            Luminosity                & 0.45  & 0.45 \\
\cline{1-3}
            Total                     & 1.65  & 1.90 \\

      \hline
    \end{tabular}
  \end{center}
\end{table}
\begin{table}
\centering
  \caption{\small\label{tab:cross_diff_tt}{Measured differential cross section (pb) 
  for the \mr {\tau^+\tau^-}
  channel for $\sqrt{s^\prime/s}>0.85$, as a function of the polar angle of the negative
   tau with respect to the incoming electron. Statistical and systematic uncertainties are combined.} }
  \vspace*{0.5cm}
  \begin{center}
    \begin{tabular}{|c|ccccccc|}
          \hline
                                                      &\multicolumn{7}{c|} {${E_{\rm cm}}$ (GeV) } \\
       $\cos\theta$            & 189          & 192          & 196          & 200          & 202    & 205       & 207 \\
                         &              & &&&&& \\
      \hline
$[-0.95,-0.80]$                   &$0.2\pm0.2$ &$-0.2\pm0.4$ &$0.8\pm0.3$ &$0.6\pm0.2$ &$1.0\pm0.3$ &$0.9\pm0.3$ &$0.2\pm0.2$\\
$[-0.80,-0.60]$                   &$0.4\pm0.2$ &$0.7\pm0.3$  &$0.8\pm0.2$ &$0.9\pm0.2$ &$0.1\pm0.3$ &$0.5\pm0.2$ &$0.4\pm0.2$\\
$[-0.60,-0.40]$                   &$0.1\pm0.2$ &$0.2\pm0.4$  &$0.0\pm0.2$ &$0.6\pm0.2$ &$0.3\pm0.3$ &$0.4\pm0.2$ &$0.4\pm0.2$\\
$[-0.40,-0.20]$                   &$0.9\pm0.2$ &$1.3\pm0.4$  &$0.5\pm0.3$ &$0.6\pm0.2$ &$0.5\pm0.3$ &$0.3\pm0.2$ &$0.6\pm0.2$\\
$[-0.20,\,\,\,\,\,0.00]$          &$0.9\pm0.2$ &$1.1\pm0.5$  &$0.9\pm0.3$ &$1.4\pm0.3$ &$0.7\pm0.4$ &$0.3\pm0.3$ &$0.3\pm0.2$\\
$[\,\,\,\,\,0.00,\,\,\,\,\,0.20]$ &$0.8\pm0.3$ &$1.7\pm0.6$  &$0.8\pm0.4$ &$1.0\pm0.3$ &$0.9\pm0.5$ &$1.3\pm0.3$ &$1.5\pm0.3$\\
$[\,\,\,\,\,0.20,\,\,\,\,\,0.40]$ &$2.0\pm0.3$ &$1.9\pm0.7$  &$2.3\pm0.4$ &$1.6\pm0.4$ &$2.1\pm0.5$ &$1.8\pm0.4$ &$1.1\pm0.3$\\
$[\,\,\,\,\,0.40,\,\,\,\,\,0.60]$ &$2.4\pm0.3$ &$2.6\pm0.8$  &$1.8\pm0.5$ &$2.0\pm0.4$ &$1.0\pm0.6$ &$1.5\pm0.4$ &$1.4\pm0.3$\\
$[\,\,\,\,\,0.60,\,\,\,\,\,0.80]$ &$2.3\pm0.4$ &$2.1\pm0.9$  &$2.5\pm0.5$ &$3.0\pm0.5$ &$4.8\pm0.7$ &$1.4\pm0.5$ &$1.6\pm0.4$\\
$[\,\,\,\,\,0.80,\,\,\,\,\,0.95]$ &$5.3\pm0.6$ &$2.1\pm1.3$  &$3.5\pm0.8$ &$3.9\pm0.7$ &$4.2\pm1.0$ &$5.3\pm0.7$ &$4.0\pm0.6$\\

      \hline
    \end{tabular}
  \end{center}
\end{table}
\begin{table}
\centering
  \caption{\small\label{tab:assy_tt}{Measured
  \mr {\tau^+\tau^-} forward-backward asymmetry
   for $\sqrt{s^\prime/s}>0.85$ and in the range $|\cos\theta|<0.95$.
  The SM prediction from {\tt ZFITTER} is also given in the last column.} }
  \vspace*{0.5cm}
  \begin{center}
    \begin{tabular}{|c|c|c|}
      \hline
          ${E_{\rm cm}}$  (GeV)      & $ A_{\rm FB}^{\rm{\tau^+\tau^-}}$      & \mr{SM} \mr{prediction} \\
     \hline
            189              & $0.598\pm 0.046\pm0.012$      & $0.570$\\
            192              & $0.489\pm 0.124\pm0.010$      & $0.567$\\
            196              & $0.543\pm 0.075\pm0.011$      & $0.563$\\
            200              & $0.445\pm 0.073\pm0.010$      & $0.560$\\
            202              & $0.654\pm 0.090\pm0.013$      & $0.557$\\
            205              & $0.593\pm 0.075\pm0.012$      & $0.555$\\
            208              & $0.568\pm 0.062\pm0.012$      & $0.554$\\
      \hline
    \end{tabular}
  \end{center}
\end{table}
%%%%%%%%%%%%%%%%%%%%%%%%%%%%%%%
%%%%%%%%%%%%%%%%%%%%%%%%%%%%%%%
%%%%%%%%%%%%%%%%%%%%%%%%%%%%%%%
\begin{table}
\centering
  \caption{\small\label{tab:eff_back_ee}{
  Selection efficiencies and background fractions for
  the \mr {e^+e^-} exclusive channel for two angular ranges.
  The statistical uncertainties are also given.}}
  \vspace*{0.5cm}
  \begin{center}
    \begin{tabular}{|c|c|c|c|}
      \hline
    $ \cos\theta^*$ & ${E_{\rm cm}}$ & Efficiency  & Background  \\
                 &     \mr{(GeV)}    &   (\mr{\%})   &   (\mr{\%})   \\

      \hline
$[-0.9,0.9]$            &       189         & $84.1\pm 0.3$  & $7.2\pm0.2$ \\
                        &       192         & $85.7\pm 0.3$  & $6.9\pm0.2$ \\
                        &       196         & $85.7\pm 0.3$  & $5.9\pm0.2$ \\
                        &       200         & $86.3\pm 0.2$  & $7.3\pm0.2$ \\
                        &       202         & $86.5\pm 0.2$  & $8.3\pm0.2$ \\
                        &       205         & $86.8\pm 0.2$  & $6.5\pm0.2$ \\
                        &       207         & $87.2\pm 0.2$  & $6.9\pm0.2$ \\
      \hline
$[-0.9,0.7]$            &       189         & $92.4\pm 0.4$  & $8.8\pm0.5$   \\
                        &       192         & $92.8\pm 0.4$  & $8.5\pm0.5 $ \\
                        &       196         & $92.8\pm 0.4$  & $6.7\pm0.4$   \\
                        &       200         & $93.9\pm 0.4$  & $8.6\pm0.5$   \\
                        &       202         & $94.4\pm 0.4$  & $9.7\pm0.5$   \\
                        &       205         & $93.8\pm 0.4$  & $8.1\pm0.5$   \\
                        &       207         & $93.8\pm 0.4$  & $8.7\pm0.5$   \\

      \hline
    \end{tabular}
  \end{center}
\end{table}
\begin{table}
\centering
  \caption{\small\label{tab:cross_ee}{Measured \mr {e^+e^-} exclusive cross sections
   over two angular ranges.
  The numbers of selected events and
  the predicted SM cross sections from {\tt BHWIDE} are also given.}}
  \vspace*{0.5cm}
  \begin{center}
    \begin{tabular}{|c|c|c|c|c|}
      \hline
  $ \cos\theta^*$   &  ${E_{\rm cm}}$ & Number & $\sigma_{\rm {e^+e^-}}$   & \mr{SM} \mr{prediction} \\
               &  \mr{(GeV)}  & of events &    (\mr{pb})        &         (\mr{pb})           \\

      \hline
$[-0.9,0.9]$               &  189     & 14473  & $91.7\pm0.9\pm 0.6$  & $94.8\pm0.9 $ \\
                           &  192     & 2321   & $87.4\pm2.0\pm0.8$   & $91.6\pm0.9 $ \\
                           &  196     & 6416   & $87.3\pm 1.2\pm0.9$  & $88.2\pm0.8 $ \\
                           &  200     & 6596   & $81.9\pm 1.1\pm0.7$  & $83.9\pm0.8 $ \\
                           &  202     & 3238   & $82.6\pm 1.6\pm0.6$  & $82.3\pm0.8 $ \\
                           &  205     & 6226   & $81.9\pm 1.2\pm0.7$  & $79.6\pm0.7 $ \\
                           &  207     & 10030  & $79.5\pm 0.9\pm0.6$  & $78.7\pm0.7 $ \\
      \hline
$[-0.9,0.7] $              &  189     & 3286   & $18.6\pm 0.4\pm0.2$  & $19.2\pm0.3 $ \\
                           &  192     & 504    & $17.1\pm 0.9\pm0.2$  & $18.3\pm0.3 $ \\
                           &  196     & 1482   & $18.4\pm 0.5\pm0.2$  & $18.0\pm0.3 $ \\
                           &  200     & 1468   & $16.4\pm 0.5\pm0.2$  & $17.4\pm0.3 $ \\
                           &  202     & 742    & $17.1\pm 0.7\pm0.2$  & $16.9\pm0.3 $ \\
                           &  205     & 1358   & $16.1\pm 0.5\pm0.2$  & $15.8\pm0.3 $ \\
                           &  207     & 2262   & $16.1\pm 0.4\pm0.2$  & $16.0\pm0.3 $ \\
      \hline
    \end{tabular}
  \end{center}
\end{table}
\begin{table}
\centering
  \caption{\small\label{tab:syst_cross_ee}{
   Contributions to the systematic uncertainties (in \%)
   on the measured \mr {e^+e^-} cross sections, averaged among the centre-of-mass energies.} }
  \vspace*{0.5cm}
  \begin{center}
    \begin{tabular}{|l|cc|}

      \hline
      Source         & \multicolumn{2}{c|}{$\cos\theta^*$ } \\
                     & $[-0.9,0.9]$ & $[-0.9,0.7]$ \\
      \hline
    MC statistics              & 0.33   & 0.61 \\
     Detector response         & 0.36   & 0.15 \\
     Background contamination  & 0.23   & 0.27 \\
     Luminosity                & 0.46   & 0.46  \\
\cline{1-3}
    Total                      & 0.71   & 0.82 \\

\hline
    \end{tabular}
  \end{center}
\end{table}
{\small{
\begin{table}
\centering
  \caption{\small\label{tab:cross_diff_ee}{Measured
  differential cross section (pb) for the \mr {e^+e^-}
  channel for $\sqrt{s^\prime/s}>0.85$, as a function of polar angle. Statistical and systematic uncertainties are combined.} }
  \vspace*{0.5cm}
  \begin{center}
    \begin{tabular}{|c|ccccccc|}
      \hline
                                                          &\multicolumn{7}{c|}   {${E_{\rm cm}}$ (GeV) } \\
      $\cos\theta^*$           & 189          & 192          & 196          & 200          & 202    & 205       & 207 \\
                         &              & &&&&& \\
      \hline
$[-0.95,-0.80]$                   & $0.7\pm0.2$  & $0.5\pm0.4$   & $2.0\pm0.5$  & $0.8\pm0.3$   & $0.7\pm0.4$   & $1.1\pm0.3$  & $1.0\pm0.3$\\
$[-0.80,-0.60]$                   & $1.5\pm0.2$  & $1.8\pm0.7$   & $1.3\pm0.4$  & $1.3\pm0.3$   & $1.3\pm0.5$   & $1.0\pm0.3$  & $1.1\pm0.2$\\
$[-0.60,-0.40]$                   & $1.9\pm0.3$  & $2.0\pm0.6$   & $2.1\pm0.4$  & $1.5\pm0.3$   & $1.3\pm0.5$   & $1.4\pm0.3$  & $1.9\pm0.3$\\
$[-0.40,-0.20]$                   & $2.3\pm0.3$  & $1.3\pm0.6$   & $2.4\pm0.5$  & $2.2\pm0.4$   & $3.0\pm0.7$   & $2.5\pm0.4$  & $1.7\pm0.3$\\
$[-0.20,\,\,\,\,\,0.00]$          & $4.2\pm0.4$  & $3.8\pm0.9$   & $3.6\pm0.5$  & $3.7\pm0.5$   & $4.1\pm0.8$   & $2.8\pm0.5$  & $3.3\pm0.4$\\
$[\,\,\,\,\,0.00,\,\,\,\,\,0.20]$ & $5.6\pm0.5$  & $4.5\pm1.0$   & $5.9\pm0.7$  & $5.8\pm0.7$   & $5.5\pm0.9$   & $5.5\pm0.6$  & $5.3\pm0.5$\\
$[\,\,\,\,\,0.20,\,\,\,\,\,0.40]$ & $11.8\pm0.7$ & $11.5\pm1.6$  & $12.8\pm1.0$ & $8.8\pm0.8$   & $12.1\pm1.4$  & $9.0\pm0.9$  & $8.7\pm0.7$\\
$[\,\,\,\,\,0.40,\,\,\,\,\,0.60]$ & $30\pm1$ & $27\pm2$  & $27\pm1$ & $26\pm1$  & $26\pm2$  & $26\pm1$ & $28\pm1$\\
$[\,\,\,\,\,0.60,\,\,\,\,\,0.80]$ & $120\pm2$& $112\pm5$ & $112\pm3$& $104\pm2$ & $109\pm4$ & $105\pm3$& $100\pm2$\\
$[\,\,\,\,\,0.80,\,\,\,\,\,0.95]$ & $374\pm5$& $362\pm11$& $355\pm7$& $339\pm6$ & $331\pm9$ & $338\pm6$ &$328\pm5$\\
\hline
    \end{tabular}
  \end{center}
\end{table}
}}

%% file: tab-had-events.tex
%hadronic events
\begin{table}
\centering
 \caption{{\small {Numbers of selected hadronic events
for the heavy-quark measurements.}}}\label{tab:had-events} \vspace*{0.5cm}
  \begin{center}
    \begin{tabular}{|c|c|}
      \hline
      ${E_{\rm cm}}$ (GeV) & Number of events  \\
      \hline
      189   &  2952 \\
      192   & 485 \\
      196   & 1256 \\
      200   & 1279 \\
      202   & 611 \\
      205   & 1128 \\
      207   & 1814 \\
      \hline
    \end{tabular}
  \end{center}
\end{table}

%% file: tab-bmod.tex
%b modeling
\begin{table}
\centering
\caption{\small Uncertainties on b-quark physics
 modeling~\cite{pdg,lep-hf,aleph-cmod}.
\label{tab:bcmod1}} \vspace*{0.5cm}
\begin{center}
\begin{tabular}{|l|c|}
      \hline
      Source & Uncertainty (\%)        \\
      \hline
        B-hadron fractions:    &       \\
        \mr{B^+}               & 3.3    \\
        \mr{B^0}               & 3.3    \\
        \mr{B_{s}^{+}}         & 13.1   \\
        \mr{\Lambda_{b}}       & 17.2   \\
        \hline
        Semileptonic decays   & 8.0    \\
        \hline
${\rm B}^0/\overline{{\rm B}^0}$ mixing & 5.0    \\
        \hline
        Multiplicity           & 1.2   \\
        \hline
        Lifetime:              &       \\
        \mr{B^+}               & 1.7   \\
        \mr{B^0}               & 2.1   \\
        \mr{B_{s}^{+}}         & 4.1   \\
        \mr{\Lambda_{b}}       & 6.5   \\
      \hline
    \end{tabular}
  \end{center}
  \end{table}

%% file: tab-cmod.tex
%c modeling 
\begin{table}
\centering
\caption{\small Uncertainties on c-quark physics
 modeling~\cite{pdg,lep-hf,aleph-cmod}.
\label{tab:bcmod2}} \vspace*{0.5cm}
\begin{center}
\begin{tabular}{|l|c|}
      \hline
  Source  &     Uncertainty (\%)\\
      \hline
      D-hadron fractions:   &  \\
\mr{D^+}      & 10.2 \\
\mr{D^*}      & 6.5 \\
\mr{D^0}      & 3.9 \\
\mr{D_s^+}    & 31.0 \\
\mr{\Lambda_c}& 27.7 \\
\hline
 Multiplicity & 4.3 \\
\hline
 Lifetime:    & \\
\mr{D^+}      & 1.9 \\
\mr{D^0}      & 0.7 \\
\hline
Branching ratios: & \\
\mr{D^+ \rightarrow \ell X}      & 12.0 \\
\mr{D^0 \rightarrow \ell X}      & 11.0 \\
\mr{D^+ \rightarrow K X} & 7.5 \\
\mr{D^0 \rightarrow K X} & 11.5 \\
     \hline
    \end{tabular}
  \end{center}
      \end{table}

%% file: tab-rb-syst-comb.tex
%Rb Systematics Combined
\begin{table}[hbtp]
  \caption{\label{tab:rb-syst}{\small{Contributions to the systematic 
  uncertainties on the measured values of
  $R_{\rm b}$, averaged among the centre-of-mass energies.}}}
  \begin{center}
    \begin{tabular}{|l|c|}
      \hline
      Source & Uncertainty \\

      \hline
      b tagging: & \\
      \ \ jet reconstruction & 0.0029 \\
      \ \ track selection   & 0.0055 \\
      \ \ smearing procedure & 0.0015 \\
      \ \ hemisphere correlations & 0.0003 \\
      b physics         &0.0005 \\
      Radiative hadronic background &0.0013  \\
      udsc background   &0.0022  \\
      MC statistics     &0.0002  \\
              \hline
      Total        &0.0069  \\

      \hline
    \end{tabular}
  \end{center}
\end{table}

%% file: tab-rb-meas.tex
%Rb measurements
\begin{table}[hbtp]
\caption{\label{tab:rb-final} \small {Measured values of $R_{\rm b}$ 
(with their statistical and systematic uncertainties), as a function
 of the centre-of-mass energy, for 
 $\sqrt{s^\prime}/s>0.9$ and $|\cos\theta|<0.95$.
  The SM prediction from {\tt ZFITTER} is given in the last column.}}
  \begin{center}
    \begin{tabular}{|c|c|c|}
      \hline
      ${E_{\rm cm}}$ (GeV) &  $R_{\rm b}$  & SM prediction \\
      \hline
      189   & 0.159 \plmo 0.016 \plmo 0.007 & 0.1654 \\
      192   & 0.144 \plmo 0.027 \plmo 0.007 & 0.1649 \\
      196   & 0.148 \plmo 0.020 \plmo 0.007 & 0.1642 \\
      200   & 0.173 \plmo 0.021 \plmo 0.008 & 0.1636\\
      202   & 0.128 \plmo 0.024 \plmo 0.006 & 0.1633 \\
      205   & 0.135 \plmo 0.019 \plmo 0.006 & 0.1528 \\
      207   & 0.146 \plmo 0.016 \plmo 0.007 & 0.1526\\
      \hline
    \end{tabular}
  \end{center}
\end{table}

%% file: tab-rc-syst-comb.tex
%Rc Systematics
\begin{table}[hbtp]
  \caption{\label{tab:rc-syst}{{\small {Contributions to the systematic 
  uncertainties on the
  measured values of $R_{\rm c}$, averaged among the centre-of-mass energies. }}}}
  \begin{center}
    \begin{tabular}{|l|c|}
      \hline
      Source & Uncertainty \\

      \hline
MC statistics
      &0.0017  \\
Luminosity
      &0.0022 \\
Pre-selection
      &0.0054  \\
Detector response
      &0.0018  \\
uds correction
      & 0.0074 \\
udsc selection
      & 0.0075  \\
b rejection
      & 0.0022  \\
      c modelling :    &  \\
      hadron fractions
& 0.0004 \\
      lifetime
& 0.0001 \\
      multiplicity
& 0.0007 \\
      branching ratios
& 0.0005 \\
      \hline
      Total
      & 0.0125 \\

      \hline
    \end{tabular}
  \end{center}
\end{table}

%% file: tab-rc-meas.tex
%Afbb measurements
\begin{table}[hbtp]
\caption{\label{tab:rc-meas} {\small {Measured values of $R_{\rm c}$ 
(with their statistical and systematic uncertainties), as a function
 of the centre-of-mass energy, for $\sqrt{s^\prime}/s>0.9$ 
 and $|\cos\theta|<0.95$.
  The SM prediction from {\tt ZFITTER} is given in the last column.}}}
  \begin{center}
    \begin{tabular}{|c|c|c|}
      \hline
      ${E_{\rm cm}}$ (GeV) &  $R_{\rm c}$  & SM prediction\\
      \hline
      189   & 0.245 \plmo 0.023 \plmo 0.013 & 0.2525\\
      192   & 0.283 \plmo 0.059 \plmo 0.015 & 0.2533\\
      196   & 0.287 \plmo 0.033 \plmo 0.012 & 0.2544\\
      200   & 0.258 \plmo 0.035 \plmo 0.013 & 0.2554\\
      202   & 0.307 \plmo 0.050 \plmo 0.013 & 0.2560\\
      205   & 0.299 \plmo 0.037 \plmo 0.013 & 0.2567\\
      207   & 0.280 \plmo 0.029 \plmo 0.013 & 0.2571 \\
      \hline
    \end{tabular}
  \end{center}
\end{table}

%% file: tab-afbb-syst-comb.tex
%Afbb Systematics
\begin{table}[hbtp]
\centering
  \caption{\label{tab:afbb-syst}{\small {Contributions to the systematic
  uncertainties on the measured values of 
  ${A_{\rm FB}^{\rm b}}$  over the angular range $|\cos\theta|<0.9$,
  averaged among the centre-of-mass energies. }} }
  \vspace*{0.5cm}
  \begin{center}
    \begin{tabular}{|l|c|}
      \hline
      Source & Uncertainty \\

      \hline
      MC statistics           & 0.0064  \\
      Luminosity              & 0.0003   \\
      Pre-selection           & 0.0015 \\
      Detector response       & 0.0062  \\
      c background            & 0.0089  \\
      uds background          & 0.0015  \\
      Jet charge              & 0.0713 \\
      b modeling :           &  \\
      Hadron fractions         & 0.0026 \\
      Leptonic branching ratio & 0.0034  \\
      Multiplicity            & 0.0046  \\
      Mixing                  & 0.0030  \\
      \hline
      Total               & 0.0727 \\

      \hline
    \end{tabular}
  \end{center}
\end{table}

%% file: tab-afbb-meas.tex
%Afbb measurements
\begin{table}[hbtp]
\centering
 \caption{\label{tab:afbb-meas}{ \small {Measured values of
$A_{\rm FB}^{\rm b}$ as a function of centre-of-mass energy, 
 together with the statistical and systematic
 uncertainties, for $\sqrt{s^\prime}/s>0.9$  and $|\cos\theta|<0.95$.
 The SM prediction from {\tt ZFITTER} is given in the last column.}}}
\vspace*{0.5cm}
  \begin{center}
    \begin{tabular}{|c|c|c|}
      \hline
      ${E_{\rm cm}}$ (GeV) &  $A_{\rm FB}^{\rm b}$ & SM prediction   \\
      \hline
      189   &    0.335 \plmo 0.167 \plmo 0.066 & 0.569 \\
      192   &    0.566 \plmo 0.599 \plmo 0.108 & 0.571 \\
      196   &    0.205 \plmo 0.243 \plmo 0.041 & 0.574 \\
      200   &    0.605 \plmo 0.206 \plmo 0.116 & 0.576 \\
      202   &    0.678 \plmo 0.476 \plmo 0.139 & 0.578 \\
      205   &    0.642 \plmo 0.350 \plmo 0.079 & 0.579 \\
      207   & $-$0.263 \plmo 0.240 \plmo 0.053 & 0.580 \\
      \hline
    \end{tabular}
  \end{center}
\end{table}

%% file: tab-coef.tex
\begin{table}[hbtp]
\caption{\label{tab:qfb-coef} Coefficients of the linear constraints between the deviations
$\Delta\sigma_{\rm q}$ and $\Delta A_{\rm FB}^{\rm q}$ of the cross sections and asymmetries
from the SM, and the measured values of $\Delta=\langle Q_{\rm FB}^{\rm depl}\rangle-
\langle Q_{\rm FB}^{\rm MC}\rangle $, for the two extreme centre-of-mass energies.}
  \begin{center}
    \begin{tabular}{|c|ccccc|ccccc|}
      \hline
      ${E_{\rm cm}}$& \multicolumn{5}{c|}{ 
                ${\partial\langle Q_{\rm FB}\rangle}/{\partial\sigma_{\rm q}}$ (pb$^{-1}$)} & 
                      \multicolumn{5}{c|}{ 
                ${\partial\langle Q_{\rm FB}\rangle}/{\partial A_{\rm FB}^{\rm q}}$}\\ 
      (GeV)           & u & d & s & c & b & u & d & s & c & b \\
      \hline
      189   &  33.3 &  $-$26.6 &  $-$31.3 &  15.6 &  $-$5.6 &
              295.1 & $-$112.4 & $-$138.0 & 156.1 & $-$23.1 \\
      207   &  40.7 &  $-$31.1 &  $-$36.3 &  20.9 &  $-$6.8 &
              286.3 & $-$106.4 & $-$128.1 & 159.6 & $-$22.7 \\
      \hline
    \end{tabular}
  \end{center}
\end{table}

%% file: tab-qfb-syst-comb.tex
%Afbb Systematics
\begin{table}[hbtp]
\centering
  \caption{\label{tab:qfb-syst}{\small {Contributions to the 
  systematic uncertainties (multiplied by $10^4$) 
  on the measured values of  \Qfb,
  averaged among the centre-of-mass energies.}}}
  \vspace*{0.5cm}
  \begin{center}
    \begin{tabular}{|l|c|}
      \hline
      Source & Uncertainty $(\times 10^4)$ \\

      \hline
      MC statistics
      &2.10 \\
      Luminosity
      &0.01 \\
      Pre-selection
      &1.45 \\
      Detector response
      &0.58  \\
      b rejection
      &0.72  \\
      Jet charge
      &20.26  \\
      D fraction
      &1.26 \\
      Multiplicity
      & 1.16  \\
      \hline
      Total   & 20.52 \\

      \hline
    \end{tabular}
  \end{center}
\end{table}

%% file: tab-qfb-meas.tex
%Afbb measurements
\begin{table}[hbtp]
\centering \caption{\label{tab:qfb-meas} {\small {Measured values of
${\Delta = \langle Q_{\rm FB}^{\rm depl}\rangle - \langle Q_{\rm FB}^{\rm MC}\rangle}$ as a function of 
centre-of-mass energy.}}}
 \vspace*{0.5cm}
  \begin{center}
    \begin{tabular}{|c|c|}
     \hline
      ${E_{\rm cm}}$ (GeV) &  \mr{\Delta \times 10^{4}}  \\
      \hline
      189   &  $-$28.80  \plmo 31.58 \plmo 28.77 \\
      192   &     34.46  \plmo 81.78 \plmo 15.97 \\
      196   &     43.79  \plmo 44.74 \plmo 17.37 \\
      200   & $-$137.16  \plmo 50.47 \plmo 15.47 \\
      202   &    104.47  \plmo 67.97 \plmo 13.23 \\
      205   &      0.91  \plmo 50.83 \plmo 18.15 \\
      207   &     68.17  \plmo 38.25 \plmo 19.79 \\
      \hline
    \end{tabular}
  \end{center}
\end{table}

%% file: tab-ci-models.tex
\begin{table}
    \begin{center}
        \caption{\label{tab:ci-models} \small {Four-fermion interaction models
	considered in this paper.}}
\vspace*{0.5cm}
\begin{tabular}{|c|cccc|}
  \hline
    Model & $\eta_{\rm LL}$ & $\eta_{\rm RR}$ & $\eta_{\rm LR}$ & $\eta_{\rm RL}$ \\
    \hline
LL & 1 & 0 & 0 & 0 \\
RR & 0 & 1 & 0 & 0 \\
VV & 1 & 1& 1& 1 \\
AA &  1& 1& $-$1& $-$1 \\
LR & 0 & 0& 1& 0\\
RL & 0 & 0 & 0 & 1\\
LL+RR & 1 & 1& 0 & 0\\
LR+RL & 0 & 0 & 1& 1 \\
\hline

\end{tabular}
\end{center}
\end{table}

%% file: tab-ci-lepton.tex
% dileptons contact term
\begin{table}
\centering
  \caption{\small\label{tab:contact_lepton}{Limits on contact
  interactions coupling to
  di-lepton final states. The $68$\%~C.L. range is given for
  the fitted variable $\epsilon$,
   while $95\%$~C.L. lower limits are given for $\Lambda^{\pm}$.
  The results for the \mr {e^+e^-\to\ell^+\ell^-} process
  assume lepton universality of the contact interactions. } }
  \vspace*{0.5cm}
  \begin{center}
    \begin{tabular}{|c|c|c|c|}
      \hline
       Model & $[\epsilon^-,\epsilon^+] ({\rm TeV}^{-2})$ & $\Lambda^- ({\rm TeV}) $
       &        $\Lambda^+ ({\rm TeV})$  \\
      \hline
      \mr {e^+e^- \rightarrow e^+e^-} &    &    &    \\
          LL             &  $[-0.005,+0.038]$  & $7.0$  & $4.5$    \\
          RR             &  $[-0.005,+0.039]$  & $6.8$  & $4.4$    \\
          VV             &  $[-0.006,+0.002]$  & $12.5$ & $10.3$   \\
          AA             &  $[-0.010,+0.001]$  & $10.6$ & $8.3$   \\
          LR, RL             &  $[-0.012,+0.011]$  & $6.9$  & $6.5$    \\
         LL+RR           &  $[-0.018,+0.002]$  & $9.8$ & $6.4$    \\
         LR+RL           &  $[-0.006,+0.010]$  & $9.6$ & $9.5$    \\
      \hline
      \mr {e^+e^- \rightarrow \mu^+\mu^-} &    &    &   \\
          LL             &  $[-0.001,+0.017]$  & $9.5$  & $6.6$    \\
          RR             &  $[-0.013,+0.019]$  & $9.1$  & $6.3$    \\
          VV             &  $[-0.001,+0.007]$  & $15.9$ & $10.5$   \\
          AA             &  $[-0.002,+0.006]$  & $12.6$ & $10.5$   \\
          LR, RL             &  $[-0.210,+0.018]$  & $2.0$  & $6.1$    \\
         LL+RR           &  $[-0.001,+0.009]$  & $13.2$ & $9.0$    \\
         LR+RL           &  $[-0.002,+0.006]$  & $11.9$ & $10.1$    \\
      \hline
      \mr {e^+e^- \rightarrow \tau^+\tau^-} &    &    &  \\
          LL             &  $[-0.021,+0.001]$  & $5.8$  & $7.9$    \\
          RR             &  $[-0.024,+0.001]$  & $5.5$  & $7.6$    \\
          VV             &  $[-0.008,+0.000]$  & $9.3$ & $12.8$   \\
          AA             &  $[-0.008,+0.003]$  & $9.0$ & $10.5$   \\
          LR, RL             &  $[-0.213,+0.000]$  & $2.1$  & $6.4$    \\
         LL+RR           &  $[-0.011,+0.001]$  & $8.1$ & $10.8$    \\
         LR+RL           &  $[-0.016,+0.003]$  & $2.1$ & $8.7$    \\
      \hline
      \mr {e^+e^- \rightarrow \ell^+\ell^-} &    &    &  \\
          LL             &  $[-0.001,+0.011]$  & $10.3$  & $7.9$    \\
          RR             &  $[-0.002,+0.012]$  & $9.8$  & $7.7$    \\
          VV             &  $[-0.001,+0.003]$  & $17.1$ & $14.0$   \\
          AA             &  $[-0.001,+0.004]$  & $14.8$ & $12.2$   \\
          LR, RL             &  $[-0.006,+0.008]$  & $8.5$  & $8.2$    \\
         LL+RR           &  $[-0.001,+0.005]$  & $14.2$ & $11.0$    \\
         LR+RL           &  $[-0.003,+0.004]$  & $12.1$ & $11.5$    \\
      \hline
    \end{tabular}
  \end{center}
\end{table}

%% file: tab-ci-hadron.tex
\begin{table}
  \centering
  \caption{\small {Limits on contact interactions coupling
   to hadronic final states. The 68\%~C.L. range is given for $\epsilon$,
   while 95\%~C.L. lower limits are given for \mr{\Lambda^{\pm}}.
    The results for the \cc\ and \bb\ final states
   assume that the contact interactions affect only
   c or b quarks.}}\label{tab:ci-hadron}
  \vspace*{0.5cm}
  \begin{tabular}{|c|c|cc|cc|}
    \hline
    Model & \mr{[\epsilon^-,\epsilon^+](TeV^{-2})}
    & \mr{\Lambda^-}(TeV) & \mr{\Lambda^+}(TeV) & \mr{\Lambda^-}(TeV) & \mr{\Lambda^+}(TeV)\\
    \hline
    \mr {e^+e^- \rightarrow c\bar c } & &   &  & \multicolumn{2}{c|}{\small Including hadron measurements} \\
    LL & $[-0.036,-0.006]$ & 4.4 & 5.8   & 5.6 & 9.4 \\
    RR & $[-0.045,+0.402]$ & 3.8 & 1.5   & 4.8 & 6.9 \\
    VV & $[-0.018,-0.002]$ & 6.1 & 9.1   & 7.9 & 12.0 \\
    AA & $[-0.024,-0.005]$ & 5.4 & 8.4   & 6.5 & 11.2 \\
    LR & $[-0.026,+0.183]$ & 3.4 & 2.1   & 3.8 & 2.2 \\
    RL & $[-0.067,+0.103]$ & 2.9 & 2.5   & 3.1 & 2.7 \\
    LL+RR & $[-0.022,-0.003]$ & 5.5 & 8.7  & 7.2 & 12.2 \\
    LR+RL & $[-0.024,+0.119]$ & 3.9 & 2.6  & 4.5 & 2.7 \\
    \hline
    \mr {e^+e^- \rightarrow b\bar b } & & & & &\\
    LL    & $[-0.027,-0.007]$ & 4.9 & 9.4   & &  \\
    RR    & $[-0.130,-0.031]$ & 2.6 & 6.5   & &  \\
    VV    & $[-0.035,-0.013]$ & 4.5 & 10.9  & &  \\
    AA    & $[-0.016,-0.003]$ & 5.7 & 11.3  & &  \\
    LR    & $[-0.103,+0.033]$ & 2.8 & 3.9   & &  \\
    RL    & $[-0.032,+0.019]$ & 4.6 & 2.4   & &  \\
    LL+RR & $[-0.019,-0.005]$ & 5.8 & 11.1  & &  \\
    LR+RL & $[-0.031,+0.056]$ & 4.4 & 3.5   & & \\
    \hline
    \mr {e^+e^- \rightarrow q\bar q} &  &  & & \multicolumn{2}{c|}{\small Including heavy-flavour measurements}\\
     LL   & $[-0.011,+0.002]$  &  8.0 &  9.7   & 7.2 & 12.9   \\
     RR   & $[-0.021,+0.001]$  &  5.6 &  7.6   & 5.3 & 10.2  \\
     VV   & $[-0.008,+0.000]$  &  9.0 & 12.2  & 8.3 & 16.9   \\
     AA   & $[-0.006,+0.001]$  & 10.6 & 12.9 & 9.6 & 15.9  \\
     LR   & $[-0.004,+0.042]$  &  5.2 &  4.1  & 5.1 & 4.3   \\
     RL   & $[-0.015,+0.008]$  &  6.0 &  3.8   & 6.0 & 8.2   \\
     LL+RR & $[-0.008,+0.001]$ &  9.3 & 12.3  & 8.6 & 16.3 \\
     LR+RL & $[-0.006,+0.060]$ &  7.0 &  3.6   & 6.8 & 3.7  \\
    \hline
  \end{tabular}
\end{table}

%% file: tab-sneu-models.tex
\begin{table}
  \centering
  \small\caption{For the R-parity violating
  models considered in the analysis, and
  for each di-lepton channel,  the involved
  coupling  and the type of exchanged sneutrino
  in the $s$ or $t$ channel. }\label{tab:sneu-models}
  \begin{tabular}{|c| c| c| c|}
    \hline
    \mr{\lambda^2} & \ee & \mumu & \tautau \\
    \hline
    \mr{\lambda_{121}^2} & \mr{\tilde \nu_{\mu}}($s,t$) &  &  \\
    \mr{\lambda_{131}^2} & \mr{\tilde \nu_{\tau}} ($s,t$) &  &  \\
    \mr{\lambda_{121}\lambda_{233}}&  &  & \mr{\tilde \nu_{\mu}} ($s$) \\
    \mr{\lambda_{131}\lambda_{232}}&   & \mr{\tilde \nu_{\tau}} ($s$) & \\
    \hline
  \end{tabular}
\end{table}

%% file: tab-lq-limits.tex
% lq table
\begin{table}[htbp]
\caption{ \small \label{tab:lq-limits}{The $95\%$~C.L. lower
limits (in GeV/$c^2$) on the mass  of leptoquarks
of various species, coupling to the 
first, second or third generation of quarks with strength $g=e$.
 A dash indicates that no limit can be
set, while NA denotes leptoquarks coupling only to top quarks and hence not
visible at LEP.} }
\vspace*{0.5cm}
 \begin{center}
\begin{tabular}{|c|c|c|c|c|c|c|c|}
\hline
% & \multicolumn{7}{c}{Limits on scalar leptoquark mass (GeV/$c^2$)} \\

Quark generation & \SL{0} & \SR{0} & \SBR{0} & \SL{\half} & \SR{\half} & \SBL{\half} & \SL{1} \\
\hline
1st &  490  & 211    &   189   &  182    &  194   &   -  &  474 \\

2nd &  544  &  103 &     194 &  161   &   185  &   -  &   517\\

3rd &  NA  &  NA   &    336  &   NA  &   220  &  -   & 769  \\
\hline
%\multicolumn{8}{c}{\null}\\
% & \multicolumn{7}{c}{Limits on vector leptoquark mass (GeV/$c^2$)} \\

  & \VL{0} & \VR{0} & \VBR{0} & \VL{\half} & \VR{\half} & \VBL{\half} & \VL{1} \\
\hline
1st &  581  & 155    &    407 &  254   & 223    & 175     &  629   \\

2nd & 581   & 157    & 395     &  253   & 207    & 163    &  601   \\

3rd & 540   & 194    &   NA   & 320    &  177    &   NA   &   540  \\
\hline
\end{tabular}
\end{center}
\end{table}

%% file: tab-zprime.tex
\begin{table} [hbt] 
\caption{\small \label{tab-zprimlim}
95\,\%\,CL lower limits on the Z$^{\prime}$ mass in the five considered models.}
\begin{center} \begin{tabular} {|c|c|}
\hline
    Model        & m$_{\rm Z^\prime}$ limit~(GeV) \\
\hline % -----------------------------------------------
    $E_6 (\chi)$  &  680  \\
    $E_6 (\psi)$  &  410  \\
    $E_6 (\eta)$  &  350  \\
    $E_6 (I)$     &  510  \\
    LR symmetric  &  600  \\
\hline % ----------------------------------------------

\end{tabular}\end{center}

\end{table}